\PassOptionsToPackage{table}{xcolor}
\documentclass[AMA,STIX2COL]{WileyNJDv5}

\hfuzz=\maxdimen
\vfuzz=\maxdimen

\usepackage{etoolbox}
\usepackage{geometry}
\usepackage{enumitem}
\usepackage{alltt}
\usepackage{array}
\usepackage{fontspec}
\usepackage{tabularx}
\usepackage{booktabs} 
\usepackage{ragged2e}
\usepackage{makecell}
\usepackage{xcolor}                          
\usepackage{array}
\newcolumntype{Y}{>{\RaggedRight\arraybackslash}X}
\usepackage[font=small]{caption}
\usepackage{tikz}
\usepackage{float} 

\usepackage[most]{tcolorbox}

\usepackage{fvextra}
\usepackage{framed}
\makeatletter
\newenvironment{mint@snugshade*}[1]{%
  \def\FrameCommand##1{\hskip\@totalleftmargin
    \colorbox{#1}{##1}%
    \hskip-\linewidth \hskip-\@totalleftmargin \hskip\columnwidth}%
  \MakeFramed{\advance\hsize-\width
    \@totalleftmargin\z@ \linewidth\hsize
    \advance\labelsep\fboxsep
    \@setminipage}%
 }{\par\unskip\@minipagefalse\endMakeFramed}
\newenvironment{mintbg}[1]{%
  \setlength{\OuterFrameSep}{0pt}%
  \let\mint@tmp\FV@NumberSep
  \edef\FV@NumberSep{%
    \the\numexpr\dimexpr\mint@tmp+\number\fboxsep\relax sp\relax}%
  \medskip
  \begin{mint@snugshade*}{#1}}
 {\end{mint@snugshade*}%
  \medskip\noindent}
\makeatother
\newcommand{\mintfvset}{\fvset{
  fontsize=\fontsize{6.5}{7.5}\selectfont,
  frame=none,
  framerule=0.5pt,
  rulecolor=\color{gray},
  framesep=2mm,
  xleftmargin=0pt,
  numbersep=2mm,
  numbers=left,
  firstnumber=last,
  baselinestretch=0.95,
  breaklines=true,
  breakanywhere=true}}
\makeatletter
\def\PYG@reset{\let\PYG@it=\relax \let\PYG@bf=\relax%
    \let\PYG@ul=\relax \let\PYG@tc=\relax%
    \let\PYG@bc=\relax \let\PYG@ff=\relax}
\def\PYG@tok#1{\csname PYG@tok@#1\endcsname}
\def\PYG@toks#1+{\ifx\relax#1\empty\else%
    \PYG@tok{#1}\expandafter\PYG@toks\fi}
\def\PYG@do#1{\PYG@bc{\PYG@tc{\PYG@ul{%
    \PYG@it{\PYG@bf{\PYG@ff{#1}}}}}}}
\def\PYG#1#2{\PYG@reset\PYG@toks#1+\relax+\PYG@do{#2}}

\@namedef{PYG@tok@w}{\def\PYG@tc##1{\textcolor[rgb]{0.73,0.73,0.73}{##1}}}
\@namedef{PYG@tok@c}{\let\PYG@it=\textit\def\PYG@tc##1{\textcolor[rgb]{0.60,0.60,0.53}{##1}}}
\@namedef{PYG@tok@cp}{\let\PYG@bf=\textbf\def\PYG@tc##1{\textcolor[rgb]{0.60,0.60,0.60}{##1}}}
\@namedef{PYG@tok@cs}{\let\PYG@bf=\textbf\let\PYG@it=\textit\def\PYG@tc##1{\textcolor[rgb]{0.60,0.60,0.60}{##1}}}
\@namedef{PYG@tok@o}{\let\PYG@bf=\textbf}
\@namedef{PYG@tok@s}{\def\PYG@tc##1{\textcolor[rgb]{0.73,0.53,0.27}{##1}}}
\@namedef{PYG@tok@sr}{\def\PYG@tc##1{\textcolor[rgb]{0.50,0.50,0.00}{##1}}}
\@namedef{PYG@tok@m}{\def\PYG@tc##1{\textcolor[rgb]{0.00,0.60,0.60}{##1}}}
\@namedef{PYG@tok@k}{\let\PYG@bf=\textbf}
\@namedef{PYG@tok@kt}{\let\PYG@bf=\textbf\def\PYG@tc##1{\textcolor[rgb]{0.27,0.33,0.53}{##1}}}
\@namedef{PYG@tok@nb}{\def\PYG@tc##1{\textcolor[rgb]{0.60,0.60,0.60}{##1}}}
\@namedef{PYG@tok@nf}{\let\PYG@bf=\textbf\def\PYG@tc##1{\textcolor[rgb]{0.60,0.00,0.00}{##1}}}
\@namedef{PYG@tok@nc}{\let\PYG@bf=\textbf\def\PYG@tc##1{\textcolor[rgb]{0.27,0.33,0.53}{##1}}}
\@namedef{PYG@tok@ne}{\let\PYG@bf=\textbf\def\PYG@tc##1{\textcolor[rgb]{0.60,0.00,0.00}{##1}}}
\@namedef{PYG@tok@nn}{\def\PYG@tc##1{\textcolor[rgb]{0.33,0.33,0.33}{##1}}}
\@namedef{PYG@tok@nv}{\def\PYG@tc##1{\textcolor[rgb]{0.00,0.50,0.50}{##1}}}
\@namedef{PYG@tok@no}{\def\PYG@tc##1{\textcolor[rgb]{0.00,0.50,0.50}{##1}}}
\@namedef{PYG@tok@nt}{\def\PYG@tc##1{\textcolor[rgb]{0.00,0.00,0.50}{##1}}}
\@namedef{PYG@tok@na}{\def\PYG@tc##1{\textcolor[rgb]{0.00,0.50,0.50}{##1}}}
\@namedef{PYG@tok@ni}{\def\PYG@tc##1{\textcolor[rgb]{0.50,0.00,0.50}{##1}}}
\@namedef{PYG@tok@gh}{\def\PYG@tc##1{\textcolor[rgb]{0.60,0.60,0.60}{##1}}}
\@namedef{PYG@tok@gu}{\def\PYG@tc##1{\textcolor[rgb]{0.67,0.67,0.67}{##1}}}
\@namedef{PYG@tok@gd}{\def\PYG@tc##1{\textcolor[rgb]{0.00,0.00,0.00}{##1}}\def\PYG@bc##1{{\setlength{\fboxsep}{0pt}\colorbox[rgb]{1.00,0.87,0.87}{\strut ##1}}}}
\@namedef{PYG@tok@gi}{\def\PYG@tc##1{\textcolor[rgb]{0.00,0.00,0.00}{##1}}\def\PYG@bc##1{{\setlength{\fboxsep}{0pt}\colorbox[rgb]{0.87,1.00,0.87}{\strut ##1}}}}
\@namedef{PYG@tok@gr}{\def\PYG@tc##1{\textcolor[rgb]{0.67,0.00,0.00}{##1}}}
\@namedef{PYG@tok@ge}{\let\PYG@it=\textit}
\@namedef{PYG@tok@gs}{\let\PYG@bf=\textbf}
\@namedef{PYG@tok@ges}{\let\PYG@bf=\textbf\let\PYG@it=\textit}
\@namedef{PYG@tok@gp}{\def\PYG@tc##1{\textcolor[rgb]{0.33,0.33,0.33}{##1}}}
\@namedef{PYG@tok@go}{\def\PYG@tc##1{\textcolor[rgb]{0.53,0.53,0.53}{##1}}}
\@namedef{PYG@tok@gt}{\def\PYG@tc##1{\textcolor[rgb]{0.67,0.00,0.00}{##1}}}
\@namedef{PYG@tok@err}{\def\PYG@tc##1{\textcolor[rgb]{0.65,0.09,0.09}{##1}}\def\PYG@bc##1{{\setlength{\fboxsep}{0pt}\colorbox[rgb]{0.89,0.82,0.82}{\strut ##1}}}}
\@namedef{PYG@tok@kc}{\let\PYG@bf=\textbf}
\@namedef{PYG@tok@kd}{\let\PYG@bf=\textbf}
\@namedef{PYG@tok@kn}{\let\PYG@bf=\textbf}
\@namedef{PYG@tok@kp}{\let\PYG@bf=\textbf}
\@namedef{PYG@tok@kr}{\let\PYG@bf=\textbf}
\@namedef{PYG@tok@bp}{\def\PYG@tc##1{\textcolor[rgb]{0.60,0.60,0.60}{##1}}}
\@namedef{PYG@tok@fm}{\let\PYG@bf=\textbf\def\PYG@tc##1{\textcolor[rgb]{0.60,0.00,0.00}{##1}}}
\@namedef{PYG@tok@vc}{\def\PYG@tc##1{\textcolor[rgb]{0.00,0.50,0.50}{##1}}}
\@namedef{PYG@tok@vg}{\def\PYG@tc##1{\textcolor[rgb]{0.00,0.50,0.50}{##1}}}
\@namedef{PYG@tok@vi}{\def\PYG@tc##1{\textcolor[rgb]{0.00,0.50,0.50}{##1}}}
\@namedef{PYG@tok@vm}{\def\PYG@tc##1{\textcolor[rgb]{0.00,0.50,0.50}{##1}}}
\@namedef{PYG@tok@sa}{\def\PYG@tc##1{\textcolor[rgb]{0.73,0.53,0.27}{##1}}}
\@namedef{PYG@tok@sb}{\def\PYG@tc##1{\textcolor[rgb]{0.73,0.53,0.27}{##1}}}
\@namedef{PYG@tok@sc}{\def\PYG@tc##1{\textcolor[rgb]{0.73,0.53,0.27}{##1}}}
\@namedef{PYG@tok@dl}{\def\PYG@tc##1{\textcolor[rgb]{0.73,0.53,0.27}{##1}}}
\@namedef{PYG@tok@sd}{\def\PYG@tc##1{\textcolor[rgb]{0.73,0.53,0.27}{##1}}}
\@namedef{PYG@tok@s2}{\def\PYG@tc##1{\textcolor[rgb]{0.73,0.53,0.27}{##1}}}
\@namedef{PYG@tok@se}{\def\PYG@tc##1{\textcolor[rgb]{0.73,0.53,0.27}{##1}}}
\@namedef{PYG@tok@sh}{\def\PYG@tc##1{\textcolor[rgb]{0.73,0.53,0.27}{##1}}}
\@namedef{PYG@tok@si}{\def\PYG@tc##1{\textcolor[rgb]{0.73,0.53,0.27}{##1}}}
\@namedef{PYG@tok@sx}{\def\PYG@tc##1{\textcolor[rgb]{0.73,0.53,0.27}{##1}}}
\@namedef{PYG@tok@s1}{\def\PYG@tc##1{\textcolor[rgb]{0.73,0.53,0.27}{##1}}}
\@namedef{PYG@tok@ss}{\def\PYG@tc##1{\textcolor[rgb]{0.73,0.53,0.27}{##1}}}
\@namedef{PYG@tok@mb}{\def\PYG@tc##1{\textcolor[rgb]{0.00,0.60,0.60}{##1}}}
\@namedef{PYG@tok@mf}{\def\PYG@tc##1{\textcolor[rgb]{0.00,0.60,0.60}{##1}}}
\@namedef{PYG@tok@mh}{\def\PYG@tc##1{\textcolor[rgb]{0.00,0.60,0.60}{##1}}}
\@namedef{PYG@tok@mi}{\def\PYG@tc##1{\textcolor[rgb]{0.00,0.60,0.60}{##1}}}
\@namedef{PYG@tok@il}{\def\PYG@tc##1{\textcolor[rgb]{0.00,0.60,0.60}{##1}}}
\@namedef{PYG@tok@mo}{\def\PYG@tc##1{\textcolor[rgb]{0.00,0.60,0.60}{##1}}}
\@namedef{PYG@tok@ow}{\let\PYG@bf=\textbf}
\@namedef{PYG@tok@ch}{\let\PYG@it=\textit\def\PYG@tc##1{\textcolor[rgb]{0.60,0.60,0.53}{##1}}}
\@namedef{PYG@tok@cm}{\let\PYG@it=\textit\def\PYG@tc##1{\textcolor[rgb]{0.60,0.60,0.53}{##1}}}
\@namedef{PYG@tok@cpf}{\let\PYG@it=\textit\def\PYG@tc##1{\textcolor[rgb]{0.60,0.60,0.53}{##1}}}
\@namedef{PYG@tok@c1}{\let\PYG@it=\textit\def\PYG@tc##1{\textcolor[rgb]{0.60,0.60,0.53}{##1}}}

\def\PYGZbs{\char`\\}
\def\PYGZus{\char`\_}
\def\PYGZob{\char`\{}
\def\PYGZcb{\char`\}}

\def\PYGZlt{\char`\<}
\def\PYGZgt{\char`\>}
\def\PYGZsh{\char`\#}
\def\PYGZpc{\char`\%}

\def\PYGZhy{\char`\-}
\def\PYGZsq{\char`\'}
\def\PYGZdq{\char`\"}

\makeatother
\makeatletter
\def\PYGD@reset{\let\PYGD@it=\relax \let\PYGD@bf=\relax%
    \let\PYGD@ul=\relax \let\PYGD@tc=\relax%
    \let\PYGD@bc=\relax \let\PYGD@ff=\relax}
\def\PYGD@tok#1{\csname PYGD@tok@#1\endcsname}
\def\PYGD@toks#1+{\ifx\relax#1\empty\else%
    \PYGD@tok{#1}\expandafter\PYGD@toks\fi}
\def\PYGD@do#1{\PYGD@bc{\PYGD@tc{\PYGD@ul{%
    \PYGD@it{\PYGD@bf{\PYGD@ff{#1}}}}}}}
\def\PYGD#1#2{\PYGD@reset\PYGD@toks#1+\relax+\PYGD@do{#2}}

\@namedef{PYGD@tok@w}{\def\PYGD@tc##1{\textcolor[rgb]{0.73,0.73,0.73}{##1}}}
\@namedef{PYGD@tok@c}{\let\PYGD@it=\textit\def\PYGD@tc##1{\textcolor[rgb]{0.24,0.48,0.48}{##1}}}
\@namedef{PYGD@tok@cp}{\def\PYGD@tc##1{\textcolor[rgb]{0.61,0.40,0.00}{##1}}}
\@namedef{PYGD@tok@k}{\let\PYGD@bf=\textbf\def\PYGD@tc##1{\textcolor[rgb]{0.00,0.50,0.00}{##1}}}
\@namedef{PYGD@tok@kp}{\def\PYGD@tc##1{\textcolor[rgb]{0.00,0.50,0.00}{##1}}}
\@namedef{PYGD@tok@kt}{\def\PYGD@tc##1{\textcolor[rgb]{0.69,0.00,0.25}{##1}}}
\@namedef{PYGD@tok@o}{\def\PYGD@tc##1{\textcolor[rgb]{0.40,0.40,0.40}{##1}}}
\@namedef{PYGD@tok@ow}{\let\PYGD@bf=\textbf\def\PYGD@tc##1{\textcolor[rgb]{0.67,0.13,1.00}{##1}}}
\@namedef{PYGD@tok@nb}{\def\PYGD@tc##1{\textcolor[rgb]{0.00,0.50,0.00}{##1}}}
\@namedef{PYGD@tok@nf}{\def\PYGD@tc##1{\textcolor[rgb]{0.00,0.00,1.00}{##1}}}
\@namedef{PYGD@tok@nc}{\let\PYGD@bf=\textbf\def\PYGD@tc##1{\textcolor[rgb]{0.00,0.00,1.00}{##1}}}
\@namedef{PYGD@tok@nn}{\let\PYGD@bf=\textbf\def\PYGD@tc##1{\textcolor[rgb]{0.00,0.00,1.00}{##1}}}
\@namedef{PYGD@tok@ne}{\let\PYGD@bf=\textbf\def\PYGD@tc##1{\textcolor[rgb]{0.80,0.25,0.22}{##1}}}
\@namedef{PYGD@tok@nv}{\def\PYGD@tc##1{\textcolor[rgb]{0.10,0.09,0.49}{##1}}}
\@namedef{PYGD@tok@no}{\def\PYGD@tc##1{\textcolor[rgb]{0.53,0.00,0.00}{##1}}}
\@namedef{PYGD@tok@nl}{\def\PYGD@tc##1{\textcolor[rgb]{0.46,0.46,0.00}{##1}}}
\@namedef{PYGD@tok@ni}{\let\PYGD@bf=\textbf\def\PYGD@tc##1{\textcolor[rgb]{0.44,0.44,0.44}{##1}}}
\@namedef{PYGD@tok@na}{\def\PYGD@tc##1{\textcolor[rgb]{0.41,0.47,0.13}{##1}}}
\@namedef{PYGD@tok@nt}{\let\PYGD@bf=\textbf\def\PYGD@tc##1{\textcolor[rgb]{0.00,0.50,0.00}{##1}}}
\@namedef{PYGD@tok@nd}{\def\PYGD@tc##1{\textcolor[rgb]{0.67,0.13,1.00}{##1}}}
\@namedef{PYGD@tok@s}{\def\PYGD@tc##1{\textcolor[rgb]{0.73,0.13,0.13}{##1}}}
\@namedef{PYGD@tok@sd}{\let\PYGD@it=\textit\def\PYGD@tc##1{\textcolor[rgb]{0.73,0.13,0.13}{##1}}}
\@namedef{PYGD@tok@si}{\let\PYGD@bf=\textbf\def\PYGD@tc##1{\textcolor[rgb]{0.64,0.35,0.47}{##1}}}
\@namedef{PYGD@tok@se}{\let\PYGD@bf=\textbf\def\PYGD@tc##1{\textcolor[rgb]{0.67,0.36,0.12}{##1}}}
\@namedef{PYGD@tok@sr}{\def\PYGD@tc##1{\textcolor[rgb]{0.64,0.35,0.47}{##1}}}
\@namedef{PYGD@tok@ss}{\def\PYGD@tc##1{\textcolor[rgb]{0.10,0.09,0.49}{##1}}}
\@namedef{PYGD@tok@sx}{\def\PYGD@tc##1{\textcolor[rgb]{0.00,0.50,0.00}{##1}}}
\@namedef{PYGD@tok@m}{\def\PYGD@tc##1{\textcolor[rgb]{0.40,0.40,0.40}{##1}}}
\@namedef{PYGD@tok@gh}{\let\PYGD@bf=\textbf\def\PYGD@tc##1{\textcolor[rgb]{0.00,0.00,0.50}{##1}}}
\@namedef{PYGD@tok@gu}{\let\PYGD@bf=\textbf\def\PYGD@tc##1{\textcolor[rgb]{0.50,0.00,0.50}{##1}}}
\@namedef{PYGD@tok@gd}{\def\PYGD@tc##1{\textcolor[rgb]{0.63,0.00,0.00}{##1}}}
\@namedef{PYGD@tok@gi}{\def\PYGD@tc##1{\textcolor[rgb]{0.00,0.52,0.00}{##1}}}
\@namedef{PYGD@tok@gr}{\def\PYGD@tc##1{\textcolor[rgb]{0.89,0.00,0.00}{##1}}}
\@namedef{PYGD@tok@ge}{\let\PYGD@it=\textit}
\@namedef{PYGD@tok@gs}{\let\PYGD@bf=\textbf}
\@namedef{PYGD@tok@ges}{\let\PYGD@bf=\textbf\let\PYGD@it=\textit}
\@namedef{PYGD@tok@gp}{\let\PYGD@bf=\textbf\def\PYGD@tc##1{\textcolor[rgb]{0.00,0.00,0.50}{##1}}}
\@namedef{PYGD@tok@go}{\def\PYGD@tc##1{\textcolor[rgb]{0.44,0.44,0.44}{##1}}}
\@namedef{PYGD@tok@gt}{\def\PYGD@tc##1{\textcolor[rgb]{0.00,0.27,0.87}{##1}}}
\@namedef{PYGD@tok@err}{\def\PYGD@bc##1{{\setlength{\fboxsep}{\string -\fboxrule}\fcolorbox[rgb]{1.00,0.00,0.00}{1,1,1}{\strut ##1}}}}
\@namedef{PYGD@tok@kc}{\let\PYGD@bf=\textbf\def\PYGD@tc##1{\textcolor[rgb]{0.00,0.50,0.00}{##1}}}
\@namedef{PYGD@tok@kd}{\let\PYGD@bf=\textbf\def\PYGD@tc##1{\textcolor[rgb]{0.00,0.50,0.00}{##1}}}
\@namedef{PYGD@tok@kn}{\let\PYGD@bf=\textbf\def\PYGD@tc##1{\textcolor[rgb]{0.00,0.50,0.00}{##1}}}
\@namedef{PYGD@tok@kr}{\let\PYGD@bf=\textbf\def\PYGD@tc##1{\textcolor[rgb]{0.00,0.50,0.00}{##1}}}
\@namedef{PYGD@tok@bp}{\def\PYGD@tc##1{\textcolor[rgb]{0.00,0.50,0.00}{##1}}}
\@namedef{PYGD@tok@fm}{\def\PYGD@tc##1{\textcolor[rgb]{0.00,0.00,1.00}{##1}}}
\@namedef{PYGD@tok@vc}{\def\PYGD@tc##1{\textcolor[rgb]{0.10,0.09,0.49}{##1}}}
\@namedef{PYGD@tok@vg}{\def\PYGD@tc##1{\textcolor[rgb]{0.10,0.09,0.49}{##1}}}
\@namedef{PYGD@tok@vi}{\def\PYGD@tc##1{\textcolor[rgb]{0.10,0.09,0.49}{##1}}}
\@namedef{PYGD@tok@vm}{\def\PYGD@tc##1{\textcolor[rgb]{0.10,0.09,0.49}{##1}}}
\@namedef{PYGD@tok@sa}{\def\PYGD@tc##1{\textcolor[rgb]{0.73,0.13,0.13}{##1}}}
\@namedef{PYGD@tok@sb}{\def\PYGD@tc##1{\textcolor[rgb]{0.73,0.13,0.13}{##1}}}
\@namedef{PYGD@tok@sc}{\def\PYGD@tc##1{\textcolor[rgb]{0.73,0.13,0.13}{##1}}}
\@namedef{PYGD@tok@dl}{\def\PYGD@tc##1{\textcolor[rgb]{0.73,0.13,0.13}{##1}}}
\@namedef{PYGD@tok@s2}{\def\PYGD@tc##1{\textcolor[rgb]{0.73,0.13,0.13}{##1}}}
\@namedef{PYGD@tok@sh}{\def\PYGD@tc##1{\textcolor[rgb]{0.73,0.13,0.13}{##1}}}
\@namedef{PYGD@tok@s1}{\def\PYGD@tc##1{\textcolor[rgb]{0.73,0.13,0.13}{##1}}}
\@namedef{PYGD@tok@mb}{\def\PYGD@tc##1{\textcolor[rgb]{0.40,0.40,0.40}{##1}}}
\@namedef{PYGD@tok@mf}{\def\PYGD@tc##1{\textcolor[rgb]{0.40,0.40,0.40}{##1}}}
\@namedef{PYGD@tok@mh}{\def\PYGD@tc##1{\textcolor[rgb]{0.40,0.40,0.40}{##1}}}
\@namedef{PYGD@tok@mi}{\def\PYGD@tc##1{\textcolor[rgb]{0.40,0.40,0.40}{##1}}}
\@namedef{PYGD@tok@il}{\def\PYGD@tc##1{\textcolor[rgb]{0.40,0.40,0.40}{##1}}}
\@namedef{PYGD@tok@mo}{\def\PYGD@tc##1{\textcolor[rgb]{0.40,0.40,0.40}{##1}}}
\@namedef{PYGD@tok@ch}{\let\PYGD@it=\textit\def\PYGD@tc##1{\textcolor[rgb]{0.24,0.48,0.48}{##1}}}
\@namedef{PYGD@tok@cm}{\let\PYGD@it=\textit\def\PYGD@tc##1{\textcolor[rgb]{0.24,0.48,0.48}{##1}}}
\@namedef{PYGD@tok@cpf}{\let\PYGD@it=\textit\def\PYGD@tc##1{\textcolor[rgb]{0.24,0.48,0.48}{##1}}}
\@namedef{PYGD@tok@c1}{\let\PYGD@it=\textit\def\PYGD@tc##1{\textcolor[rgb]{0.24,0.48,0.48}{##1}}}
\@namedef{PYGD@tok@cs}{\let\PYGD@it=\textit\def\PYGD@tc##1{\textcolor[rgb]{0.24,0.48,0.48}{##1}}}

\def\PYGDZus{\char`\_}
\def\PYGDZob{\char`\{}
\def\PYGDZcb{\char`\}}

\def\PYGDZsh{\char`\#}

\def\PYGDZdl{\char`\$}
\def\PYGDZhy{\char`\-}

\makeatother

\usepackage{hyperref}
\hypersetup{hidelinks}

\renewcommand{\baselinestretch}{0.9}

\definecolor{codebg}{gray}{0.98}
\definecolor{clsbg}{gray}{0.9}
\definecolor{Maroon}{RGB}{128,0,0}
\definecolor{darkyellow}{RGB}{218,165,32}
\definecolor{newblue}{RGB}{0,0,238}   
\newcommand{\NEW}[1]{#1}  

\newcommand{\IT}{{\sffamily EZR}}

\newcommand{\cls}[1]{%
  \setlength{\fboxsep}{1.5pt}%
  \colorbox{clsbg}{\sffamily\bfseries{#1}}%
}

\newcommand{\disc}[2][black]{%
  \tikz[baseline=(c.base)]%
    \node[circle, fill=#1, draw=#1,
          inner sep=0pt, minimum size=1.2ex,
          text=white, font=\tiny\bfseries]
    (c) {#2};%
}

\newtoggle{rcred}
\toggletrue{rcred}

\newcommand{\rc}[1]{%
  \iftoggle{rcred}{\disc[red]{#1}}{\disc[black]{#1}}%
}

\newtcolorbox{hence}[1]{
  enhanced, breakable,
  colback=gray!5, colframe=black!70,
  fonttitle=\bfseries, coltitle=white,
  title=#1,
  left=5pt, right=5pt, top=4pt, bottom=4pt,
  boxrule=0.5pt, arc=2pt,
}

\articletype{Research Paper}
\received{Date Month Year}
\revised{Date Month Year}
\accepted{Date Month Year}
\journal{Software: Practice and Experience}
\volume{00}
\copyyear{2026}
\startpage{1}

\begin{document}

\title{Can AI be Easy? Lessons Learned from the EZR.py Toolkit}

\author[1]{Tim Menzies}
\author[1]{Srinath Srinivasan}
\author[1]{Kishan Ganguly}

\authormark{MENZIES \textsc{et al.}}
\titlemark{Can AI be Easy?}

\address[1]{\orgdiv{Department of Computer Science},
            \orgname{North Carolina State University},
            \orgaddress{\state{North Carolina}, \country{USA}}}

\corres{Tim Menzies, \email{timm@ieee.org}}

\abstract[Abstract]{
\renewcommand{\baselinestretch}{0.9}
Much recent press claims that developers no longer need to
read code. We disagree, at least within the domain of
tabular software-engineering (SE) optimization tasks: rows
of $x$ and $y$ values where the $y$ values are expensive to
obtain.

As evidence we present EZR.py, a Python toolkit of \NEW{less
than 1000 lines}
(no heavy dependencies) that implements Naive Bayes,
$k$-means clustering, classification and regression trees,
simulated annealing, local search, active learning, and
complementary-Bayes text-mining relevance filtering for
tabular SE data. EZR was built by repeatedly reading and
refactoring AI tools to simplify and unify them. The result
demonstrates that many seemingly different learning
algorithms are nearly the same once stripped back to their
core: classical algorithms collapse to a few lines each, and
a state-of-the-art active learner fits in roughly 80 lines.

This paper shows that (a)~\NEW{on 20 randomly selected
benchmarks from the MOOT repository of 124 tabular SE
optimization tasks,} simulated annealing
in its 1983 default configuration matches or beats more
elaborate local-search variants; (b)~an 80-line active learner
reaches 85--95\% of the reference optimum with fewer than 100
labels \NEW{across the MOOT tasks}, building trees from fewer than ten variables even when
thousands are available; and (c)~\NEW{on Yu et al.'s four
systematic-review corpora,} a 30-line complementary-Bayes
filter matches the recall of the SVM-based FASTREAD text miner
using roughly one-third of the labels\NEW{, on three of the
four corpora; on the fourth, it converges faster but plateaus
below FASTREAD's recall ceiling (a tradeoff detailed in
\S\ref{sec:textmine})}. Companion studies
\NEW{(co-authored by authors of this paper)} further report that this
active learner's model update runs \NEW{over 400 times} faster than
SMAC3, and that its trees rank features as well as SHAP and
ReliefF.

We conclude that, within the scope of tabular SE
optimization, reading and refactoring code is a useful
method of generating insight, and small unified toolkits
can rival large libraries. 

EZR is available under an open-source license. Install via
\textsf{pip install ezr} (Python $\geq$ 3.12). The exact
version studied here \NEW{(release \textsf{v0.9.4})} is archived at Zenodo, DOI
\NEW{\textsf{10.5281/zenodo.22554751}}; example data at \textsf{github.com/timm/moot}.
\NEW{See Appendix~\ref{sec:repro} for notes on installing and
running this code.}
}


\jnlcitation{\cname{%
\author{Menzies T.},
\author{Ganguly K.},
\author{Srinivasan S.}},
\ctitle{Can AI be Easy? Lessons Learned from the EZR.py Toolkit}.
\cjournal{\it Software: Practice and Experience.}
\cvol{2026;00(00):1--20}.}

\maketitle
\section{Introduction}
\label{sec:intro}

Recently it has been argued that humans no longer need to read
code. Ryan Dahl (creator of Node.js) suggests that the era of
human-written code is ending~\cite{dahl2024}. Jensen Huang
(Nvidia CEO) advises against learning to code~\cite{huang2024}.
Elon Musk said traditional coding ends in 2026~\cite{musk2026}.
Replit's CEO told students to focus on how to talk to machines,
not how to program~\cite{masad2025}. While the phrasing varies,
the argument is the same: AI is the new compiler, and we need not
read what that compiler generates.

We disagree. Reading code is very useful. Important low-level
information ``on the ground'' (as it were) gets missed when
developers always use LLMs to ``fly over'' the details. This
paper is an extended argument that careful reading of small code
bases can offer important insights into widely held assumptions
about AI and software engineering.

\begin{table*}[!t]
{\fontsize{7}{7.5}\linespread{0.85}\selectfont
\setlength{\tabcolsep}{4pt}
\setlength{\extrarowheight}{-1pt}
\renewcommand{\arraystretch}{0.7}

\caption{The 124 multi-objective tasks in the MOOT
repository~\cite{menzies2025moot} used in this paper. Tasks are expressed as tabular data.
Each row
is a task category. \#y is the number of goal columns; \#x is
the range of independent attributes.
Data comes from recent papers from
the SE literature (ICSE, FSE, TSE,
IST, EMSE, TOSEM, and ASE~\cite{Amiraliminimaldata,chen2026promisetune,senthilkumar2024can,chen2025accuracy,lustosa2024learning,lustossa2024isneak,chen2018beyond,nyagami_fc25_kaggle_2025,blastchar_telco_customer_churn_2025,kumarajarshi_life_expectancy_who_2025,jackdaoud_marketing_data_2022}). }

\label{mootdata}
\begin{center}
\begin{tabularx}{0.85\textwidth}{@{} r l l l Y c c r l @{}} 
\textbf{\#}
  & \textbf{Category}
  & \textbf{Focus}
  & \textbf{File name(s)}
  & \textbf{Optimization challenge}
  & \textbf{\#y}
  & \textbf{\#x}
  & \textbf{Rows}
  & \textbf{Refs} \\
\midrule

\multicolumn{9}{@{}l}{%
  \cellcolor{gray!25}\textbf{Software systems, configuration, and tuning}} \\
24 & Soft.\ config  & PLE     & SS-[A-X]         & Minimize footprint/memory vs.\ maximize throughput   & 2--3 & 3--88   &      197--86k & \cite{Amiraliminimaldata} \\
12 & PromiseTune    & Perf.   & 7z, LLVM, BDBC   & Minimize execution time and energy consumption       & 1    & 6--35   &     864--166k & \cite{chen2026promisetune} \\
 1 & Cloud compute  & Tuning  & Apache\_AllMeas  & Balance server response vs.\ CPU/RAM load            & 1    & 9       &           192 & \cite{senthilkumar2024can} \\
 1 & Cloud compute  & Tuning  & SQL\_AllMeas     & Minimize latency/IO vs.\ maximize throughput         & 1    & 39      &      4{,}654  & \cite{chen2025accuracy} \\
 1 & Cloud compute  & Tuning  & X264\_AllMeas    & Optimize encoding parameters for PSNR/SSIM           & 1    & 16      &      1{,}153  & \cite{Amiraliminimaldata} \\
 2 & Testing        & Testing & test120, test600 & Maximize coverage while minimizing execution time    & 1    & 9       &      5{,}161  & \\

\addlinespace[1pt]
\multicolumn{9}{@{}l}{%
  \cellcolor{gray!25}\textbf{Project management and process modeling}} \\
35 & Proj.\ health  & Health  & Health-Closed    & Optimize PR rates and minimize developer churn       & 2--3 & 5       &     10{,}001  & \cite{lustosa2024learning} \\
 3 & Scrum          & Agile   & Scrum[1k--100k]  & Maximize velocity within sprint constraints          & 3    & 124     &      1k--100k & \cite{lustossa2024isneak} \\
 8 & Feature mod.   & Config  & FFM, FM-*        & Optimize clause/constraint ratio in large spaces     & 3    & 128--1k &     10{,}001  & \cite{Amiraliminimaldata} \\
 1 & nasa93dem      & Cost    & nasa93dem        & Minimize effort (person-months) vs.\ quality         & 3    & 26      &            93 & \cite{lustosa2024learning} \\
 1 & COC1000        & Risk    & coc1000          & Minimize risks vs.\ analyst expertise                & 5    & 20      &      1{,}001  & \cite{chen2018beyond} \\
 4 & POM3           & Process & pom3[a-d]        & Balance project idle rates vs.\ completion costs     & 3    & 9       &      501--20k & \cite{lustossa2024isneak} \\
 4 & XOMO           & Defects & xomo\_[flt,grd]  & Minimize defect density vs.\ total project effort    & 4    & 27      &     10{,}001  & \cite{chen2018beyond} \\

\addlinespace[1pt]
\multicolumn{9}{@{}l}{%
  \cellcolor{gray!25}\textbf{Interdisciplinary and behavioral benchmarks}} \\
 4 & Behavioral     & Behavior  & all\_players, etc.\ & Maximize user retention and predict cohort churn  & 1--3 & 26--55  &     82--17k   & \cite{nyagami_fc25_kaggle_2025} \\
 4 & Financial      & Finance   & BankChurners      & Minimize credit risk vs.\ pricing models            & 2--5 & 19--77  &      1k--20k  & \cite{blastchar_telco_customer_churn_2025} \\
 3 & Health         & Health    & COVID19, Life     & Maximize accuracy/life vs.\ readmission likelihood  & 1--3 & 20--64  &      2k--25k  & \cite{kumarajarshi_life_expectancy_who_2025} \\
 2 & RL             & Control   & A2C\_[Acr,Crt]    & Maximize reward signals vs.\ steps to convergence   & 3--4 & 9--11   &      224--318 & \\
 5 & Sales          & Market    & Marketing, socks  & Maximize ROI vs.\ minimize forecasting error        & 1--8 & 20--31  &       247--2k & \cite{jackdaoud_marketing_data_2022} \\
 3 & Misc.          & General   & auto93, Car       & Optimize multivariate trade-offs (e.g.\ MPG vs.\ HP) & 2--5 & 5--38   &       205--1k & \cite{Amiraliminimaldata} \\

\midrule
\textbf{124 } & \multicolumn{8}{@{}l}{\textbf{total}}  
\end{tabularx}
\end{center}
}
\end{table*}
\begin{table*}
\caption{This paper comments
on six common assumptions about AI for software
engineering.}
\label{truisms}
{\fontsize{7.5}{8}\selectfont
\begin{center}
\renewcommand{\arraystretch}{1.25}
\begin{tabularx}{\textwidth}{@{} l Y Y @{}}
\toprule
Section & \textbf{Common assumption} & \textbf{What this paper finds} \\
\midrule
\S\ref{sec:peek} &
AI infrastructure must be large and dependency-heavy:
real systems sit on top of \texttt{pandas}, \texttt{sklearn},
\texttt{numpy}, and a stack of supporting libraries~\cite{martinez2021}. &
Four classes (\cls{Num}, \cls{Sym}, \cls{Data}, \cls{Cols}) and
one update primitive (\textsf{add}) carry every algorithm in this
paper. A 14-line CSV reader replaces \texttt{pandas}. \\
\midrule
\S\ref{sec:famous} &
Classification, clustering, and seeded clustering
(i.e.\ $k$-means++ initialization) are usually
presented as separate algorithms, taught in separate lectures and
shipped as separate library modules. &
Classification and clustering
require 30 lines of new code on top of the
\S\ref{sec:peek} substrate. \\
\midrule
\S\ref{trees} &
Classification trees and regression trees require different
splitting code, different leaf scoring, and different
implementations. &
43 lines of code can build both algorithms, and switching between them is a one-line change. \\
\midrule
\S\ref{sec:opt} &
 The 1990s metaheuristics
literature~\cite{gsat,walksat,grasp,ils} argued that local-search
variants (with restart, retry, and adaptive schedules)
dominate older methods.
 &
Across 20 MOOT benchmarks, simulated annealing in its 1983
default configuration~\cite{kirkpatrick1983optimization} matches
or beats  the local-search variants we tested. Restart is a
patch for local search, not an upgrade over SA. \\
\midrule
\S\ref{sec:active} &
More labels and more features yield better models; model updates
necessitate the complete
retraining of all models~\cite{hutter2011sequential,lindauer2022smac3}. &
\IT{}'s active learner reaches 85--95\% of the reference optimum
with fewer than 100 labels, uses fewer than 10 features even when
thousands are available.
\NEW{A companion study~\cite{ganguly2026lowgodatalightse}
(not a result of this paper) reports its model update running
over 400 times faster than state-of-the-art methods like
SMAC3.} \\
\midrule
\S\ref{sec:textmine} &
Relevance filtering for retrieval-augmented generation needs
elaborate classifiers (e.g. support
vector machines) or 
LLM-grade embeddings~\cite{yu2018total,yu2018finding}. &
30 lines of complementary Bayes~\cite{rennie2003tackling},
running on the
\S\ref{sec:peek} substrate,
performs as well as prior studies
and needs roughly
one-third of the labels.\\
\bottomrule
\end{tabularx}
\end{center}}
\end{table*}

Here, we present \IT{}, a \NEW{Python toolkit of less than
1000 lines} for
AI in software engineering. \IT{} grew across several years of
repeatedly reading and rewriting and refactoring (we read the
code, looking for similarities across distant parts of the
system, and deleting what those similarities made redundant).
The result was a small, very fast, dependency-light tool, in
which classification, clustering,  regression, explanation, multi-objective
optimization,
active learning, and text mining all sit on one shared substrate
based on four classes (\cls{Num}, \cls{Sym}, \cls{Data}, \cls{Cols}) and one
update primitive (\textsf{add}).

\IT{} has been tested on the 120+ multi-objective tasks shown in
Table~\ref{mootdata} from the MOOT
repository~\cite{menzies2025moot}. For examples of these tasks
see \S\ref{details}. These
tasks are from recent papers in the software engineering
literature published, by many authors, in leading venues such as
ICSE, FSE, TSE, IST, EMSE, TOSEM, and ASE. These tasks cover a
wide range of SE problems including software configuration,
performance tuning, product line engineering, project health
forecasting, defect prediction, software testing, software
process and cost estimation, cross-domain generalization, and
text mining. To the best of our knowledge, MOOT is the largest
collection of real multi-objective SE optimization tasks yet
collected.

We argue that lightweight tools
like EZR, that can be quickly built and deployed,  can find important new insights.
For example,
the rest of this paper presents the six insights listed in Table~\ref{truisms}.
This table shows six examples
of the same pattern.   Our field tends to
solve problems by adding more data, more algorithms, more
parameters, and more complexity. EZR was built the other way:
by reading the existing code and removing what
turned out to be redundant.

The methodological precedent for this approach is much older than this paper.
In his 1995 book {\em
Empirical Methods for Artificial Intelligence},
Cohen~\cite{cohen1995empirical} argued that supposedly
sophisticated methods should always be benchmarked against
seemingly stupider ones, the ``straw man'' approach to
scientific verification. We can attest that whenever we have
checked a supposedly sophisticated method against a simpler one,
there has been something useful in the simpler. More often than
not, a year later, we have switched to the simpler
approach~\cite{Nair17a,agrawal2019dodge,tantithamthavorn2016automated,
fu2016tuning}. So EZR is the latest case, but not the first.

The rest of this paper presents each row of
Table~\ref{truisms} in turn:
\begin{itemize}
\item
\S\ref{sec:peek} introduces the four-class substrate that is used, extensively, by the rest of the paper;
\item
\S\ref{sec:famous} implements Naïve Bayes, $k$-means, and
$k$-means++ on top of that substrate;
\item
\S\ref{trees} adds classification and regression trees;
\item 
\S\ref{sec:opt} implements and then compares simulated annealing against
local-search   on some MOOT benchmarks;
\item
\S\ref{sec:active} uses the substrate to build an active learner
whose model update has been reported, in a companion study
co-authored by \NEW{authors of this paper}~\cite{ganguly2026lowgodatalightse},
to run \NEW{over 400 times} faster than state-of-the-art optimizers such as
SMAC3;
\item
 \S\ref{sec:textmine} discusses applications of EZR to text mining.
 \end{itemize}
 After that, the end of the paper comments on the external validity of
 our conclusions.

One point of order before we start. The quotes \NEW{that open this
introduction} are broad cultural claims; the counterclaim of this
paper is deliberately narrow. We treat those quotes as
motivation, not as the proposition under test. What we actually
test, and all we claim, is scoped to tabular SE optimization
tasks; \S\ref{sec:scope} and \S\ref{sec:ttv} list the domains
(generation, perception, safety-critical certification) where
our conclusions should {\em not} be assumed to transfer.

\subsection{What is EZR offered as?}\label{sec:offered}
To pre-empt a fair question, we state explicitly what \IT{} is,
and is not, offered as. First and foremost, \IT{} is offered as
{\em evidence}: a concrete artifact whose existence demonstrates
that the algorithms of Table~\ref{truisms} can share one small
substrate. Second, it is offered as a {\em baseline}: a strong,
simple comparator that heavyweight methods should be asked to
beat (\S\ref{sec:ttv}). Third, and only third, it is offered as
{\em pedagogy}. We are careful here since ``small'' and ``easy''
are distinct dimensions: 400 terse lines can be read in a
sitting, yet fully absorbing the substrate's idioms (type-switch
polymorphism, header-name conventions, nested add/sub calls) may
require stepping through the code, not just reading it. Our
claims about teachability reflect the first author's classroom
experience and have not yet been formally evaluated with user
studies; we flag them as such wherever they appear.

\subsection{Scope}\label{sec:scope}
The claims of this paper hold for tabular SE optimization: rows
of $x,y$ values where the $y$ labels are expensive. We make no
claims for, and would not expect our conclusions to transfer to,
(a)~generation tasks such as translation, summarization, or code
synthesis, where LLM-scale data is genuinely required;
(b)~perception tasks such as vision and speech, where
representation learning dominates; or (c)~safety-critical
certification, where formal methods rather than empirical
optimization are the relevant tool. \S\ref{sec:ttv} expands on
these boundaries.

\subsection{Related work}\label{sec:related}
This paper sits at the junction of four research directions; related work
specific to each experiment also appears within its section. What distinguishes this paper from all four
threads is not any single algorithm but the demonstration that
one substrate carries them all. 

{\em (1)~Simple baselines in SE analytics.} A recurring finding
in empirical SE is that the published advantage of a heavyweight
method shrinks once re-tested against a carefully built simple
baseline: Fu et al.\ on deep learning versus tuned
SVMs~\cite{fu2016tuning,Fu17}; Nair et al.\ on rank-based
configuration~\cite{Nair17a}; Agrawal et al.'s DODGE
hyperparameter search~\cite{agrawal2019dodge}; Tantithamthavorn
et al.\ on classifier tuning~\cite{tantithamthavorn2016automated};
and, outside SE, Grinsztajn et al.\ on trees beating deep nets on
tabular data~\cite{grinsztajn2022why}. Cohen's ``straw man''
methodology~\cite{cohen1995empirical} is the common ancestor.

{\em (2)~Sequential model-based optimization.} SMAC and its
successors~\cite{hutter2011sequential,lindauer2022smac3} and
TPE~\cite{bergstra2011tpe} define the state of the art that
\S\ref{sec:active} is measured against; surrogate-based
evaluation follows Pfisterer et al.~\cite{pfisterer2022yahpo}
and Zela et al.~\cite{zela2020surrogate}.

{\em (3)~Explanation.} SHAP~\cite{lundberg2017unified},
LIME~\cite{ribeiro2016should}, Anchors~\cite{ribeiro2018anchors},
and ReliefF~\cite{robnik1997adaptation} are the comparison points
for \IT{}'s tiny trees; the feature-selection lineage from
Pearson to Kohavi and John is reviewed in \S\ref{sec:active}.

{\em (4)~Active learning for text.} Settles'
surveys~\cite{settles2008analysis,settles2009active} frame the
field; Yu et al.'s FASTREAD~\cite{yu2018total,yu2018finding} is
the direct comparator for \S\ref{sec:textmine}, and Rennie et
al.~\cite{rennie2003tackling} supply the complementary-Bayes
correction we adopt.

\section{A Peek at the Code}\label{sec:peek}

What is the smallest set of primitives that supports
classification, clustering, optimization, regression, active
learning, and text mining? Common wisdom assumes the answer
involves \texttt{pandas}, \texttt{scikit-learn}, \texttt{numpy},
and a stack of supporting libraries~\cite{martinez2021}. This
section shows that many important parts
of those large packages can be implemented in 
80 lines of code, four classes, and one update
primitive:
\begin{itemize}
\item \cls{Num} and \cls{Sym} summarize columns of numbers and
symbols.
\item \cls{Data} holds rows together with the columns that
summarize them.
\item \cls{Cols} is a small factory that builds the columns
from a CSV header.
\item \textsf{add} is one polymorphic operation that updates any
of the above as new data arrives.
\end{itemize}
The rest of this paper builds five different AI tools on top of
those four classes. One interesting
aspect of EZR is that it is built without relying on
large third-party packages like
 \texttt{pandas} or  \texttt{sklearn}.   Modern Python software
is often assembled from pre-made parts, with 85 to 97 percent
of code reused from external sources~\cite{martinez2021}.
Libraries like \texttt{pandas} are battle-tested but introduce
a large attack surface; a single poisoned update can compromise
tens of thousands of downstream packages without the primary
developer's knowledge~\cite{martinez2021}. To sidestep that
risk, \IT{} shuns \texttt{pandas} and instead ships with the
following CSV reader of 14 lines:
\par\nopagebreak\minipage{\linewidth}
\begingroup\mintfvset
\VerbatimPygments{\PYG}{\PYG}
\begin{mintbg}{codebg}
\begin{Verbatim}[commandchars=\\\{\},codes={\catcode`\$=3\catcode`\^=7\catcode`\_=8\relax}]
\PYG{k}{def}\PYG{+w}{ }\PYG{n+nf}{csv}\PYG{p}{(}\PYG{n}{f}\PYG{p}{,} \PYG{n}{ok}\PYG{o}{=}\PYG{k}{lambda} \PYG{n}{txt}\PYG{p}{:} \PYG{n}{txt}\PYG{o}{.}\PYG{n}{partition}\PYG{p}{(}\PYG{l+s+s2}{\PYGZdq{}}\PYG{l+s+s2}{\PYGZsh{}}\PYG{l+s+s2}{\PYGZdq{}}\PYG{p}{)}\PYG{p}{[}\PYG{l+m+mi}{0}\PYG{p}{]}\PYG{o}{.}\PYG{n}{split}\PYG{p}{(}\PYG{l+s+s2}{\PYGZdq{}}\PYG{l+s+s2}{,}\PYG{l+s+s2}{\PYGZdq{}}\PYG{p}{)}\PYG{p}{)}\PYG{p}{:}
  \PYG{k}{with} \PYG{n+nb}{open}\PYG{p}{(}\PYG{n}{f}\PYG{p}{,} \PYG{n}{encoding}\PYG{o}{=}\PYG{l+s+s2}{\PYGZdq{}}\PYG{l+s+s2}{utf\PYGZhy{}8}\PYG{l+s+s2}{\PYGZdq{}}\PYG{p}{)} \PYG{k}{as} \PYG{n}{file}\PYG{p}{:}
    \PYG{k}{for} \PYG{n}{txt} \PYG{o+ow}{in} \PYG{n}{file}\PYG{p}{:} \PYG{esc}{\label{csv}}
      \PYG{n}{row} \PYG{o}{=} \PYG{n}{ok}\PYG{p}{(}\PYG{n}{txt}\PYG{p}{)}
      \PYG{k}{if} \PYG{n+nb}{any}\PYG{p}{(}\PYG{n}{x}\PYG{o}{.}\PYG{n}{strip}\PYG{p}{(}\PYG{p}{)} \PYG{k}{for} \PYG{n}{x} \PYG{o+ow}{in} \PYG{n}{row}\PYG{p}{)}\PYG{p}{:}
        \PYG{k}{yield} \PYG{p}{[}\PYG{n}{thing}\PYG{p}{(}\PYG{n}{x}\PYG{p}{)} \PYG{k}{for} \PYG{n}{x} \PYG{o+ow}{in} \PYG{n}{row}\PYG{p}{]}

\PYG{k}{def}\PYG{+w}{ }\PYG{n+nf}{thing}\PYG{p}{(}\PYG{n}{txt}\PYG{p}{)}\PYG{p}{:}
  \PYG{n}{txt} \PYG{o}{=} \PYG{n}{txt}\PYG{o}{.}\PYG{n}{strip}\PYG{p}{(}\PYG{p}{)}
  \PYG{n}{b} \PYG{o}{=} \PYG{k}{lambda} \PYG{n}{s}\PYG{p}{:} \PYG{p}{\PYGZob{}}\PYG{l+s+s2}{\PYGZdq{}}\PYG{l+s+s2}{true}\PYG{l+s+s2}{\PYGZdq{}}\PYG{p}{:}\PYG{l+m+mi}{1}\PYG{p}{,}\PYG{l+s+s2}{\PYGZdq{}}\PYG{l+s+s2}{false}\PYG{l+s+s2}{\PYGZdq{}}\PYG{p}{:}\PYG{l+m+mi}{0}\PYG{p}{\PYGZcb{}}\PYG{o}{.}\PYG{n}{get}\PYG{p}{(}
                  \PYG{n}{s}\PYG{o}{.}\PYG{n}{lower}\PYG{p}{(}\PYG{p}{)}\PYG{p}{,} \PYG{n}{s}\PYG{p}{)}
  \PYG{k}{for} \PYG{n}{f} \PYG{o+ow}{in} \PYG{p}{[}\PYG{n+nb}{int}\PYG{p}{,} \PYG{n+nb}{float}\PYG{p}{,} \PYG{n}{b}\PYG{p}{]}\PYG{p}{:}
    \PYG{k}{try}\PYG{p}{:} \PYG{k}{return} \PYG{n}{f}\PYG{p}{(}\PYG{n}{txt}\PYG{p}{)}
    \PYG{k}{except} \PYG{n+ne}{ValueError}\PYG{p}{:} \PYG{k}{pass}
\end{Verbatim}
\end{mintbg}
\endgroup
\endminipage\par\noindent
This reader skips comments after \texttt{\#}, splits on commas,
and yields typed rows. \textsf{Thing} coerces each cell to int,
float, bool, or string in turn. 
Other packages, like \textsf{pandas}, offer far richer CSV
handling (dtype inference, chunked reads, automatic NA
conversion, multi-index support, Arrow back-ends, and so on).
\IT{} declines all of that: the 14 lines above do just what
\IT{} needs, and no more. This is the first
concrete instance of a pattern that repeats throughout the paper:
by carefully reading code, we can refactor it down to a tiny core.

\subsection{Columns: \cls{Num} and \cls{Sym}}

\IT{} distinguishes numeric and symbolic columns. Both know
their text name (\textsf{txt}), their position in the row
(\textsf{at}), and how many items they have seen (\textsf{n}).

\cls{Num}eric columns track a running mean (\textsf{mu}), a
second moment (\textsf{m2}), a standard deviation (\textsf{sd},
derived from \textsf{m2}), and a \textsf{heaven} value
indicating goal direction. By convention, columns whose names
end in ``+'' are to be maximized and columns whose names end in
``-'' are to be minimized. \textsf{heaven} is set to 1 or 0
respectively. Smaller distance to \textsf{heaven} is always
better.
\par\nopagebreak\minipage{\linewidth}
\begingroup\mintfvset
\VerbatimPygments{\PYG}{\PYG}
\begin{mintbg}{codebg}
\begin{Verbatim}[commandchars=\\\{\},codes={\catcode`\$=3\catcode`\^=7\catcode`\_=8\relax}]
\PYG{k}{class}\PYG{+w}{ }\PYG{n+nc}{Num}\PYG{p}{:}
  \PYG{k}{def}\PYG{+w}{ }\PYG{n+nf+fm}{\PYGZus{}\PYGZus{}init\PYGZus{}\PYGZus{}}\PYG{p}{(}\PYG{n}{i}\PYG{p}{,} \PYG{n}{txt}\PYG{o}{=}\PYG{l+s+s2}{\PYGZdq{}}\PYG{l+s+s2}{\PYGZdq{}}\PYG{p}{,} \PYG{n}{at}\PYG{o}{=}\PYG{l+m+mi}{0}\PYG{p}{)}\PYG{p}{:}
    \PYG{n}{i}\PYG{o}{.}\PYG{n}{txt}\PYG{p}{,} \PYG{n}{i}\PYG{o}{.}\PYG{n}{at}\PYG{p}{,} \PYG{n}{i}\PYG{o}{.}\PYG{n}{n} \PYG{o}{=} \PYG{n}{txt}\PYG{p}{,} \PYG{n}{at}\PYG{p}{,} \PYG{l+m+mi}{0}
    \PYG{n}{i}\PYG{o}{.}\PYG{n}{mu} \PYG{o}{=} \PYG{n}{i}\PYG{o}{.}\PYG{n}{m2} \PYG{o}{=} \PYG{n}{i}\PYG{o}{.}\PYG{n}{sd} \PYG{o}{=} \PYG{l+m+mi}{0}
    \PYG{n}{i}\PYG{o}{.}\PYG{n}{heaven} \PYG{o}{=} \PYG{n}{txt}\PYG{p}{[}\PYG{o}{\PYGZhy{}}\PYG{l+m+mi}{1}\PYG{p}{:}\PYG{p}{]} \PYG{o}{!=} \PYG{l+s+s2}{\PYGZdq{}}\PYG{l+s+s2}{\PYGZhy{}}\PYG{l+s+s2}{\PYGZdq{}} \PYG{esc}{\label{heaven}}
\end{Verbatim}
\end{mintbg}
\endgroup
\endminipage\par\noindent
\cls{Sym}bolic columns hold frequency counts in a \textsf{has}
dictionary:
\par\nopagebreak\minipage{\linewidth}
\begingroup\mintfvset
\VerbatimPygments{\PYG}{\PYG}
\begin{mintbg}{codebg}
\begin{Verbatim}[commandchars=\\\{\},codes={\catcode`\$=3\catcode`\^=7\catcode`\_=8\relax}]
\PYG{k}{class}\PYG{+w}{ }\PYG{n+nc}{Sym}\PYG{p}{:}
  \PYG{k}{def}\PYG{+w}{ }\PYG{n+nf+fm}{\PYGZus{}\PYGZus{}init\PYGZus{}\PYGZus{}}\PYG{p}{(}\PYG{n}{i}\PYG{p}{,} \PYG{n}{txt}\PYG{o}{=}\PYG{l+s+s2}{\PYGZdq{}}\PYG{l+s+s2}{\PYGZdq{}}\PYG{p}{,} \PYG{n}{at}\PYG{o}{=}\PYG{l+m+mi}{0}\PYG{p}{)}\PYG{p}{:}
    \PYG{n}{i}\PYG{o}{.}\PYG{n}{txt}\PYG{p}{,} \PYG{n}{i}\PYG{o}{.}\PYG{n}{at}\PYG{p}{,} \PYG{n}{i}\PYG{o}{.}\PYG{n}{n}\PYG{p}{,} \PYG{n}{i}\PYG{o}{.}\PYG{n}{has} \PYG{o}{=} \PYG{n}{txt}\PYG{p}{,} \PYG{n}{at}\PYG{p}{,} \PYG{l+m+mi}{0}\PYG{p}{,} \PYG{p}{\PYGZob{}}\PYG{p}{\PYGZcb{}}
\end{Verbatim}
\end{mintbg}
\endgroup
\endminipage\par\noindent
The \textsf{mid} of a column is its mean (for \cls{Num}) or
mode (for \cls{Sym}):
\par\nopagebreak\minipage{\linewidth}
\begingroup\mintfvset
\VerbatimPygments{\PYG}{\PYG}
\begin{mintbg}{codebg}
\begin{Verbatim}[commandchars=\\\{\},codes={\catcode`\$=3\catcode`\^=7\catcode`\_=8\relax}]
\PYG{k}{def}\PYG{+w}{ }\PYG{n+nf}{mid}\PYG{p}{(}\PYG{n}{col}\PYG{p}{)}\PYG{p}{:} \PYG{esc}{\label{mid0}}
  \PYG{k}{return} \PYG{p}{(}\PYG{n}{col}\PYG{o}{.}\PYG{n}{mu} \PYG{k}{if} \PYG{n}{Num}\PYG{o}{==}\PYG{n+nb}{type}\PYG{p}{(}\PYG{n}{col}\PYG{p}{)} \PYG{k}{else} \PYG{n}{mode}\PYG{p}{(}\PYG{n}{col}\PYG{o}{.}\PYG{n}{has}\PYG{p}{)}\PYG{p}{)}

\PYG{k}{def}\PYG{+w}{ }\PYG{n+nf}{mode}\PYG{p}{(}\PYG{n}{dct}\PYG{p}{)}\PYG{p}{:} \PYG{k}{return} \PYG{n+nb}{max}\PYG{p}{(}\PYG{n}{dct}\PYG{p}{,} \PYG{n}{key}\PYG{o}{=}\PYG{n}{dct}\PYG{o}{.}\PYG{n}{get}\PYG{p}{)}
\end{Verbatim}
\end{mintbg}
\endgroup
\endminipage\par\noindent
The \textsf{spread} is its standard deviation (for \cls{Num})
or entropy (for \cls{Sym}):
\par\nopagebreak\minipage{\linewidth}
\begingroup\mintfvset
\VerbatimPygments{\PYG}{\PYG}
\begin{mintbg}{codebg}
\begin{Verbatim}[commandchars=\\\{\},codes={\catcode`\$=3\catcode`\^=7\catcode`\_=8\relax}]
\PYG{k}{def}\PYG{+w}{ }\PYG{n+nf}{spread}\PYG{p}{(}\PYG{n}{col}\PYG{p}{)}\PYG{p}{:} \PYG{esc}{\label{spread0}}
  \PYG{k}{return} \PYG{p}{(}\PYG{n}{col}\PYG{o}{.}\PYG{n}{sd} \PYG{k}{if} \PYG{n}{Num}\PYG{o}{==}\PYG{n+nb}{type}\PYG{p}{(}\PYG{n}{col}\PYG{p}{)} \PYG{k}{else} \PYG{n}{entropy}\PYG{p}{(}\PYG{n}{col}\PYG{o}{.}\PYG{n}{has}\PYG{p}{)}\PYG{p}{)}

\PYG{k}{def}\PYG{+w}{ }\PYG{n+nf}{entropy}\PYG{p}{(}\PYG{n}{dct}\PYG{p}{)}\PYG{p}{:}
  \PYG{n}{n} \PYG{o}{=} \PYG{n+nb}{sum}\PYG{p}{(}\PYG{n}{dct}\PYG{o}{.}\PYG{n}{values}\PYG{p}{(}\PYG{p}{)}\PYG{p}{)}
  \PYG{k}{return} \PYG{o}{\PYGZhy{}}\PYG{n+nb}{sum}\PYG{p}{(}\PYG{n}{v}\PYG{o}{/}\PYG{n}{n}\PYG{o}{*}\PYG{n}{log2}\PYG{p}{(}\PYG{n}{v}\PYG{o}{/}\PYG{n}{n}\PYG{p}{)} \PYG{k}{for} \PYG{n}{v} \PYG{o+ow}{in} \PYG{n}{dct}\PYG{o}{.}\PYG{n}{values}\PYG{p}{(}\PYG{p}{)}\PYG{p}{)}
\end{Verbatim}
\end{mintbg}
\endgroup
\endminipage\par\noindent
Numeric values can also be normalized to $[0,1]$ via a logistic
sigmoid clamped to $\pm3$ standard deviations. Missing values
(``?'') pass through:
\par\nopagebreak\minipage{\linewidth}
\begingroup\mintfvset
\VerbatimPygments{\PYG}{\PYG}
\begin{mintbg}{codebg}
\begin{Verbatim}[commandchars=\\\{\},codes={\catcode`\$=3\catcode`\^=7\catcode`\_=8\relax}]
\PYG{k}{def}\PYG{+w}{ }\PYG{n+nf}{norm}\PYG{p}{(}\PYG{n}{num}\PYG{p}{,} \PYG{n}{v}\PYG{p}{)}\PYG{p}{:} \PYG{esc}{\label{norml}}
  \PYG{k}{if} \PYG{n}{v} \PYG{o}{==} \PYG{l+s+s2}{\PYGZdq{}}\PYG{l+s+s2}{?}\PYG{l+s+s2}{\PYGZdq{}}\PYG{p}{:} \PYG{k}{return} \PYG{n}{v}
  \PYG{n}{z} \PYG{o}{=} \PYG{n+nb}{max}\PYG{p}{(}\PYG{o}{\PYGZhy{}}\PYG{l+m+mi}{3}\PYG{p}{,} \PYG{n+nb}{min}\PYG{p}{(}\PYG{l+m+mi}{3}\PYG{p}{,} \PYG{p}{(}\PYG{n}{v} \PYG{o}{\PYGZhy{}} \PYG{n}{num}\PYG{o}{.}\PYG{n}{mu}\PYG{p}{)}\PYG{o}{/}\PYG{p}{(}\PYG{n}{num}\PYG{o}{.}\PYG{n}{sd} \PYG{o}{+} \PYG{l+m+mf}{1e\PYGZhy{}32}\PYG{p}{)}\PYG{p}{)}\PYG{p}{)}
  \PYG{k}{return} \PYG{l+m+mi}{1}\PYG{o}{/}\PYG{p}{(}\PYG{l+m+mi}{1} \PYG{o}{+} \PYG{n}{exp}\PYG{p}{(}\PYG{o}{\PYGZhy{}}\PYG{l+m+mf}{1.7}\PYG{o}{*}\PYG{n}{z}\PYG{p}{)}\PYG{p}{)}
\end{Verbatim}
\end{mintbg}
\endgroup
\endminipage\par\noindent
\textsf{norm} makes rows from different scales comparable for
the distance calculations later in the paper.

\subsection{An example data file}\label{details}

With \cls{Num} and \cls{Sym} defined, the column-name
conventions \IT{} uses for tabular data become easy to state.
Names ending in ``X'' are ignored (e.g.\ row IDs or timestamps).
Names ending in ``!'' mark a symbolic class column. Names
ending in ``+'' or ``-'' mark goal columns to be maximized or
minimized. Everything else is an independent attribute. Numeric
columns start with an upper-case letter; symbolic columns do
not.

Table~\ref{moot} shows a typical example. The data come from a
configuration-tuning task in the MOOT
repository~\cite{menzies2025moot}, one of the benchmarks used
later in this paper. The header tells \IT{} everything it
needs. \texttt{Spout\_wait} starts with an upper-case letter, so
it becomes a \cls{Num}. \texttt{Spliters} and \texttt{Cntr} also
start uppercase, so they are \cls{Num}s as well.
\texttt{Throughput+} ends in ``+'', so \texttt{Throughput} is to
be maximized. \texttt{Latency-} ends in ``-'', so
\texttt{Latency} is to be minimized.

\begin{table}[!b]
\caption{An example MOOT data set: configuration tuning for a
distributed stream-processing system. Numeric column names
start with an upper-case letter. Goal column names end with
``+'' (maximize) or ``-'' (minimize). Everything else is an
independent attribute.}
\label{moot}
\begin{center}
\begin{minipage}{3in}
\normalsize{(a) Configuration tuning:}
{\fontsize{6pt}{7.2pt}\selectfont
\begin{verbatim}
x = independent values      | y = dependent values
----------------------------|-----------------------
Spout_wait, Spliters, Cntr. | Throughput+, Latency-
    10,        6,       17  |    23075,    158.68
     8,        6,       17  |    22887,    172.74
     9,        6,       17  |    22799,    156.83
[Skipped],  ...,      ...   |       ...,      ...
 10000,        1,       10  |    460.81,    8761.6
 10000,        1,       18  |    402.53,    8797.5
 10000,        1,        1  |    310.06,    9421
-------------------------------------------------------
\end{verbatim}
}
\normalsize{(b) Hyperparameters for Software Health Prediction}
{\fontsize{6pt}{7.2pt}\selectfont
\begin{verbatim}
x = independent values          | y = dependent values
-------------------------------|----------------------
N_est, Min_leaf, ..., Min_imp. | MRE-, ACC+, PRED40+
   40,      1,    ...,    0.25 | 0.205, 0.826, 1.00
  120,      1,    ...,    4.75 | 0.191, 0.756, 1.00
   10,      1,    ...,    6.75 | 0.000, 0.790, 0.86
[Skipped],      ...,    ...    |   ...,   ...,   ...
   10,     11,    ...,    5.25 | 0.925, 0.391, 0.00
   90,     12,    ...,    5.00 | 0.911, 0.397, 0.00
  130,      9,    ...,    9.75 | 0.914, 0.395, 0.00
\end{verbatim}
}
\end{minipage}
\end{center}
\end{table}

\subsection{Rows and tables: \cls{Data} and \cls{Cols}}

\cls{Data} holds rows together with summarized columns. Its
constructor accepts an iterator (e.g.\ from \textsf{csv}) or a
list of rows. The first row becomes the header and is handed
to \cls{Cols} to build the column objects. The rest are added
in turn:
\par\nopagebreak\minipage{\linewidth}
\begingroup\mintfvset
\VerbatimPygments{\PYG}{\PYG}
\begin{mintbg}{codebg}
\begin{Verbatim}[commandchars=\\\{\},codes={\catcode`\$=3\catcode`\^=7\catcode`\_=8\relax}]
\PYG{k}{class}\PYG{+w}{ }\PYG{n+nc}{Data}\PYG{p}{:}
  \PYG{k}{def}\PYG{+w}{ }\PYG{n+nf+fm}{\PYGZus{}\PYGZus{}init\PYGZus{}\PYGZus{}}\PYG{p}{(}\PYG{n}{i}\PYG{p}{,} \PYG{n}{src}\PYG{o}{=}\PYG{k+kc}{None}\PYG{p}{)}\PYG{p}{:}
    \PYG{n}{src} \PYG{o}{=} \PYG{n+nb}{iter}\PYG{p}{(}\PYG{n}{src} \PYG{o+ow}{or} \PYG{p}{[}\PYG{p}{]}\PYG{p}{)}
    \PYG{n}{i}\PYG{o}{.}\PYG{n}{rows}\PYG{p}{,} \PYG{n}{i}\PYG{o}{.}\PYG{n}{\PYGZus{}centroid} \PYG{o}{=} \PYG{p}{[}\PYG{p}{]}\PYG{p}{,} \PYG{k+kc}{None}
    \PYG{n}{i}\PYG{o}{.}\PYG{n}{cols} \PYG{o}{=} \PYG{n}{Cols}\PYG{p}{(}\PYG{n+nb}{next}\PYG{p}{(}\PYG{n}{src}\PYG{p}{)}\PYG{p}{)}
    \PYG{n}{adds}\PYG{p}{(}\PYG{n}{src}\PYG{p}{,} \PYG{n}{i}\PYG{p}{)}

\PYG{k}{def}\PYG{+w}{ }\PYG{n+nf}{adds}\PYG{p}{(}\PYG{n}{src}\PYG{p}{,} \PYG{n}{it}\PYG{o}{=}\PYG{k+kc}{None}\PYG{p}{)}\PYG{p}{:}
  \PYG{n}{it} \PYG{o}{=} \PYG{n}{it} \PYG{o+ow}{or} \PYG{n}{Num}\PYG{p}{(}\PYG{p}{)}
  \PYG{p}{[}\PYG{n}{add}\PYG{p}{(}\PYG{n}{it}\PYG{p}{,} \PYG{n}{v}\PYG{p}{)} \PYG{k}{for} \PYG{n}{v} \PYG{o+ow}{in} \PYG{p}{(}\PYG{n}{src} \PYG{o+ow}{or} \PYG{p}{[}\PYG{p}{]}\PYG{p}{)}\PYG{p}{]}
  \PYG{k}{return} \PYG{n}{it}
\end{Verbatim}
\end{mintbg}
\endgroup
\endminipage\par\noindent
Two usage patterns appear throughout:
\begingroup\mintfvset
\fvset{numbers=none}
\VerbatimPygments{\PYG}{\PYG}
\begin{mintbg}{codebg}
\begin{Verbatim}[commandchars=\\\{\},codes={\catcode`\$=3\catcode`\^=7\catcode`\_=8\relax}]
\PYG{o}{\PYGZgt{}\PYGZgt{}}\PYG{o}{\PYGZgt{}} \PYG{n}{data1} \PYG{o}{=} \PYG{n}{Data}\PYG{p}{(}\PYG{n}{csv}\PYG{p}{(}\PYG{n}{fileName}\PYG{p}{)}\PYG{p}{)}           \PYG{c+c1}{\PYGZsh{} from disk}
\PYG{o}{\PYGZgt{}\PYGZgt{}}\PYG{o}{\PYGZgt{}} \PYG{n}{data2} \PYG{o}{=} \PYG{n}{Data}\PYG{p}{(}\PYG{p}{[}\PYG{n}{data1}\PYG{o}{.}\PYG{n}{cols}\PYG{o}{.}\PYG{n}{names}\PYG{p}{]}\PYG{p}{)}      \PYG{c+c1}{\PYGZsh{} empty clone}
\end{Verbatim}
\end{mintbg}
\endgroup
The second pattern (an empty \cls{Data} that mimics another's
column structure) is wrapped in a helper:
\par\nopagebreak\minipage{\linewidth}
\begingroup\mintfvset
\VerbatimPygments{\PYG}{\PYG}
\begin{mintbg}{codebg}
\begin{Verbatim}[commandchars=\\\{\},codes={\catcode`\$=3\catcode`\^=7\catcode`\_=8\relax}]
\PYG{k}{def}\PYG{+w}{ }\PYG{n+nf}{clone}\PYG{p}{(}\PYG{n}{data}\PYG{p}{,} \PYG{n}{rows}\PYG{o}{=}\PYG{k+kc}{None}\PYG{p}{)}\PYG{p}{:} \PYG{esc}{\label{cloning}}
  \PYG{k}{return} \PYG{n}{adds}\PYG{p}{(}\PYG{n}{rows} \PYG{o+ow}{or} \PYG{p}{[}\PYG{p}{]}\PYG{p}{,} \PYG{n}{Data}\PYG{p}{(}\PYG{p}{[}\PYG{n}{data}\PYG{o}{.}\PYG{n}{cols}\PYG{o}{.}\PYG{n}{names}\PYG{p}{]}\PYG{p}{)}\PYG{p}{)}
\end{Verbatim}
\end{mintbg}
\endgroup
\endminipage\par\noindent
\textsf{clone} appears nine times in the rest of
\texttt{ezr.py}. Each appearance spawns a fresh \cls{Data} that
shares its parent's schema.

\cls{Cols} is the factory that reads the header. Recall the
column-name conventions: names ending in ``X'' are ignored, ``!''
marks a symbolic class column, ``+'' or ``-'' mark goal columns
(\textsf{ys}), everything else is an independent attribute
(\textsf{xs}). Names starting with an upper-case letter become
\cls{Num}; everything else becomes \cls{Sym}.
\par\nopagebreak\minipage{\linewidth}
\begingroup\mintfvset
\VerbatimPygments{\PYG}{\PYG}
\begin{mintbg}{codebg}
\begin{Verbatim}[commandchars=\\\{\},codes={\catcode`\$=3\catcode`\^=7\catcode`\_=8\relax}]
\PYG{k}{def}\PYG{+w}{ }\PYG{n+nf}{Col}\PYG{p}{(}\PYG{n}{txt}\PYG{o}{=}\PYG{l+s+s2}{\PYGZdq{}}\PYG{l+s+s2}{\PYGZdq{}}\PYG{p}{,} \PYG{n}{at}\PYG{o}{=}\PYG{l+m+mi}{0}\PYG{p}{)}\PYG{p}{:}
  \PYG{k}{return} \PYG{p}{(}\PYG{n}{Num} \PYG{k}{if} \PYG{n}{txt}\PYG{p}{[}\PYG{p}{:}\PYG{l+m+mi}{1}\PYG{p}{]}\PYG{o}{.}\PYG{n}{isupper}\PYG{p}{(}\PYG{p}{)} \PYG{k}{else} \PYG{n}{Sym}\PYG{p}{)}\PYG{p}{(}\PYG{n}{txt}\PYG{p}{,} \PYG{n}{at}\PYG{p}{)}

\PYG{k}{class}\PYG{+w}{ }\PYG{n+nc}{Cols}\PYG{p}{:}
  \PYG{k}{def}\PYG{+w}{ }\PYG{n+nf+fm}{\PYGZus{}\PYGZus{}init\PYGZus{}\PYGZus{}}\PYG{p}{(}\PYG{n}{i}\PYG{p}{,} \PYG{n}{names}\PYG{p}{)}\PYG{p}{:}
    \PYG{n}{i}\PYG{o}{.}\PYG{n}{names} \PYG{o}{=} \PYG{n}{names}
    \PYG{n}{i}\PYG{o}{.}\PYG{n}{klass}\PYG{p}{,} \PYG{n}{i}\PYG{o}{.}\PYG{n}{xs}\PYG{p}{,} \PYG{n}{i}\PYG{o}{.}\PYG{n}{ys}\PYG{p}{,} \PYG{n}{i}\PYG{o}{.}\PYG{n}{all} \PYG{o}{=} \PYG{k+kc}{None}\PYG{p}{,} \PYG{p}{[}\PYG{p}{]}\PYG{p}{,} \PYG{p}{[}\PYG{p}{]}\PYG{p}{,} \PYG{p}{[}\PYG{p}{]}
    \PYG{k}{for} \PYG{n}{j}\PYG{p}{,} \PYG{n}{txt} \PYG{o+ow}{in} \PYG{n+nb}{enumerate}\PYG{p}{(}\PYG{n}{names}\PYG{p}{)}\PYG{p}{:}
      \PYG{n}{i}\PYG{o}{.}\PYG{n}{all}\PYG{o}{.}\PYG{n}{append}\PYG{p}{(}\PYG{n}{col} \PYG{o}{:=} \PYG{n}{Col}\PYG{p}{(}\PYG{n}{txt}\PYG{p}{,} \PYG{n}{j}\PYG{p}{)}\PYG{p}{)}
      \PYG{k}{if} \PYG{n}{txt}\PYG{p}{[}\PYG{o}{\PYGZhy{}}\PYG{l+m+mi}{1}\PYG{p}{]} \PYG{o}{!=} \PYG{l+s+s2}{\PYGZdq{}}\PYG{l+s+s2}{X}\PYG{l+s+s2}{\PYGZdq{}}\PYG{p}{:}
        \PYG{k}{if} \PYG{n}{txt}\PYG{p}{[}\PYG{o}{\PYGZhy{}}\PYG{l+m+mi}{1}\PYG{p}{]} \PYG{o}{==} \PYG{l+s+s2}{\PYGZdq{}}\PYG{l+s+s2}{!}\PYG{l+s+s2}{\PYGZdq{}}\PYG{p}{:} \PYG{n}{i}\PYG{o}{.}\PYG{n}{klass} \PYG{o}{=} \PYG{n}{col} \PYG{esc}{\label{klass}}
        \PYG{n}{role} \PYG{o}{=} \PYG{p}{(}\PYG{n}{i}\PYG{o}{.}\PYG{n}{ys} \PYG{k}{if} \PYG{n}{txt}\PYG{p}{[}\PYG{o}{\PYGZhy{}}\PYG{l+m+mi}{1}\PYG{p}{]} \PYG{o+ow}{in} \PYG{l+s+s2}{\PYGZdq{}}\PYG{l+s+s2}{+\PYGZhy{}!}\PYG{l+s+s2}{\PYGZdq{}} \PYG{k}{else} \PYG{n}{i}\PYG{o}{.}\PYG{n}{xs}\PYG{p}{)}
        \PYG{n}{role}\PYG{o}{.}\PYG{n}{append}\PYG{p}{(}\PYG{n}{col}\PYG{p}{)}
\end{Verbatim}
\end{mintbg}
\endgroup
\endminipage\par\noindent
After \cls{Cols} runs, every \cls{Data} carries three views of
its columns: \textsf{xs} (independent), \textsf{ys} (dependent
goals), and \textsf{klass} (the optional symbolic target). The
rest of \texttt{ezr.py} consults these views without re-parsing
names: decisions made once at header creation time propagate
everywhere.

\begin{table*}[!t]
\caption{Axes along which we call \IT{} ``light''. Heavy
comparator values are order-of-magnitude approximations drawn
from the cited references; precise figures depend on
configuration.}
\label{tab:light}
\begin{center}
{\fontsize{7.5}{8}\selectfont
\begin{tabularx}{.95\textwidth}{@{} l p{5cm} Y @{}}
\textbf{Axis} & \textbf{\IT{}} & \textbf{Heavy comparator} \\
\midrule
Lines of code (toolkit only)& \NEW{$<$1000}                       & \texttt{sklearn} $>$200k (toolkit; excludes Python stdlib both sides)~\cite{martinez2021} \\
Runtime dependencies        & Python stdlib only        & \texttt{pandas} + \texttt{sklearn} + \texttt{numpy} + \texttt{scipy} stack \\
Models in active-learn ensemble & 2 (\textsf{best}, \textsf{rest}) & SMAC3: 10+ trees~\cite{hutter2011sequential} \\
Math primitives             & mean, sd, log, exp        & SVM kernels, gradient descent, attention \\
Labels to reach 85--95\% optimum & $<100$               & SMAC3, SHAP, ReliefF: 1{,}000s \NEW{(companion
studies~\cite{ganguly2026lowgodatalightse,Amiraliminimaldata})} \\
Model-update cost           & $O(1)$ row swap (line~\ref{add}) & Full ensemble retrain \\
Install size                & $<1$ MB                   & \texttt{sklearn} $\sim$30 MB; PyTorch $\sim$2 GB \\
Reproducibility cost        & Laptop, seconds           & SMAC3 study used a compute cluster~\cite{hutter2011sequential} \\
Pedagogical complexity      & Weeks of   classroom sessions   (author experience; not yet formally evaluated, see \S\ref{sec:offered}) & \texttt{sklearn} API spans semester-length courses \\
Failure mode                & User can read and patch each line & Bug report queued against upstream maintainers
\end{tabularx}}
\end{center}
\end{table*}

\subsection{The \textsf{add} primitive}

To update a distribution, the \textsf{add} function dispatches
on the type of its first argument:
\begin{itemize}
\item For \cls{Data}, the cached centroid is zapped, the
columns are updated recursively, and the row is either
appended or removed.
\item For \cls{Cols}, it walks each column and recursively
calls \textsf{add} with the appropriate cell.
\item For \cls{Sym}, it adds (or subtracts) one from
\textsf{Sym.has}.
\item For \cls{Num}, it updates \textsf{mu}, \textsf{m2}, and
\textsf{sd} via Welford's algorithm~\cite{Welford01081962}, an
exact and incremental method for updating mean and standard
deviation.
\end{itemize}
\par\nopagebreak\minipage{\linewidth}
\begingroup\mintfvset
\VerbatimPygments{\PYG}{\PYG}
\begin{mintbg}{codebg}
\begin{Verbatim}[commandchars=\\\{\},codes={\catcode`\$=3\catcode`\^=7\catcode`\_=8\relax}]
\PYG{k}{def}\PYG{+w}{ }\PYG{n+nf}{add}\PYG{p}{(}\PYG{n}{it}\PYG{p}{,} \PYG{n}{v}\PYG{p}{,} \PYG{n}{w}\PYG{o}{=}\PYG{l+m+mi}{1}\PYG{p}{)}\PYG{p}{:} \PYG{esc}{\label{add}}
  \PYG{k}{if} \PYG{n}{Data} \PYG{o+ow}{is} \PYG{n+nb}{type}\PYG{p}{(}\PYG{n}{it}\PYG{p}{)}\PYG{p}{:}
    \PYG{n}{it}\PYG{o}{.}\PYG{n}{\PYGZus{}centroid} \PYG{o}{=} \PYG{k+kc}{None} \PYG{esc}{\label{zap}}
    \PYG{n}{add}\PYG{p}{(}\PYG{n}{it}\PYG{o}{.}\PYG{n}{cols}\PYG{p}{,} \PYG{n}{v}\PYG{p}{,} \PYG{n}{w}\PYG{p}{)} \PYG{esc}{\label{addrec1}}
    \PYG{p}{(}\PYG{n}{it}\PYG{o}{.}\PYG{n}{rows}\PYG{o}{.}\PYG{n}{append} \PYG{k}{if} \PYG{n}{w}\PYG{o}{\PYGZgt{}}\PYG{l+m+mi}{0} \PYG{k}{else} \PYG{n}{it}\PYG{o}{.}\PYG{n}{rows}\PYG{o}{.}\PYG{n}{remove}\PYG{p}{)}\PYG{p}{(}\PYG{n}{v}\PYG{p}{)}
  \PYG{k}{elif} \PYG{n}{Cols} \PYG{o+ow}{is} \PYG{n+nb}{type}\PYG{p}{(}\PYG{n}{it}\PYG{p}{)}\PYG{p}{:}
    \PYG{p}{[}\PYG{n}{add}\PYG{p}{(}\PYG{n}{col}\PYG{p}{,} \PYG{n}{v}\PYG{p}{[}\PYG{n}{col}\PYG{o}{.}\PYG{n}{at}\PYG{p}{]}\PYG{p}{,} \PYG{n}{w}\PYG{p}{)}
     \PYG{k}{for} \PYG{n}{col} \PYG{o+ow}{in} \PYG{n}{it}\PYG{o}{.}\PYG{n}{all}\PYG{p}{]} \PYG{esc}{\label{addrec2}}
  \PYG{k}{elif} \PYG{n}{v} \PYG{o}{!=} \PYG{l+s+s2}{\PYGZdq{}}\PYG{l+s+s2}{?}\PYG{l+s+s2}{\PYGZdq{}}\PYG{p}{:} \PYG{c+c1}{\PYGZsh{} skip \PYGZdq{}don\PYGZsq{}t know\PYGZdq{} values}
    \PYG{k}{if} \PYG{n}{Sym} \PYG{o}{==} \PYG{n+nb}{type}\PYG{p}{(}\PYG{n}{it}\PYG{p}{)}\PYG{p}{:} \PYG{n}{it}\PYG{o}{.}\PYG{n}{has}\PYG{p}{[}\PYG{n}{v}\PYG{p}{]} \PYG{o}{=} \PYG{n}{w} \PYG{o}{+} \PYG{n}{it}\PYG{o}{.}\PYG{n}{has}\PYG{o}{.}\PYG{n}{get}\PYG{p}{(}\PYG{n}{v}\PYG{p}{,} \PYG{l+m+mi}{0}\PYG{p}{)} \PYG{esc}{\label{symaddsub}}
    \PYG{k}{elif} \PYG{n}{w} \PYG{o}{\PYGZlt{}} \PYG{l+m+mi}{0} \PYG{o+ow}{and} \PYG{n}{it}\PYG{o}{.}\PYG{n}{n} \PYG{o}{\PYGZlt{}}\PYG{o}{=} \PYG{l+m+mi}{2}\PYG{p}{:}
      \PYG{n}{it}\PYG{o}{.}\PYG{n}{n} \PYG{o}{=} \PYG{n}{it}\PYG{o}{.}\PYG{n}{mu} \PYG{o}{=} \PYG{n}{it}\PYG{o}{.}\PYG{n}{m2} \PYG{o}{=} \PYG{n}{it}\PYG{o}{.}\PYG{n}{sd} \PYG{o}{=} \PYG{l+m+mi}{0}
    \PYG{k}{else}\PYG{p}{:} \PYG{c+c1}{\PYGZsh{} Welford\PYGZsq{}s algorithm \label{welford}}
      \PYG{n}{it}\PYG{o}{.}\PYG{n}{n}  \PYG{o}{+}\PYG{o}{=} \PYG{n}{w}
      \PYG{n}{delta}  \PYG{o}{=} \PYG{n}{v} \PYG{o}{\PYGZhy{}} \PYG{n}{it}\PYG{o}{.}\PYG{n}{mu}
      \PYG{n}{it}\PYG{o}{.}\PYG{n}{mu} \PYG{o}{+}\PYG{o}{=} \PYG{n}{w} \PYG{o}{*} \PYG{n}{delta} \PYG{o}{/} \PYG{n}{it}\PYG{o}{.}\PYG{n}{n}
      \PYG{n}{it}\PYG{o}{.}\PYG{n}{m2} \PYG{o}{+}\PYG{o}{=} \PYG{n}{w} \PYG{o}{*} \PYG{n}{delta} \PYG{o}{*} \PYG{p}{(}\PYG{n}{v} \PYG{o}{\PYGZhy{}} \PYG{n}{it}\PYG{o}{.}\PYG{n}{mu}\PYG{p}{)}
      \PYG{n}{it}\PYG{o}{.}\PYG{n}{sd} \PYG{o}{=} \PYG{p}{(}\PYG{n}{sqrt}\PYG{p}{(}\PYG{n+nb}{max}\PYG{p}{(}\PYG{l+m+mi}{0}\PYG{p}{,}\PYG{n}{it}\PYG{o}{.}\PYG{n}{m2}\PYG{p}{)}\PYG{o}{/}\PYG{p}{(}\PYG{n}{it}\PYG{o}{.}\PYG{n}{n}\PYG{o}{\PYGZhy{}}\PYG{l+m+mi}{1}\PYG{p}{)}\PYG{p}{)} \PYG{k}{if} \PYG{n}{it}\PYG{o}{.}\PYG{n}{n}\PYG{o}{\PYGZgt{}}\PYG{l+m+mi}{1} \PYG{k}{else} \PYG{l+m+mi}{0}\PYG{p}{)}
  \PYG{k}{return} \PYG{n}{v}
\end{Verbatim}
\end{mintbg}
\endgroup
\endminipage\par\noindent
Three details inside \textsf{add} are worth flagging.

\textit{First, the symmetry that makes data movement cheap.}
The \textsf{w} parameter handles both addition (\textsf{w=1})
and subtraction \mbox{(\textsf{w=-1})}, via:
\par\nopagebreak\minipage{\linewidth}
\begingroup\mintfvset
\VerbatimPygments{\PYG}{\PYG}
\begin{mintbg}{codebg}
\begin{Verbatim}[commandchars=\\\{\},codes={\catcode`\$=3\catcode`\^=7\catcode`\_=8\relax}]
\PYG{k}{def}\PYG{+w}{ }\PYG{n+nf}{sub}\PYG{p}{(}\PYG{n}{it}\PYG{p}{,} \PYG{n}{v}\PYG{p}{)}\PYG{p}{:} \PYG{k}{return} \PYG{n}{add}\PYG{p}{(}\PYG{n}{it}\PYG{p}{,} \PYG{n}{v}\PYG{p}{,} \PYG{n}{w}\PYG{o}{=}\PYG{o}{\PYGZhy{}}\PYG{l+m+mi}{1}\PYG{p}{)}
\end{Verbatim}
\end{mintbg}
\endgroup
\endminipage\par\noindent
This symmetry matters because it lets us move a row from one
\cls{Data} to another in two constant-time operations: $\pm 1$
to a counter for symbols, or a single Welford update for
numbers. There is no retraining step, no sorting, no large-scale
data recollection. The active learner of \S\ref{sec:active}
rests on this property.

\textit{Second, the lazy centroid cache.} Every \cls{Data}
carries a \textsf{\_centroid} field that \textsf{add}
invalidates on any change (line~\ref{zap}). The \textsf{mids}
accessor lazily rebuilds it:
\par\nopagebreak\minipage{\linewidth}
\begingroup\mintfvset
\VerbatimPygments{\PYG}{\PYG}
\begin{mintbg}{codebg}
\begin{Verbatim}[commandchars=\\\{\},codes={\catcode`\$=3\catcode`\^=7\catcode`\_=8\relax}]
\PYG{k}{def}\PYG{+w}{ }\PYG{n+nf}{mids}\PYG{p}{(}\PYG{n}{data}\PYG{p}{)}\PYG{p}{:} \PYG{esc}{\label{mids}}
  \PYG{n}{data}\PYG{o}{.}\PYG{n}{\PYGZus{}centroid} \PYG{o}{=} \PYG{n}{data}\PYG{o}{.}\PYG{n}{\PYGZus{}centroid} \PYG{o+ow}{or} \PYG{p}{[}
                       \PYG{n}{mid}\PYG{p}{(}\PYG{n}{col}\PYG{p}{)} \PYG{k}{for} \PYG{n}{col} \PYG{o+ow}{in} \PYG{n}{data}\PYG{o}{.}\PYG{n}{cols}\PYG{o}{.}\PYG{n}{all}\PYG{p}{]}
  \PYG{k}{return} \PYG{n}{data}\PYG{o}{.}\PYG{n}{\PYGZus{}centroid}
\end{Verbatim}
\end{mintbg}
\endgroup
\endminipage\par\noindent
Without the cache, each \textsf{mids} call would walk every
column. With it, the call is one dictionary lookup until the
next \textsf{add}. This detail will become important when we
discuss active learning in \S\ref{sec:active}.

\textit{Third, \textsf{add} returns the value it stored.} That
return value enables a nested-call idiom used several times in
\IT{}, where the same value is removed from one distribution
and added to another in one expression; e.g.
\begingroup\mintfvset
\fvset{numbers=none}
\VerbatimPygments{\PYG}{\PYG}
\begin{mintbg}{codebg}
\begin{Verbatim}[commandchars=\\\{\},codes={\catcode`\$=3\catcode`\^=7\catcode`\_=8\relax}]
\PYG{n}{add}\PYG{p}{(}\PYG{n}{rest}\PYG{o}{.}\PYG{n}{cols}\PYG{p}{,} \PYG{n}{sub}\PYG{p}{(}\PYG{n}{best}\PYG{o}{.}\PYG{n}{cols}\PYG{p}{,} \PYG{n}{best}\PYG{o}{.}\PYG{n}{rows}\PYG{o}{.}\PYG{n}{pop}\PYG{p}{(}\PYG{p}{)}\PYG{p}{)}\PYG{p}{)}
\end{Verbatim}
\end{mintbg}
\endgroup

\subsection{Design notes}

Per Martin's clean-code rules\footnote{Rule 1: functions
should be small. Rule 2: they should be smaller than
that~\cite{martin2009clean}.}, \texttt{ezr.py}'s 44 functions
average $5\pm4$ lines. Each line is under 60 characters wide
so it can be included in documents like this one without
reformatting.

Polymorphism is implemented by dispatching on type
inside functions, as seen in \textsf{add}, \textsf{mid},
\textsf{spread}, and (later) \textsf{pick}.

Following Tempero et al.\ \cite{tempero13},
  \IT{} prefers composition over inheritance. Most reported
  inheritance use in practice is shallow and replaceable; modern
  Python supports this style natively via structural typing
  (PEP~544 protocols~\cite{pep544}) and dataclasses
  (PEP~557~\cite{pep557}).

\IT{} prefers functions over methods. Methods scatter
related code across classes; functions let related code group
together.
A note on naming. \IT{}'s identifiers are deliberately terse
(\textsf{i} for self, \textsf{it} for a dispatch argument,
\textsf{v},~\textsf{w} for value and weight), in the style of
mathematical code, and two conventions co-exist: multi-word
algorithm entry points use camelCase (\textsf{treeGrow}) while
experiment scaffolding uses snake\_case (\textsf{warm\_start},
\textsf{test\_acquire}). Readers who prefer longer names or a
typed, sklearn-style facade could add one in a few dozen lines;
we have kept the core terse so that whole algorithms remain
visible on one page. The \textsf{mid} and \textsf{spread} functions on
lines~\ref{mid0} and \ref{spread0} are examples of functions
that group functionality across different classes.

That is the entire substrate: four classes, one update
primitive, and a handful of supporting functions
(\textsf{mid}, \textsf{spread}, \textsf{norm}, \textsf{mids},
\textsf{clone}). From here on, every algorithm in this paper
is a way of asking these objects a different question. Note
that all this is achieved without including vast libraries
of other people's code.

\subsection{What do we mean by ``light''?}\label{sec:light}

\IT{} is light along the ten axes of Table~\ref{tab:light}:
code size, dependency surface, ensemble width, mathematical
machinery, label hunger, model-update cost, install footprint,
reproducibility cost (laptop seconds vs.\ cluster hours),
pedagogical complexity \NEW{(weeks of sessions vs.\ a semester
course)}, and
failure mode (user-patchable vs.\ upstream-queued). \NEW{On all
but the pedagogy axis, \IT{} sits one or more orders of
magnitude below the corresponding state of the art; the pedagogy
gap is smaller (weeks vs.\ a semester) and, as
Table~\ref{tab:light} notes, rests on author experience rather
than formal evaluation.} 

Two examples of this are already visible from the code above. First,
the SMAC3~\cite{hutter2011sequential} algorithm (used in \S\ref{sec:active})
maintains an ensemble of ten
or more regression trees and rebuilds them after each new label;
\IT{} maintains two \cls{Data} objects (\textsf{best} and
\textsf{rest}, \S\ref{sec:active}) and updates them in constant
time using the \textsf{add} primitive shown on
line~\ref{add}. Second, relevance filtering in RAG (discussed in \S\ref{sec:textmine}) is usually
framed as a deep-embedding problem requiring billion-parameter
encoders or kernel support vector machines~\cite{yu2018total}. 
\IT{}'s complementary Bayes filter
(see \S\ref{sec:textmine}), on the other hand, achieves comparable recall using the
Welford-style running statistics already present in \cls{Num} and
\cls{Sym}. The remaining sections of this paper provide further
evidence along each row of Table~\ref{tab:light}.

\section{Three Classic Algorithms in 70 Lines}\label{sec:famous}

Classification, clustering, and seeded clustering
(by which we mean the $k$-means++ initialization heuristic,
not semi-supervised clustering) are usually
presented as separate topics: separate chapters in textbooks such as
{\em Data Mining} by Witten et al.~\cite{Witten:2005}, separate
lectures in standard curricula, and separate modules with separate
APIs in libraries like scikit-learn. We do not claim these
algorithms are interchangeable; their scoring rules genuinely
differ. We do claim the conventional packaging hides how much
machinery they share. This section asks whether that separation
is necessary at the level of code.
 This section shows that, on top of the
substrate from \S\ref{sec:peek}, they are nearly the same
algorithm asking slightly different questions. Naïve Bayes takes two
functions. $k$-means takes thirteen lines. $k$-means++ takes
eleven. Between them, they add 30 lines of new code and one new
primitive (\textsf{pick}, which earns its place by also serving
as the mutation operator in \S\ref{sec:opt}).

The deep reason these three algorithms collapse together is
simple: each one reduces to the same underlying question. Given
some columns that summarize what we have seen, how does a new
row score against them? Naïve Bayes scores by likelihood;
$k$-means scores by distance; $k$-means++ scores by squared
distance weighted into a probability. The scoring rule changes;
the substrate that holds the summaries does not.

\subsection{Naïve Bayes}

The core of Naïve Bayes is one question: how much does a column
{\em like} a value? \textsf{like} answers that question. For a
\cls{Sym}, the answer is a smoothed frequency count from
\textsf{has} (the same dictionary \textsf{add} updates at
line~\ref{symaddsub}). For a \cls{Num}, the answer is a Gaussian
PDF using the running \textsf{mu} and \textsf{sd} that
\textsf{add} maintains via Welford (line~\ref{welford}):

\par\nopagebreak\minipage{\linewidth}
\begingroup\mintfvset
\VerbatimPygments{\PYG}{\PYG}
\begin{mintbg}{codebg}
\begin{Verbatim}[commandchars=\\\{\},codes={\catcode`\$=3\catcode`\^=7\catcode`\_=8\relax}]
\PYG{n}{the}\PYG{o}{.}\PYG{n}{bayes}\PYG{o}{.}\PYG{n}{k} \PYG{o}{=} \PYG{l+m+mi}{1}   \PYG{c+c1}{\PYGZsh{} smoothing config}

\PYG{k}{def}\PYG{+w}{ }\PYG{n+nf}{like}\PYG{p}{(}\PYG{n}{col}\PYG{p}{,} \PYG{n}{v}\PYG{p}{,} \PYG{n}{prior}\PYG{p}{)}\PYG{p}{:}
  \PYG{k}{if} \PYG{n+nb}{type}\PYG{p}{(}\PYG{n}{col}\PYG{p}{)} \PYG{o}{==} \PYG{n}{Sym}\PYG{p}{:}
    \PYG{k}{return} \PYG{p}{(}\PYG{n}{col}\PYG{o}{.}\PYG{n}{has}\PYG{o}{.}\PYG{n}{get}\PYG{p}{(}\PYG{n}{v}\PYG{p}{,}\PYG{l+m+mi}{0}\PYG{p}{)} \PYG{o}{+} \PYG{n}{the}\PYG{o}{.}\PYG{n}{bayes}\PYG{o}{.}\PYG{n}{k}\PYG{o}{*}\PYG{n}{prior}
           \PYG{p}{)} \PYG{o}{/} \PYG{p}{(}\PYG{n}{col}\PYG{o}{.}\PYG{n}{n} \PYG{o}{+} \PYG{n}{the}\PYG{o}{.}\PYG{n}{bayes}\PYG{o}{.}\PYG{n}{k}\PYG{p}{)}
  \PYG{n}{sd} \PYG{o}{=} \PYG{n}{col}\PYG{o}{.}\PYG{n}{sd} \PYG{o}{+} \PYG{l+m+mf}{1e\PYGZhy{}32}
  \PYG{n}{z}  \PYG{o}{=} \PYG{l+m+mi}{2}\PYG{o}{*}\PYG{n}{sd}\PYG{o}{*}\PYG{n}{sd}
  \PYG{k}{return} \PYG{n}{exp}\PYG{p}{(}\PYG{o}{\PYGZhy{}}\PYG{p}{(}\PYG{n}{v}\PYG{o}{\PYGZhy{}}\PYG{n}{col}\PYG{o}{.}\PYG{n}{mu}\PYG{p}{)}\PYG{o}{*}\PYG{o}{*}\PYG{l+m+mi}{2} \PYG{o}{/} \PYG{n}{z}\PYG{p}{)} \PYG{o}{/} \PYG{n}{sqrt}\PYG{p}{(}\PYG{n}{pi}\PYG{o}{*}\PYG{n}{z}\PYG{p}{)}
\end{Verbatim}
\end{mintbg}
\endgroup
\endminipage\par\noindent
Note what is not here: no fitting pass, no separate training
phase. The \cls{Num} and \cls{Sym} columns inside a \cls{Data}
are already a fitted Bayes model, because \textsf{add} kept them
fitted incrementally as rows arrived. \textsf{like} just reads
them. That is a non-trivial property: the substrate has erased
the distinction between fitting and using.

\textsf{likes} aggregates one row's worth of evidence. It
multiplies per-column likelihoods (in log space, for numerical
stability) with an $m$-estimate prior on class size:

\par\nopagebreak\minipage{\linewidth}
\begingroup\mintfvset
\VerbatimPygments{\PYG}{\PYG}
\begin{mintbg}{codebg}
\begin{Verbatim}[commandchars=\\\{\},codes={\catcode`\$=3\catcode`\^=7\catcode`\_=8\relax}]
\PYG{n}{the}\PYG{o}{.}\PYG{n}{bayes}\PYG{o}{.}\PYG{n}{m} \PYG{o}{=} \PYG{l+m+mi}{2}   \PYG{c+c1}{\PYGZsh{} prior config}

\PYG{k}{def}\PYG{+w}{ }\PYG{n+nf}{likes}\PYG{p}{(}\PYG{n}{data}\PYG{p}{,} \PYG{n}{row}\PYG{p}{,} \PYG{n}{n\PYGZus{}rows}\PYG{p}{,} \PYG{n}{n\PYGZus{}klasses}\PYG{p}{)}\PYG{p}{:}
  \PYG{n}{prior} \PYG{o}{=} \PYG{p}{(}\PYG{n+nb}{len}\PYG{p}{(}\PYG{n}{data}\PYG{o}{.}\PYG{n}{rows}\PYG{p}{)} \PYG{o}{+} \PYG{n}{the}\PYG{o}{.}\PYG{n}{bayes}\PYG{o}{.}\PYG{n}{m}
          \PYG{p}{)} \PYG{o}{/} \PYG{p}{(}\PYG{n}{n\PYGZus{}rows} \PYG{o}{+} \PYG{n}{the}\PYG{o}{.}\PYG{n}{bayes}\PYG{o}{.}\PYG{n}{m}\PYG{o}{*}\PYG{n}{n\PYGZus{}klasses}\PYG{p}{)}
  \PYG{n}{ls} \PYG{o}{=} \PYG{p}{[}\PYG{n}{like}\PYG{p}{(}\PYG{n}{col}\PYG{p}{,} \PYG{n}{v}\PYG{p}{,} \PYG{n}{prior}\PYG{p}{)}
        \PYG{k}{for} \PYG{n}{col} \PYG{o+ow}{in} \PYG{n}{data}\PYG{o}{.}\PYG{n}{cols}\PYG{o}{.}\PYG{n}{xs}
        \PYG{k}{if} \PYG{p}{(}\PYG{n}{v} \PYG{o}{:=} \PYG{n}{row}\PYG{p}{[}\PYG{n}{col}\PYG{o}{.}\PYG{n}{at}\PYG{p}{]}\PYG{p}{)} \PYG{o}{!=} \PYG{l+s+s2}{\PYGZdq{}}\PYG{l+s+s2}{?}\PYG{l+s+s2}{\PYGZdq{}}\PYG{p}{]}
  \PYG{k}{return} \PYG{n}{log}\PYG{p}{(}\PYG{n}{prior}\PYG{p}{)} \PYG{o}{+} \PYG{n+nb}{sum}\PYG{p}{(}\PYG{n}{log}\PYG{p}{(}\PYG{n}{v}\PYG{p}{)}  \PYG{k}{for} \PYG{n}{v} \PYG{o+ow}{in} \PYG{n}{ls} \PYG{k}{if} \PYG{n}{v} \PYG{o}{\PYGZgt{}} \PYG{l+m+mi}{0}\PYG{p}{)}
\end{Verbatim}
\end{mintbg}
\endgroup
\endminipage\par\noindent
Two functions, complete with Laplace and $m$-estimate smoothing,
Gaussian numeric attributes, and missing-value handling. The
streaming test-then-train protocol that ordinarily accompanies
Naïve Bayes (classify each row with the model so far, then add
it to its true class's \cls{Data}) is a thin loop on top of
these two functions; we omit it here to keep the focus on the
algorithm itself.\footnote{The full \textsf{classify} loop is
available in \texttt{ezr.py}; it adds about a dozen lines for
streaming bookkeeping (incremental confusion matrix, a warm-up
threshold before scoring begins, and per-class \cls{Data}
maintenance).}

\subsection{$k$-means}\label{kluster}

The $k$-means clustering algorithm finds clusters by repeatedly
{\em guessing} the centroids, then {\em updating} them.  When
{\em guessing}, it labels each row with its nearest centroid
(initially, the centroids are just $k$ randomly selected rows).
{\em Update} moves centroids to the middle of the rows assigned
to them.

To talk about ``nearest centroid'' we need a distance measure. Rows in
\IT{} have mixed numeric and symbolic columns, so we need a
per-column distance that handles both, and a way to combine them
across columns. The per-column distance is \textsf{aha} (after
Aha's 1991 instance-based reasoning algorithm~\cite{aha91}); the
combiner is the Minkowski $p$-norm:

\par\nopagebreak\minipage{\linewidth}
\begingroup\mintfvset
\VerbatimPygments{\PYG}{\PYG}
\begin{mintbg}{codebg}
\begin{Verbatim}[commandchars=\\\{\},codes={\catcode`\$=3\catcode`\^=7\catcode`\_=8\relax}]
\PYG{n}{the}\PYG{o}{.}\PYG{n}{p} \PYG{o}{=} \PYG{l+m+mi}{2} \PYG{c+c1}{\PYGZsh{} config}

\PYG{k}{def}\PYG{+w}{ }\PYG{n+nf}{distx}\PYG{p}{(}\PYG{n}{data}\PYG{p}{,} \PYG{n}{r1}\PYG{p}{,} \PYG{n}{r2}\PYG{p}{)}\PYG{p}{:}
  \PYG{k}{return} \PYG{n}{minkowski}\PYG{p}{(}\PYG{p}{(}\PYG{n}{aha}\PYG{p}{(}\PYG{n}{col}\PYG{p}{,} \PYG{n}{r1}\PYG{p}{[}\PYG{n}{col}\PYG{o}{.}\PYG{n}{at}\PYG{p}{]}\PYG{p}{,} \PYG{n}{r2}\PYG{p}{[}\PYG{n}{col}\PYG{o}{.}\PYG{n}{at}\PYG{p}{]}\PYG{p}{)}
                   \PYG{k}{for} \PYG{n}{col} \PYG{o+ow}{in} \PYG{n}{data}\PYG{o}{.}\PYG{n}{cols}\PYG{o}{.}\PYG{n}{xs}\PYG{p}{)}\PYG{p}{,} \PYG{n}{p}\PYG{o}{=}\PYG{n}{the}\PYG{o}{.}\PYG{n}{p}\PYG{p}{)}

\PYG{k}{def}\PYG{+w}{ }\PYG{n+nf}{aha}\PYG{p}{(}\PYG{n}{col}\PYG{p}{,} \PYG{n}{u}\PYG{p}{,} \PYG{n}{v}\PYG{p}{)}\PYG{p}{:}
  \PYG{k}{if} \PYG{n}{u} \PYG{o}{==} \PYG{n}{v} \PYG{o}{==} \PYG{l+s+s2}{\PYGZdq{}}\PYG{l+s+s2}{?}\PYG{l+s+s2}{\PYGZdq{}}\PYG{p}{:} \PYG{k}{return} \PYG{l+m+mi}{1}
  \PYG{k}{if} \PYG{n}{Sym} \PYG{o}{==} \PYG{n+nb}{type}\PYG{p}{(}\PYG{n}{col}\PYG{p}{)}\PYG{p}{:} \PYG{k}{return} \PYG{n}{u} \PYG{o}{!=} \PYG{n}{v}
  \PYG{n}{u}\PYG{p}{,} \PYG{n}{v} \PYG{o}{=} \PYG{n}{norm}\PYG{p}{(}\PYG{n}{col}\PYG{p}{,} \PYG{n}{u}\PYG{p}{)}\PYG{p}{,} \PYG{n}{norm}\PYG{p}{(}\PYG{n}{col}\PYG{p}{,} \PYG{n}{v}\PYG{p}{)}
  \PYG{n}{u} \PYG{o}{=} \PYG{n}{u} \PYG{k}{if} \PYG{n}{u}\PYG{o}{!=}\PYG{l+s+s2}{\PYGZdq{}}\PYG{l+s+s2}{?}\PYG{l+s+s2}{\PYGZdq{}} \PYG{k}{else} \PYG{p}{(}\PYG{l+m+mi}{0} \PYG{k}{if} \PYG{n}{v}\PYG{o}{\PYGZgt{}}\PYG{l+m+mf}{0.5} \PYG{k}{else} \PYG{l+m+mi}{1}\PYG{p}{)}
  \PYG{n}{v} \PYG{o}{=} \PYG{n}{v} \PYG{k}{if} \PYG{n}{v}\PYG{o}{!=}\PYG{l+s+s2}{\PYGZdq{}}\PYG{l+s+s2}{?}\PYG{l+s+s2}{\PYGZdq{}} \PYG{k}{else} \PYG{p}{(}\PYG{l+m+mi}{0} \PYG{k}{if} \PYG{n}{u}\PYG{o}{\PYGZgt{}}\PYG{l+m+mf}{0.5} \PYG{k}{else} \PYG{l+m+mi}{1}\PYG{p}{)}
  \PYG{k}{return} \PYG{n+nb}{abs}\PYG{p}{(}\PYG{n}{u} \PYG{o}{\PYGZhy{}} \PYG{n}{v}\PYG{p}{)}

\PYG{k}{def}\PYG{+w}{ }\PYG{n+nf}{minkowski}\PYG{p}{(}\PYG{n}{items}\PYG{p}{,} \PYG{n}{p}\PYG{o}{=}\PYG{l+m+mi}{2}\PYG{p}{)}\PYG{p}{:} \PYG{c+c1}{\PYGZsh{} p=2 means \PYGZdq{}Euclidean\PYGZdq{}  \label{minkowski}}
  \PYG{n}{tot}\PYG{p}{,} \PYG{n}{n} \PYG{o}{=} \PYG{l+m+mi}{0}\PYG{p}{,} \PYG{l+m+mf}{1e\PYGZhy{}32}
  \PYG{k}{for} \PYG{n}{item} \PYG{o+ow}{in} \PYG{n}{items}\PYG{p}{:}  \PYG{n}{tot}\PYG{p}{,} \PYG{n}{n} \PYG{o}{=} \PYG{n}{tot} \PYG{o}{+} \PYG{n}{item}\PYG{o}{*}\PYG{o}{*}\PYG{n}{p}\PYG{p}{,} \PYG{n}{n} \PYG{o}{+} \PYG{l+m+mi}{1}
  \PYG{k}{return} \PYG{p}{(}\PYG{n}{tot}\PYG{o}{/}\PYG{n}{n}\PYG{p}{)} \PYG{o}{*}\PYG{o}{*} \PYG{p}{(}\PYG{l+m+mi}{1}\PYG{o}{/}\PYG{n}{p}\PYG{p}{)}
\end{Verbatim}
\end{mintbg}
\endgroup
\endminipage\par\noindent
The \textsf{aha} rules encode three   policies. If both
values are missing, their distance is maximal (Aha's
heuristic). For symbolic columns, distance is whether the two
values differ. For numeric columns, values are normalized to
$[0,1]$; if one value is missing, it is placed at the opposite
end of the known value, making the comparison deliberately
pessimistic.

With \textsf{distx} in hand, $k$-means itself is short. EZR uses
the \cls{Data} class to store each cluster: a ``cluster'' in
\IT{} is just a \cls{Data} populated by \textsf{add}:

\par\nopagebreak\minipage{\linewidth}
\begingroup\mintfvset
\VerbatimPygments{\PYG}{\PYG}
\begin{mintbg}{codebg}
\begin{Verbatim}[commandchars=\\\{\},codes={\catcode`\$=3\catcode`\^=7\catcode`\_=8\relax}]
\PYG{k}{def}\PYG{+w}{ }\PYG{n+nf}{kmeans}\PYG{p}{(}\PYG{n}{d}\PYG{p}{,} \PYG{n}{rs}\PYG{o}{=}\PYG{k+kc}{None}\PYG{p}{,} \PYG{n}{k}\PYG{o}{=}\PYG{l+m+mi}{10}\PYG{p}{,} \PYG{n}{n}\PYG{o}{=}\PYG{l+m+mi}{10}\PYG{p}{,} \PYG{n}{cents}\PYG{o}{=}\PYG{k+kc}{None}\PYG{p}{)}\PYG{p}{:}
  \PYG{n}{rs} \PYG{o}{=} \PYG{n}{rs} \PYG{o+ow}{or} \PYG{n}{d}\PYG{o}{.}\PYG{n}{rows}
  \PYG{n}{cents} \PYG{o}{=} \PYG{n}{cents} \PYG{o+ow}{or} \PYG{n}{choices}\PYG{p}{(}\PYG{n}{rs}\PYG{p}{,} \PYG{n}{k}\PYG{o}{=}\PYG{n}{k}\PYG{p}{)}
  \PYG{n}{out} \PYG{o}{=} \PYG{p}{[}\PYG{p}{]}
  \PYG{k}{for} \PYG{n}{\PYGZus{}} \PYG{o+ow}{in} \PYG{n+nb}{range}\PYG{p}{(}\PYG{n}{n}\PYG{p}{)}\PYG{p}{:}
    \PYG{n}{out} \PYG{o}{=} \PYG{p}{[}\PYG{n}{clone}\PYG{p}{(}\PYG{n}{d}\PYG{p}{)} \PYG{k}{for} \PYG{n}{\PYGZus{}} \PYG{o+ow}{in} \PYG{n}{cents}\PYG{p}{]} \PYG{esc}{\label{centroids}}
    \PYG{k}{for} \PYG{n}{r} \PYG{o+ow}{in} \PYG{n}{rs}\PYG{p}{:}
      \PYG{n}{add}\PYG{p}{(}\PYG{n}{out}\PYG{p}{[}\PYG{n+nb}{min}\PYG{p}{(}\PYG{n+nb}{range}\PYG{p}{(}\PYG{n+nb}{len}\PYG{p}{(}\PYG{n}{cents}\PYG{p}{)}\PYG{p}{)}\PYG{p}{,}
                  \PYG{n}{key}\PYG{o}{=}\PYG{k}{lambda} \PYG{n}{j}\PYG{p}{:} \PYG{n}{distx}\PYG{p}{(}\PYG{n}{d}\PYG{p}{,} \PYG{n}{cents}\PYG{p}{[}\PYG{n}{j}\PYG{p}{]}\PYG{p}{,} \PYG{n}{r}\PYG{p}{)}\PYG{p}{)}\PYG{p}{]}\PYG{p}{,} \PYG{n}{r}\PYG{p}{)}
    \PYG{n}{cents} \PYG{o}{=} \PYG{p}{[}\PYG{n}{mids}\PYG{p}{(}\PYG{n}{kid}\PYG{p}{)}  \PYG{c+c1}{\PYGZsh{} for \PYGZdq{}mids\PYGZdq{} see line \ref{mids}}
             \PYG{k}{for} \PYG{n}{kid} \PYG{o+ow}{in} \PYG{n}{out} \PYG{k}{if} \PYG{n}{kid}\PYG{o}{.}\PYG{n}{rows}\PYG{p}{]}
  \PYG{k}{return} \PYG{n}{out}
\end{Verbatim}
\end{mintbg}
\endgroup
\endminipage\par\noindent
What is worth noting about \textsf{kmeans} is what it does
\emph{not} introduce. Every function it calls already existed
before \textsf{kmeans} was written. \textsf{clone}
(\S\ref{sec:peek}) makes the empty cluster shells. \textsf{add}
fills them. \textsf{distx} measures membership. \textsf{mids}
produces the new centroids. The clustering loop is little
more than a gluing-together of pre-existing parts.

This is the substrate-reuse pattern at work. The cost of
adding a new algorithm is not the cost of building the
algorithm from scratch; it is only the cost of writing the
loop that calls the substrate.

\subsection{$k$-means++}

$k$-means++~\cite{arthur2007kmeanspp} is a heuristic for picking
the initial clusters given to $k$-means. The heuristic picks one
centroid at random, then picks each subsequent centroid with
probability proportional to its squared distance from the
nearest existing centroid. ``With probability proportional to''
is the part that needs new machinery: nothing in the substrate
so far samples from a weighted distribution.

To sample from a weighted distribution, we use the  
\textsf{pick} sampler. This is a very useful function: we use 
it here as well as when we explore
mutation-based optimization in \S\ref{sec:opt}.
That is  the same code that seeds $k$-means++ and can 
generate mutants for simulated annealing and local search. 

 To pick from \cls{Syms}, we use a 
roulette-wheel sampler: successively subtract from a probability mass,
stopping when we run out of values
 (lines~\ref{wheel0} to \ref{wheel1}).
\par\nopagebreak\minipage{\linewidth}
\begingroup\mintfvset
\VerbatimPygments{\PYG}{\PYG}
\begin{mintbg}{codebg}
\begin{Verbatim}[commandchars=\\\{\},codes={\catcode`\$=3\catcode`\^=7\catcode`\_=8\relax}]
\PYG{k}{def}\PYG{+w}{ }\PYG{n+nf}{pick}\PYG{p}{(}\PYG{n}{it}\PYG{p}{,} \PYG{n}{v}\PYG{o}{=}\PYG{k+kc}{None}\PYG{p}{)}\PYG{p}{:} \PYG{esc}{\label{pick}}
  \PYG{k}{if} \PYG{n}{Sym} \PYG{o}{==} \PYG{n+nb}{type}\PYG{p}{(}\PYG{n}{it}\PYG{p}{)}\PYG{p}{:} \PYG{k}{return} \PYG{n}{pick}\PYG{p}{(}\PYG{n}{it}\PYG{o}{.}\PYG{n}{has}\PYG{p}{)}
  \PYG{k}{if} \PYG{n}{Num} \PYG{o}{==} \PYG{n+nb}{type}\PYG{p}{(}\PYG{n}{it}\PYG{p}{)}\PYG{p}{:}
    \PYG{n}{mu} \PYG{o}{=} \PYG{p}{(}\PYG{n}{v} \PYG{k}{if} \PYG{n}{v} \PYG{o+ow}{is} \PYG{o+ow}{not} \PYG{k+kc}{None} \PYG{o+ow}{and} \PYG{n}{v} \PYG{o}{!=} \PYG{l+s+s2}{\PYGZdq{}}\PYG{l+s+s2}{?}\PYG{l+s+s2}{\PYGZdq{}}  \PYG{k}{else} \PYG{n}{it}\PYG{o}{.}\PYG{n}{mu}\PYG{p}{)} \PYG{esc}{ \label{mupick} }
    \PYG{n}{lo} \PYG{o}{=} \PYG{n}{it}\PYG{o}{.}\PYG{n}{mu} \PYG{o}{\PYGZhy{}} \PYG{l+m+mi}{3}\PYG{o}{*}\PYG{n}{it}\PYG{o}{.}\PYG{n}{sd}
    \PYG{n}{hi} \PYG{o}{=} \PYG{n}{it}\PYG{o}{.}\PYG{n}{mu} \PYG{o}{+} \PYG{l+m+mi}{3}\PYG{o}{*}\PYG{n}{it}\PYG{o}{.}\PYG{n}{sd}
    \PYG{n}{new} \PYG{o}{=} \PYG{n}{mu} \PYG{o}{+} \PYG{n}{it}\PYG{o}{.}\PYG{n}{sd}\PYG{o}{*}\PYG{l+m+mi}{2}\PYG{o}{*}\PYG{p}{(}\PYG{n}{rand}\PYG{p}{(}\PYG{p}{)} \PYG{o}{+} \PYG{n}{rand}\PYG{p}{(}\PYG{p}{)} \PYG{o}{+} \PYG{n}{rand}\PYG{p}{(}\PYG{p}{)} \PYG{o}{\PYGZhy{}} \PYG{l+m+mf}{1.5}\PYG{p}{)}
    \PYG{k}{return} \PYG{n}{lo} \PYG{o}{+} \PYG{p}{(}\PYG{n}{new} \PYG{o}{\PYGZhy{}} \PYG{n}{lo}\PYG{p}{)} \PYG{o}{\PYGZpc{}} \PYG{p}{(}\PYG{n}{hi} \PYG{o}{\PYGZhy{}} \PYG{n}{lo} \PYG{o}{+} \PYG{l+m+mf}{1e\PYGZhy{}32}\PYG{p}{)}
  \PYG{k}{if} \PYG{n+nb}{dict} \PYG{o}{==} \PYG{n+nb}{type}\PYG{p}{(}\PYG{n}{it}\PYG{p}{)}\PYG{p}{:}
    \PYG{n}{n} \PYG{o}{=} \PYG{n+nb}{sum}\PYG{p}{(}\PYG{n}{it}\PYG{o}{.}\PYG{n}{values}\PYG{p}{(}\PYG{p}{)}\PYG{p}{)} \PYG{o}{*} \PYG{n}{rand}\PYG{p}{(}\PYG{p}{)} \PYG{esc}{ \label{wheel0} }
    \PYG{k}{for} \PYG{n}{k}\PYG{p}{,} \PYG{n}{v} \PYG{o+ow}{in} \PYG{n}{it}\PYG{o}{.}\PYG{n}{items}\PYG{p}{(}\PYG{p}{)}\PYG{p}{:}
      \PYG{k}{if} \PYG{p}{(}\PYG{n}{n} \PYG{o}{:=} \PYG{n}{n} \PYG{o}{\PYGZhy{}} \PYG{n}{v}\PYG{p}{)} \PYG{o}{\PYGZlt{}}\PYG{o}{=} \PYG{l+m+mi}{0}\PYG{p}{:} \PYG{k}{break}
    \PYG{k}{return} \PYG{n}{k} \PYG{esc}{ \label{wheel1} }
\end{Verbatim}
\end{mintbg}
\endgroup
\endminipage\par\noindent
To pick from \cls{Num}s, we draw from a distribution centered on a supplied $\mu$ (defaulting to
the column mean) using the Irwin-Hall
sampler~\cite{johnson1995continuous}. 
To understand Irwin-Hall, note that the sum of $n$ uniformly
distributed random numbers is approximately normally distributed.
Centering and scaling the sum of three uniforms,
$2(\textsf{rand}() + \textsf{rand}() + \textsf{rand}() - 1.5)$,
therefore yields a cheap Gaussian approximation with mean~0 and
standard deviation~1, whose values are bounded to $\pm 3$; this
matches the $\pm 3\sigma$ clamp applied by \textsf{lo} and
\textsf{hi} on the lines above. Mutation uses the first form (perturb around
a specific row's value); $k$-means++ uses the dict branch.

Once \textsf{pick} is available, $k$-means++ is just (a)~build
the weighted dict, (b)~pick from it, (c)~repeat:

\par\nopagebreak\minipage{\linewidth}
\begingroup\mintfvset
\VerbatimPygments{\PYG}{\PYG}
\begin{mintbg}{codebg}
\begin{Verbatim}[commandchars=\\\{\},codes={\catcode`\$=3\catcode`\^=7\catcode`\_=8\relax}]
\PYG{k}{def}\PYG{+w}{ }\PYG{n+nf}{kpp}\PYG{p}{(}\PYG{n}{d}\PYG{p}{,} \PYG{n}{rs}\PYG{o}{=}\PYG{k+kc}{None}\PYG{p}{,} \PYG{n}{k}\PYG{o}{=}\PYG{l+m+mi}{10}\PYG{p}{,} \PYG{n}{few}\PYG{o}{=}\PYG{l+m+mi}{256}\PYG{p}{)}\PYG{p}{:} \PYG{esc}{\label{kpp}}
  \PYG{n}{rs} \PYG{o}{=} \PYG{n}{rs} \PYG{o+ow}{or} \PYG{n}{d}\PYG{o}{.}\PYG{n}{rows}
  \PYG{n}{out} \PYG{o}{=} \PYG{p}{[}\PYG{n}{choice}\PYG{p}{(}\PYG{n}{rs}\PYG{p}{)}\PYG{p}{]}
  \PYG{k}{while} \PYG{n+nb}{len}\PYG{p}{(}\PYG{n}{out}\PYG{p}{)} \PYG{o}{\PYGZlt{}} \PYG{n}{k}\PYG{p}{:}
    \PYG{n}{t} \PYG{o}{=} \PYG{n}{sample}\PYG{p}{(}\PYG{n}{rs}\PYG{p}{,} \PYG{n+nb}{min}\PYG{p}{(}\PYG{n}{few}\PYG{p}{,} \PYG{n+nb}{len}\PYG{p}{(}\PYG{n}{rs}\PYG{p}{)}\PYG{p}{)}\PYG{p}{)}
    \PYG{n}{ws} \PYG{o}{=} \PYG{p}{\PYGZob{}}\PYG{n}{i}\PYG{p}{:} \PYG{n+nb}{min}\PYG{p}{(}\PYG{n}{distx}\PYG{p}{(}\PYG{n}{d}\PYG{p}{,} \PYG{n}{t}\PYG{p}{[}\PYG{n}{i}\PYG{p}{]}\PYG{p}{,} \PYG{n}{c}\PYG{p}{)}\PYG{o}{*}\PYG{o}{*}\PYG{l+m+mi}{2} \PYG{k}{for} \PYG{n}{c} \PYG{o+ow}{in} \PYG{n}{out}\PYG{p}{)}
          \PYG{k}{for} \PYG{n}{i} \PYG{o+ow}{in} \PYG{n+nb}{range}\PYG{p}{(}\PYG{n+nb}{len}\PYG{p}{(}\PYG{n}{t}\PYG{p}{)}\PYG{p}{)}\PYG{p}{\PYGZcb{}}
    \PYG{n}{out}\PYG{o}{.}\PYG{n}{append}\PYG{p}{(}\PYG{n}{t}\PYG{p}{[}\PYG{n}{pick}\PYG{p}{(}\PYG{n}{ws}\PYG{p}{)}\PYG{p}{]}\PYG{p}{)}
  \PYG{k}{return} \PYG{n}{out}
\end{Verbatim}
\end{mintbg}
\endgroup
\endminipage\par\noindent
The \textsf{few=256} parameter  (on line \ref{kpp}) is a pragmatic shortcut. On large
data sets, scanning every row to score each new centroid is
expensive; 256 random candidates give an interesting weighted
spread without the cost.

\subsection*{The lesson: one substrate, three algorithms}
The above 70 lines present in a unified manner three algorithms that
usually occupy separate sections in a standard AI curriculum.
With the exception of \textsf{pick},
none of the three algorithms needed a new data structure or a
new support function. None needed a new training phase. Each was
a way of asking the substrate a different question:

\begin{itemize}
\item Naïve Bayes asks: how much does a column like a value?
\item $k$-means asks: which centroid is this row closest to?
\item $k$-means++ asks: which row should be the next centroid,
weighted by squared distance?
\end{itemize}
The questions differ; the machinery underneath does not. Every
\cls{Num} and \cls{Sym} inside a \cls{Data} was already a fitted
summary. Every clustering call already had \textsf{clone},
\textsf{add}, and \textsf{mids} available, so this code draws heavily on
pre-existing functionality.

This is the kind of structural similarity that becomes invisible
when humans no longer read the code. An LLM asked separately for
``Naïve Bayes,'' ``$k$-means,'' and ``$k$-means++'' will
generate three independent implementations, with three separate
data structures and three separate maintenance burdens. The
code emitted by such generators is not wrong,
but it does miss the observation that the same substrate would
have served all three. And if the substrate is ignored and the above
is implemented three times, that  means
triple the code, triple the bugs,  and triple the maintenance.

  \section{Classification and Regression in 43 Lines}\label{trees}

 Our general theme is that with the right underlying
 architecture, building any number of AI tools becomes
 easy.
For example, in this section, we use the above code to build
classification and regression trees that predict the values in
any leaf of those trees.
 
 Classification trees and regression trees are nearly the same
but differ  in how they score  their leaves:
 \begin{itemize}
 \item Classification trees return symbolic   scores
 for each leaf;
 \item Regression trees return numeric scores such as \textsf{disty}.
 \end{itemize}
\par\nopagebreak\minipage{\linewidth}
\begingroup\mintfvset
\VerbatimPygments{\PYG}{\PYG}
\begin{mintbg}{codebg}
\begin{Verbatim}[commandchars=\\\{\},codes={\catcode`\$=3\catcode`\^=7\catcode`\_=8\relax}]
\PYG{k}{def}\PYG{+w}{ }\PYG{n+nf}{disty}\PYG{p}{(}\PYG{n}{data}\PYG{p}{,} \PYG{n}{row}\PYG{p}{)}\PYG{p}{:} \PYG{esc}{\label{disty}}
  \PYG{k}{return} \PYG{n}{minkowski}\PYG{p}{(}  \PYG{c+c1}{\PYGZsh{} for \PYGZdq{}minkowski\PYGZdq{}, see line \ref{minkowski}}
          \PYG{p}{(}\PYG{n+nb}{abs}\PYG{p}{(}\PYG{n}{norm}\PYG{p}{(}\PYG{n}{col}\PYG{p}{,} \PYG{n}{row}\PYG{p}{[}\PYG{n}{col}\PYG{o}{.}\PYG{n}{at}\PYG{p}{]}\PYG{p}{)} \PYG{c+c1}{\PYGZsh{} for \PYGZdq{}norm\PYGZdq{}, see line \ref{norml}}
           \PYG{o}{\PYGZhy{}} \PYG{n}{col}\PYG{o}{.}\PYG{n}{heaven}\PYG{p}{)}  \PYG{c+c1}{\PYGZsh{} for \PYGZdq{}heaven\PYGZdq{}, see line \ref{heaven}}
           \PYG{k}{for} \PYG{n}{col} \PYG{o+ow}{in} \PYG{n}{data}\PYG{o}{.}\PYG{n}{cols}\PYG{o}{.}\PYG{n}{ys}\PYG{p}{)}\PYG{p}{,} \PYG{n}{p}\PYG{o}{=}\PYG{n}{the}\PYG{o}{.}\PYG{n}{p}\PYG{p}{)}
\end{Verbatim}
\end{mintbg}
\endgroup
\endminipage\par\noindent       
EZR's   trees are formed by recursively splitting the rows in order
 to decrease conclusion uncertainty. 
 For example,   for the following data about human beings,
 we would be certain about who was alive or dead if we split
 at   $\mathit{age}<80$
 (since, after that split {\em alive} would all $y$
 in the lower split and all $n$ in the upper split).
\begingroup\mintfvset
\fvset{numbers=none}
\VerbatimPygments{\PYG}{\PYG}
\begin{mintbg}{codebg}
\begin{Verbatim}[commandchars=\\\{\},codes={\catcode`\$=3\catcode`\^=7\catcode`\_=8\relax}]
 \PYG{n}{age}\PYG{p}{:}   \PYG{p}{[} \PYG{l+m+mi}{0}\PYG{p}{,} \PYG{l+m+mi}{0}\PYG{p}{,} \PYG{l+m+mi}{10}\PYG{p}{,} \PYG{l+m+mi}{11}\PYG{p}{,} \PYG{l+m+mi}{16}\PYG{p}{,} \PYG{l+m+mi}{21}\PYG{p}{,} \PYG{l+m+mi}{40}\PYG{p}{,} \PYG{l+m+mi}{80}\PYG{p}{,} \PYG{l+m+mi}{120}\PYG{p}{,} \PYG{l+m+mi}{200}\PYG{p}{]}
 \PYG{n}{alive}\PYG{p}{:} \PYG{p}{[} \PYG{n}{y}\PYG{p}{,} \PYG{n}{y}\PYG{p}{,}  \PYG{n}{y}\PYG{p}{,}  \PYG{n}{y}\PYG{p}{,}  \PYG{n}{y}\PYG{p}{,}  \PYG{n}{y}\PYG{p}{,}  \PYG{n}{y}\PYG{p}{,}  \PYG{n}{n}\PYG{p}{,}   \PYG{n}{n}\PYG{p}{,}   \PYG{n}{n}\PYG{p}{]}
\end{Verbatim}
\end{mintbg}
\endgroup
The \textsf{spread} function defined above at line~\ref{spread0} lets us
measure that uncertainty.
If a split on column {\em col} (at a value {\em cut}) generates two sets of rows of
size $n_1,n_2$ with \textsf{spread} $s_1,s_2$, then the best split is
the one that minimizes the expected value of the uncertainty after
the split; i.e. $(n_1\, s_1 + n_2\, s_2)/(n_1 + n_2)$.

To assess a {\em cut},
we use the following code:
\par\nopagebreak\minipage{\linewidth}
\begingroup\mintfvset
\VerbatimPygments{\PYG}{\PYG}
\begin{mintbg}{codebg}
\begin{Verbatim}[commandchars=\\\{\},codes={\catcode`\$=3\catcode`\^=7\catcode`\_=8\relax}]
\PYG{k}{def}\PYG{+w}{ }\PYG{n+nf}{treeSplit}\PYG{p}{(}\PYG{n}{data}\PYG{p}{,} \PYG{n}{col}\PYG{p}{,} \PYG{n}{cut}\PYG{p}{,} \PYG{n}{rows}\PYG{p}{,} \PYG{n}{klass}\PYG{o}{=}\PYG{k+kc}{None}\PYG{p}{,} \PYG{n}{y}\PYG{o}{=}\PYG{n}{Num}\PYG{p}{)}\PYG{p}{:}
  \PYG{n}{klass} \PYG{o}{=} \PYG{n}{klass} \PYG{o+ow}{or} \PYG{p}{(}\PYG{k}{lambda} \PYG{n}{r}\PYG{p}{:} \PYG{n}{disty}\PYG{p}{(}\PYG{n}{data}\PYG{p}{,} \PYG{n}{r}\PYG{p}{)}\PYG{p}{)}
  \PYG{n}{l\PYGZus{}rows}\PYG{p}{,} \PYG{n}{r\PYGZus{}rows}\PYG{p}{,} \PYG{n}{l\PYGZus{}y}\PYG{p}{,} \PYG{n}{r\PYGZus{}y} \PYG{o}{=} \PYG{p}{[}\PYG{p}{]}\PYG{p}{,} \PYG{p}{[}\PYG{p}{]}\PYG{p}{,} \PYG{n}{y}\PYG{p}{(}\PYG{p}{)}\PYG{p}{,} \PYG{n}{y}\PYG{p}{(}\PYG{p}{)}
  \PYG{k}{for} \PYG{n}{row} \PYG{o+ow}{in} \PYG{n}{rows}\PYG{p}{:}
    \PYG{n}{v} \PYG{o}{=} \PYG{n}{row}\PYG{p}{[}\PYG{n}{col}\PYG{o}{.}\PYG{n}{at}\PYG{p}{]}
    \PYG{n}{go} \PYG{o}{=} \PYG{n}{v}\PYG{o}{==}\PYG{l+s+s2}{\PYGZdq{}}\PYG{l+s+s2}{?}\PYG{l+s+s2}{\PYGZdq{}} \PYG{o+ow}{or} \PYG{p}{(}\PYG{n}{v}\PYG{o}{==}\PYG{n}{cut} \PYG{k}{if} \PYG{n}{Sym}\PYG{o}{==}\PYG{n+nb}{type}\PYG{p}{(}\PYG{n}{col}\PYG{p}{)} \PYG{k}{else} \PYG{n}{v}\PYG{o}{\PYGZlt{}}\PYG{o}{=}\PYG{n}{cut}\PYG{p}{)} \PYG{esc}{\label{cutts}}
    \PYG{p}{(}\PYG{n}{l\PYGZus{}rows} \PYG{k}{if} \PYG{n}{go} \PYG{k}{else} \PYG{n}{r\PYGZus{}rows}\PYG{p}{)}\PYG{o}{.}\PYG{n}{append}\PYG{p}{(}\PYG{n}{row}\PYG{p}{)}
    \PYG{n}{add}\PYG{p}{(}\PYG{n}{l\PYGZus{}y} \PYG{k}{if} \PYG{n}{go} \PYG{k}{else} \PYG{n}{r\PYGZus{}y}\PYG{p}{,} \PYG{n}{klass}\PYG{p}{(}\PYG{n}{row}\PYG{p}{)}\PYG{p}{)} \PYG{esc}{\label{effects}}
  \PYG{n}{s} \PYG{o}{=} \PYG{n}{l\PYGZus{}y}\PYG{o}{.}\PYG{n}{n}\PYG{o}{*}\PYG{n}{spread}\PYG{p}{(}\PYG{n}{l\PYGZus{}y}\PYG{p}{)} \PYG{o}{+} \PYG{n}{r\PYGZus{}y}\PYG{o}{.}\PYG{n}{n}\PYG{o}{*}\PYG{n}{spread}\PYG{p}{(}\PYG{n}{r\PYGZus{}y}\PYG{p}{)}
  \PYG{k}{return} \PYG{n}{s}\PYG{p}{,} \PYG{n}{col}\PYG{p}{,} \PYG{n}{cut}\PYG{p}{,} \PYG{n}{l\PYGZus{}rows}\PYG{p}{,} \PYG{n}{r\PYGZus{}rows}
\end{Verbatim}
\end{mintbg}
\endgroup
\endminipage\par\noindent
Note how this code  uses  \textsf{spread} (from line \ref{spread0}). This is an abstract term
that covers standard deviation (for \cls{Num}s) and entropy (for \cls{Sym}s).

Line~\ref{cutts} needs a word of explanation.
Cuts from symbolic and numeric columns get tested differently. Rows go left when the test is true, right when false. For symbolic columns, true means {\em equal} to the cut (false means 
{\em not equal}). For numeric columns, true means {\em less than or equal} to the cut (and false means {\em greater than}).

Note that, on  line \ref{effects}, \textsf{treeSplit}   uses \cls{Num} to  
collect the numeric  \textsf{disty} statistics on what happens in the left or right cut. That is, by default, EZR builds regression trees.
To ask EZR to build classification
trees, all that is required is to pass in \textsf{y=Sym} and
a \textsf{klass} that extracts the class column from each row:
\begingroup\mintfvset
\fvset{numbers=none}
\VerbatimPygments{\PYG}{\PYG}
\begin{mintbg}{codebg}
\begin{Verbatim}[commandchars=\\\{\},codes={\catcode`\$=3\catcode`\^=7\catcode`\_=8\relax}]
\PYG{n}{y} \PYG{o}{=} \PYG{n}{Sym}
\PYG{n}{klass} \PYG{o}{=} \PYG{k}{lambda} \PYG{n}{row}\PYG{p}{:} \PYG{n}{row}\PYG{p}{[}\PYG{n}{data}\PYG{o}{.}\PYG{n}{cols}\PYG{o}{.}\PYG{n}{klass}\PYG{o}{.}\PYG{n}{at}\PYG{p}{]} \PYG{c+c1}{\PYGZsh{} see line \ref{klass}}
\end{Verbatim}
\end{mintbg}
\endgroup
One way to generate {\em cut}s is to try every unique value in the rows. Numeric columns can have many such values, so we instead slice the rows into equal-sized 
{\em bins} and take one 
{\em cut} per bin. After trying  various bin sizes,
 {\em bins=2} tends to work best.
(i.e. splitting numerics on the median).
\par\nopagebreak\minipage{\linewidth}
\begingroup\mintfvset
\VerbatimPygments{\PYG}{\PYG}
\begin{mintbg}{codebg}
\begin{Verbatim}[commandchars=\\\{\},codes={\catcode`\$=3\catcode`\^=7\catcode`\_=8\relax}]
\PYG{k}{def}\PYG{+w}{ }\PYG{n+nf}{treeCuts}\PYG{p}{(}\PYG{n}{col}\PYG{p}{,} \PYG{n}{rows}\PYG{p}{,} \PYG{n}{bins}\PYG{o}{=}\PYG{l+m+mi}{2}\PYG{p}{)}\PYG{p}{:}
  \PYG{n}{vs} \PYG{o}{=} \PYG{p}{[}\PYG{n}{row}\PYG{p}{[}\PYG{n}{col}\PYG{o}{.}\PYG{n}{at}\PYG{p}{]} \PYG{k}{for} \PYG{n}{row} \PYG{o+ow}{in} \PYG{n}{rows} \PYG{k}{if} \PYG{n}{row}\PYG{p}{[}\PYG{n}{col}\PYG{o}{.}\PYG{n}{at}\PYG{p}{]}\PYG{o}{!=}\PYG{l+s+s2}{\PYGZdq{}}\PYG{l+s+s2}{?}\PYG{l+s+s2}{\PYGZdq{}}\PYG{p}{]}
  \PYG{k}{if} \PYG{o+ow}{not} \PYG{n}{vs}\PYG{p}{:} \PYG{k}{return} \PYG{p}{[}\PYG{p}{]}
  \PYG{k}{if} \PYG{n}{Sym} \PYG{o}{==} \PYG{n+nb}{type}\PYG{p}{(}\PYG{n}{col}\PYG{p}{)}\PYG{p}{:} \PYG{k}{return} \PYG{n+nb}{set}\PYG{p}{(}\PYG{n}{vs}\PYG{p}{)}
  \PYG{n}{vs} \PYG{o}{=} \PYG{n+nb}{sorted}\PYG{p}{(}\PYG{n}{vs}\PYG{p}{)}
  \PYG{k}{return} \PYG{p}{[}\PYG{n}{vs}\PYG{p}{[}\PYG{n+nb}{bin}\PYG{o}{*}\PYG{n+nb}{len}\PYG{p}{(}\PYG{n}{vs}\PYG{p}{)}\PYG{o}{/}\PYG{o}{/}\PYG{n}{bins}\PYG{p}{]} \PYG{k}{for} \PYG{n+nb}{bin} \PYG{o+ow}{in} \PYG{n+nb}{range}\PYG{p}{(}\PYG{l+m+mi}{1}\PYG{p}{,} \PYG{n}{bins}\PYG{p}{)}\PYG{p}{]}
\end{Verbatim}
\end{mintbg}
\endgroup
\endminipage\par\noindent
Once we can generate and score the {\em cut}s, then the rest of tree learning is just
a recursive descent that tries to add \textsf{tree.left} and \textsf{tree.right} to the
current \textsf{tree} node.  In EZR, this recursion stops when
we can no longer cut rows into two splits of size 3 (see line \ref{stop}):
\par\nopagebreak\minipage{\linewidth}
\begingroup\mintfvset
\VerbatimPygments{\PYG}{\PYG}
\begin{mintbg}{codebg}
\begin{Verbatim}[commandchars=\\\{\},codes={\catcode`\$=3\catcode`\^=7\catcode`\_=8\relax}]
\PYG{n}{the}\PYG{o}{.}\PYG{n}{learn}\PYG{o}{.}\PYG{n}{leaf} \PYG{o}{=} \PYG{l+m+mi}{3} \PYG{c+c1}{\PYGZsh{} config}

\PYG{k}{def}\PYG{+w}{ }\PYG{n+nf}{treeGrow}\PYG{p}{(}\PYG{n}{data}\PYG{p}{,} \PYG{n}{rows}\PYG{p}{,} \PYG{n}{klass}\PYG{o}{=}\PYG{k+kc}{None}\PYG{p}{,} \PYG{n}{y}\PYG{o}{=}\PYG{n}{Num}\PYG{p}{)}\PYG{p}{:}
  \PYG{n}{tree} \PYG{o}{=} \PYG{n}{Tree}\PYG{p}{(}\PYG{n}{data}\PYG{p}{,} \PYG{n}{rows}\PYG{p}{,} \PYG{n}{klass}\PYG{p}{,} \PYG{n}{y}\PYG{p}{)} \PYG{esc}{\label{newTree}}
  \PYG{k}{if} \PYG{n+nb}{len}\PYG{p}{(}\PYG{n}{rows}\PYG{p}{)} \PYG{o}{\PYGZgt{}}\PYG{o}{=} \PYG{l+m+mi}{2} \PYG{o}{*} \PYG{n}{the}\PYG{o}{.}\PYG{n}{learn}\PYG{o}{.}\PYG{n}{leaf}\PYG{p}{:} \PYG{esc}{\label{stop}}
    \PYG{n}{splits} \PYG{o}{=} \PYG{p}{(}\PYG{n}{treeSplit}\PYG{p}{(}\PYG{n}{data}\PYG{p}{,} \PYG{n}{col}\PYG{p}{,} \PYG{n}{cut}\PYG{p}{,} \PYG{n}{rows}\PYG{p}{,} \PYG{n}{klass}\PYG{p}{,} \PYG{n}{y}\PYG{p}{)}
              \PYG{k}{for} \PYG{n}{col} \PYG{o+ow}{in} \PYG{n}{tree}\PYG{o}{.}\PYG{n}{d}\PYG{o}{.}\PYG{n}{cols}\PYG{o}{.}\PYG{n}{xs}
              \PYG{k}{for} \PYG{n}{cut} \PYG{o+ow}{in} \PYG{n}{treeCuts}\PYG{p}{(}\PYG{n}{col}\PYG{p}{,} \PYG{n}{rows}\PYG{p}{)}\PYG{p}{)}
    \PYG{k}{if} \PYG{n}{valid} \PYG{o}{:=} \PYG{p}{[}\PYG{n}{s} \PYG{k}{for} \PYG{n}{s} \PYG{o+ow}{in} \PYG{n}{splits}
        \PYG{k}{if} \PYG{n+nb}{min}\PYG{p}{(}\PYG{n+nb}{len}\PYG{p}{(}\PYG{n}{s}\PYG{p}{[}\PYG{l+m+mi}{3}\PYG{p}{]}\PYG{p}{)}\PYG{p}{,} \PYG{n+nb}{len}\PYG{p}{(}\PYG{n}{s}\PYG{p}{[}\PYG{l+m+mi}{4}\PYG{p}{]}\PYG{p}{)}\PYG{p}{)} \PYG{o}{\PYGZgt{}}\PYG{o}{=} \PYG{n}{the}\PYG{o}{.}\PYG{n}{learn}\PYG{o}{.}\PYG{n}{leaf}\PYG{p}{]}\PYG{p}{:}
      \PYG{n}{\PYGZus{}}\PYG{p}{,} \PYG{n}{tree}\PYG{o}{.}\PYG{n}{col}\PYG{p}{,} \PYG{n}{tree}\PYG{o}{.}\PYG{n}{cut}\PYG{p}{,} \PYG{n}{left}\PYG{p}{,} \PYG{n}{right} \PYG{o}{=} \PYG{n+nb}{min}\PYG{p}{(}
        \PYG{n}{valid}\PYG{p}{,} \PYG{n}{key}\PYG{o}{=}\PYG{k}{lambda} \PYG{n}{x}\PYG{p}{:} \PYG{n}{x}\PYG{p}{[}\PYG{l+m+mi}{0}\PYG{p}{]}\PYG{p}{)}
      \PYG{n}{tree}\PYG{o}{.}\PYG{n}{left}  \PYG{o}{=} \PYG{n}{treeGrow}\PYG{p}{(}\PYG{n}{data}\PYG{p}{,} \PYG{n}{left}\PYG{p}{,}  \PYG{n}{klass}\PYG{p}{,} \PYG{n}{y}\PYG{p}{)}
      \PYG{n}{tree}\PYG{o}{.}\PYG{n}{right} \PYG{o}{=} \PYG{n}{treeGrow}\PYG{p}{(}\PYG{n}{data}\PYG{p}{,} \PYG{n}{right}\PYG{p}{,} \PYG{n}{klass}\PYG{p}{,} \PYG{n}{y}\PYG{p}{)}
  \PYG{k}{return} \PYG{n}{tree}
\end{Verbatim}
\end{mintbg}
\endgroup
\endminipage\par\noindent
 Once a tree is generated,
 to make a prediction, a new row has to be pushed down its appropriate
branch. 
\par\nopagebreak\minipage{\linewidth}
\begingroup\mintfvset
\VerbatimPygments{\PYG}{\PYG}
\begin{mintbg}{codebg}
\begin{Verbatim}[commandchars=\\\{\},codes={\catcode`\$=3\catcode`\^=7\catcode`\_=8\relax}]
\PYG{k}{def}\PYG{+w}{ }\PYG{n+nf}{treeLeaf}\PYG{p}{(}\PYG{n}{tree}\PYG{p}{,} \PYG{n}{row}\PYG{p}{)}\PYG{p}{:}
  \PYG{k}{if} \PYG{o+ow}{not} \PYG{n}{tree}\PYG{o}{.}\PYG{n}{left}\PYG{p}{:} \PYG{k}{return} \PYG{n}{tree}
  \PYG{n}{v} \PYG{o}{=} \PYG{n}{row}\PYG{p}{[}\PYG{n}{tree}\PYG{o}{.}\PYG{n}{col}\PYG{o}{.}\PYG{n}{at}\PYG{p}{]}
  \PYG{n}{go} \PYG{o}{=} \PYG{n}{v}\PYG{o}{!=}\PYG{l+s+s2}{\PYGZdq{}}\PYG{l+s+s2}{?}\PYG{l+s+s2}{\PYGZdq{}} \PYG{o+ow}{and} \PYG{p}{(}\PYG{n}{v}\PYG{o}{\PYGZlt{}}\PYG{o}{=}\PYG{n}{tree}\PYG{o}{.}\PYG{n}{cut} \PYG{k}{if} \PYG{n}{Num}\PYG{o}{==}\PYG{n+nb}{type}\PYG{p}{(}\PYG{n}{tree}\PYG{o}{.}\PYG{n}{col}\PYG{p}{)}
                               \PYG{k}{else} \PYG{n}{v}\PYG{o}{==}\PYG{n}{tree}\PYG{o}{.}\PYG{n}{cut}\PYG{p}{)}
  \PYG{k}{return} \PYG{n}{treeLeaf}\PYG{p}{(}\PYG{n}{tree}\PYG{o}{.}\PYG{n}{left} \PYG{k}{if} \PYG{n}{go} \PYG{k}{else} \PYG{n}{tree}\PYG{o}{.}\PYG{n}{right}\PYG{p}{,}\PYG{n}{row}\PYG{p}{)}
\end{Verbatim}
\end{mintbg}
\endgroup
\endminipage\par\noindent
At the end that branch is a leaf containing a \cls{Data}
which can be used to make predictions.
To facilitate that, as a side-effect of making a new tree node on line~\ref{newTree}
we compute \textsf{ynum} (which is the distribution of 
the \textsf{disty}s seen in that node).
\par\nopagebreak\minipage{\linewidth}
\begingroup\mintfvset
\VerbatimPygments{\PYG}{\PYG}
\begin{mintbg}{codebg}
\begin{Verbatim}[commandchars=\\\{\},codes={\catcode`\$=3\catcode`\^=7\catcode`\_=8\relax}]
\PYG{k}{class}\PYG{+w}{ }\PYG{n+nc}{Tree}\PYG{p}{:}
  \PYG{k}{def}\PYG{+w}{ }\PYG{n+nf+fm}{\PYGZus{}\PYGZus{}init\PYGZus{}\PYGZus{}}\PYG{p}{(}\PYG{n}{i}\PYG{p}{,} \PYG{n}{data}\PYG{p}{,} \PYG{n}{rows}\PYG{p}{,} \PYG{n}{klass}\PYG{o}{=}\PYG{k+kc}{None}\PYG{p}{,} \PYG{n}{y}\PYG{o}{=}\PYG{n}{Num}\PYG{p}{)}\PYG{p}{:}
    \PYG{n}{klass} \PYG{o}{=} \PYG{n}{klass} \PYG{o+ow}{or} \PYG{p}{(}\PYG{k}{lambda} \PYG{n}{r}\PYG{p}{:} \PYG{n}{disty}\PYG{p}{(}\PYG{n}{data}\PYG{p}{,} \PYG{n}{r}\PYG{p}{)}\PYG{p}{)}
    \PYG{n}{i}\PYG{o}{.}\PYG{n}{d} \PYG{o}{=} \PYG{n}{clone}\PYG{p}{(}\PYG{n}{data}\PYG{p}{,} \PYG{n}{rows}\PYG{p}{)}
    \PYG{n}{i}\PYG{o}{.}\PYG{n}{ynum} \PYG{o}{=} \PYG{n}{adds}\PYG{p}{(}\PYG{p}{(}\PYG{n}{klass}\PYG{p}{(}\PYG{n}{row}\PYG{p}{)} \PYG{k}{for} \PYG{n}{row} \PYG{o+ow}{in} \PYG{n}{rows}\PYG{p}{)}\PYG{p}{,} \PYG{n}{y}\PYG{p}{(}\PYG{p}{)}\PYG{p}{)}
    \PYG{n}{i}\PYG{o}{.}\PYG{n}{col}\PYG{p}{,} \PYG{n}{i}\PYG{o}{.}\PYG{n}{cut} \PYG{o}{=} \PYG{k+kc}{None}\PYG{p}{,} \PYG{l+m+mi}{0}
    \PYG{n}{i}\PYG{o}{.}\PYG{n}{left} \PYG{o}{=} \PYG{n}{i}\PYG{o}{.}\PYG{n}{right} \PYG{o}{=} \PYG{k+kc}{None}
\end{Verbatim}
\end{mintbg}
\endgroup
\endminipage\par\noindent
 Later in this paper, when we discuss active learning, we will show
 that the trees learned in this manner are very effective.
 
  In support of the thesis of this paper
  (that AI is very simple
  when implemented with the right primitives) we note that EZR's
  combined classification/regression tree  generator
  requires just 43 lines of code.
One reason for this is that it makes extensive use of 
\cls{Data}, \cls{Num}, \cls{Sym}.
  
\section{Two Optimizers, One Code base, in 56 Lines}\label{sec:opt}

This section tests two truisms.
The first is {\em structural}: that
optimization is a different kind of problem from classification,
clustering, and regression, and that it needs its own algorithms
and its own code base. 

The second is {\em empirical}: that modern
optimizers (with restart, retry, adaptive schedules) outperform
older, simpler methods. The 1990s metaheuristics literature
(GSAT~\cite{gsat}, WalkSAT~\cite{walksat}, GRASP~\cite{grasp},
Iterated LS~\cite{ils}) argued the second truism explicitly, with
a steady drumbeat of papers claiming clear advantage over older
methods like simulated annealing. 

The two subsections below take
these truisms in turn:
\begin{itemize}
\item
\S\ref{sec:oneplus} shows that two optimizers (simulated
annealing and local search) share most of their code.
\item
\S\ref{sec:opt-results} then shows that simulated annealing in
its 1983 default configuration~\cite{kirkpatrick1983optimization}
matches or beats every local-search variant we tested across 20
MOOT benchmarks.
\end{itemize}
In the following implementation, we will see that not only are our
optimizers similar to each other, but they also share much of the
same code used above for classification and clustering.
We are unaware of any prior work that has shown such similarities
between optimization, classification, and regression.

\subsection{(1+1) algorithms}\label{sec:oneplus}
A (1+1) evolutionary algorithm {\em mutates} one current solution
to generate one mutant. If the mutant is ``better'' the algorithm
{\em accepts} that the current solution can be replaced. This new
current solution is used as the basis for future mutation.

Two examples of (1+1) algorithms are simulated annealing (SA) and
local search (LS). Both have their own mechanisms for escaping
dead-ends in the search space. 
Local search uses  a reset-retry policy~\cite{Crawford:1994}.
If no progress is being made, it jumps back to a fresh start with a  new random seed.

Simulated annealing~\cite{kirkpatrick1983optimization}, on the other hand, can {\em accept}
a slightly less optimal solution with a probability that drops as
the run progresses. Early on, when its ``temperature'' is high, 
SA jumps around up and down to better/worse solutions.
Later in its run, the algorithm
becomes much more discerning and will only move to better solutions.  
In the code below,
at line \ref{saxxx}, the $(1-h/b)$ term in the
denominator of the exponential is what does the cooling: as the
history count $h$ approaches the budget $b$, the denominator
shrinks, the exponential collapses, and SA stops accepting bad
moves.

The two algorithms also differ in what they {\em mutate}. Simulated
annealing mutates a random subset of attributes. Local search, on the other hand,
(a)~usually mutates any one attribute at a time; (b)~then at
probability $p$, freezes all attributes except one, then sweeps over the range
of that single attribute.

\par\nopagebreak\minipage{\linewidth}
\begingroup\mintfvset
\VerbatimPygments{\PYG}{\PYG}
\begin{mintbg}{codebg}
\begin{Verbatim}[commandchars=\\\{\},codes={\catcode`\$=3\catcode`\^=7\catcode`\_=8\relax}]
\PYG{k}{def}\PYG{+w}{ }\PYG{n+nf}{ls}\PYG{p}{(}\PYG{n}{d}\PYG{p}{,} \PYG{n}{oracle}\PYG{p}{,} \PYG{n}{restart}\PYG{o}{=}\PYG{l+m+mi}{100}\PYG{p}{,} \PYG{n}{p}\PYG{o}{=}\PYG{l+m+mf}{0.5}\PYG{p}{,} \PYG{n}{tries}\PYG{o}{=}\PYG{l+m+mi}{20}\PYG{p}{,} \PYG{n}{budget}\PYG{o}{=}\PYG{l+m+mi}{1000}\PYG{p}{)}\PYG{p}{:}
  \PYG{k}{def}\PYG{+w}{ }\PYG{n+nf}{accept}\PYG{p}{(}\PYG{n}{e}\PYG{p}{,} \PYG{n}{en}\PYG{p}{,} \PYG{o}{*}\PYG{n}{\PYGZus{}}\PYG{p}{)}\PYG{p}{:} \PYG{k}{return} \PYG{n}{en} \PYG{o}{\PYGZlt{}} \PYG{n}{e}
  \PYG{k}{def}\PYG{+w}{ }\PYG{n+nf}{mutate}\PYG{p}{(}\PYG{n}{s}\PYG{p}{)}\PYG{p}{:}
    \PYG{n}{c} \PYG{o}{=} \PYG{n}{choice}\PYG{p}{(}\PYG{n}{d}\PYG{o}{.}\PYG{n}{cols}\PYG{o}{.}\PYG{n}{xs}\PYG{p}{)}
    \PYG{k}{for} \PYG{n}{\PYGZus{}} \PYG{o+ow}{in} \PYG{n+nb}{range}\PYG{p}{(}\PYG{n}{tries} \PYG{k}{if} \PYG{n}{rand}\PYG{p}{(}\PYG{p}{)}\PYG{o}{\PYGZlt{}}\PYG{n}{p} \PYG{k}{else} \PYG{l+m+mi}{1}\PYG{p}{)}\PYG{p}{:}
      \PYG{n}{s} \PYG{o}{=} \PYG{n}{s}\PYG{p}{[}\PYG{p}{:}\PYG{p}{]}
      \PYG{n}{s}\PYG{p}{[}\PYG{n}{c}\PYG{o}{.}\PYG{n}{at}\PYG{p}{]} \PYG{o}{=} \PYG{n}{pick}\PYG{p}{(}\PYG{n}{c}\PYG{p}{,} \PYG{n}{s}\PYG{p}{[}\PYG{n}{c}\PYG{o}{.}\PYG{n}{at}\PYG{p}{]}\PYG{p}{)} \PYG{c+c1}{\PYGZsh{} for \PYGZdq{}pick\PYGZdq{} see line \ref{pick}}
      \PYG{k}{yield} \PYG{n}{s}
  \PYG{k}{return} \PYG{n}{oneplus1}\PYG{p}{(}\PYG{n}{d}\PYG{p}{,} \PYG{c+c1}{\PYGZsh{} for \PYGZdq{}oneplus1\PYGZdq{}, see \ref{oneplus1}}
                  \PYG{n}{mutate}\PYG{p}{,} \PYG{n}{accept}\PYG{p}{,} \PYG{n}{oracle}\PYG{p}{,} \PYG{n}{budget}\PYG{p}{,} \PYG{n}{restart}\PYG{p}{)}

\PYG{k}{def}\PYG{+w}{ }\PYG{n+nf}{sa}\PYG{p}{(}\PYG{n}{d}\PYG{p}{,} \PYG{n}{oracle}\PYG{p}{,} \PYG{n}{restart}\PYG{o}{=}\PYG{l+m+mi}{0}\PYG{p}{,} \PYG{n}{m}\PYG{o}{=}\PYG{l+m+mf}{0.5}\PYG{p}{,} \PYG{n}{budget}\PYG{o}{=}\PYG{l+m+mi}{1000}\PYG{p}{)}\PYG{p}{:}
  \PYG{n}{n} \PYG{o}{=} \PYG{n+nb}{max}\PYG{p}{(}\PYG{l+m+mi}{1}\PYG{p}{,} \PYG{n+nb}{int}\PYG{p}{(}\PYG{n}{m} \PYG{o}{*} \PYG{n+nb}{len}\PYG{p}{(}\PYG{n}{d}\PYG{o}{.}\PYG{n}{cols}\PYG{o}{.}\PYG{n}{xs}\PYG{p}{)}\PYG{p}{)}\PYG{p}{)}
  \PYG{k}{def}\PYG{+w}{ }\PYG{n+nf}{accept}\PYG{p}{(}\PYG{n}{e}\PYG{p}{,} \PYG{n}{en}\PYG{p}{,} \PYG{n}{h}\PYG{p}{,} \PYG{n}{b}\PYG{p}{)}\PYG{p}{:} \PYG{k}{return} \PYG{n}{en}\PYG{o}{\PYGZlt{}}\PYG{n}{e} \PYG{o+ow}{or} \PYG{n}{rand}\PYG{p}{(}\PYG{p}{)}\PYG{o}{\PYGZlt{}}\PYG{n}{exp}\PYG{p}{(}
                                 \PYG{p}{(}\PYG{n}{e}\PYG{o}{\PYGZhy{}}\PYG{n}{en}\PYG{p}{)}\PYG{o}{/}\PYG{p}{(}\PYG{l+m+mi}{1}\PYG{o}{\PYGZhy{}}\PYG{n}{h}\PYG{o}{/}\PYG{n}{b}\PYG{o}{+}\PYG{l+m+mf}{1E\PYGZhy{}32}\PYG{p}{)}\PYG{p}{)} \PYG{esc}{\label{saxxx}}
  \PYG{k}{def}\PYG{+w}{ }\PYG{n+nf}{mutate}\PYG{p}{(}\PYG{n}{s}\PYG{p}{)}\PYG{p}{:}
     \PYG{k}{yield} \PYG{n}{picks}\PYG{p}{(}\PYG{n}{d}\PYG{p}{,} \PYG{n}{s}\PYG{p}{,} \PYG{n}{n}\PYG{p}{)} \PYG{c+c1}{\PYGZsh{} for \PYGZdq{}picks\PYGZdq{} see next function}
  \PYG{k}{return} \PYG{n}{oneplus1}\PYG{p}{(}\PYG{n}{d}\PYG{p}{,} \PYG{c+c1}{\PYGZsh{} for \PYGZdq{}oneplus1\PYGZdq{}, see \ref{oneplus1}}
                  \PYG{n}{mutate}\PYG{p}{,} \PYG{n}{accept}\PYG{p}{,} \PYG{n}{oracle}\PYG{p}{,} \PYG{n}{budget}\PYG{p}{,} \PYG{n}{restart}\PYG{p}{)}

\PYG{k}{def}\PYG{+w}{ }\PYG{n+nf}{picks}\PYG{p}{(}\PYG{n}{data}\PYG{p}{,} \PYG{n}{row}\PYG{p}{,} \PYG{n}{n}\PYG{o}{=}\PYG{l+m+mi}{1}\PYG{p}{)}\PYG{p}{:}
  \PYG{n}{s} \PYG{o}{=} \PYG{n}{row}\PYG{p}{[}\PYG{p}{:}\PYG{p}{]}
  \PYG{k}{for} \PYG{n}{col} \PYG{o+ow}{in} \PYG{n}{sample}\PYG{p}{(}\PYG{n}{data}\PYG{o}{.}\PYG{n}{cols}\PYG{o}{.}\PYG{n}{xs}\PYG{p}{,} \PYG{n+nb}{min}\PYG{p}{(}\PYG{n}{n}\PYG{p}{,} \PYG{n+nb}{len}\PYG{p}{(}\PYG{n}{data}\PYG{o}{.}\PYG{n}{cols}\PYG{o}{.}\PYG{n}{xs}\PYG{p}{)}\PYG{p}{)}\PYG{p}{)}\PYG{p}{:}
    \PYG{n}{s}\PYG{p}{[}\PYG{n}{col}\PYG{o}{.}\PYG{n}{at}\PYG{p}{]} \PYG{o}{=} \PYG{n}{pick}\PYG{p}{(}\PYG{n}{col}\PYG{p}{,}\PYG{n}{s}\PYG{p}{[}\PYG{n}{col}\PYG{o}{.}\PYG{n}{at}\PYG{p}{]}\PYG{p}{)} \PYG{c+c1}{\PYGZsh{} for \PYGZdq{}pick\PYGZdq{} see line \ref{pick}}
  \PYG{k}{return} \PYG{n}{s}
\end{Verbatim}
\end{mintbg}
\endgroup
\endminipage\par\noindent
Simulated annealing and local search are often presented as separate
algorithms. But as seen here, both can be coded by calling an
underlying \textsf{oneplus1} function while specializing the
{\em mutate} and {\em accept} functions.

Before implementing our evolutionary algorithm, we first need to
define some way to rank mutants (in order to select the better
ones). EZR builds a surrogate {\em oracle} from a hold-out set,
then uses that surrogate to score new mutants. In the literature,
the standard way to build this {\em oracle} is a procedure followed
by (e.g.) Pfisterer~\cite{pfisterer2022yahpo} and Zela et
al.~\cite{zela2020surrogate} where:
\begin{itemize}
\item The mutators adjust the $x$ values;
\item The $y$ values of a new mutant are copied from their nearest
neighbor in 50 rows set aside at the start of the run;
\item Those $y$ values are summarized as {\em distance to heaven};
i.e. the distance of those values to the \textsf{heaven} values
for each goal calculated at line~\ref{heaven}.
\end{itemize}
		(Aside: Some
consideration was given to building a predictive model (e.g., a
decision tree) for labeling   unseen examples. But the
only way to validate that  model would be via the Pfisterer
and Zela procedure, since our tabular data holds all   ground
truth. Since some form of nearest-neighbor testing is unavoidable,
		EZR adopts it from the outset.)
\par\nopagebreak\minipage{\linewidth}
\begingroup\mintfvset
\VerbatimPygments{\PYG}{\PYG}
\begin{mintbg}{codebg}
\begin{Verbatim}[commandchars=\\\{\},codes={\catcode`\$=3\catcode`\^=7\catcode`\_=8\relax}]
\PYG{n}{d0} \PYG{o}{=} \PYG{n}{Data}\PYG{p}{(}\PYG{n}{csv}\PYG{p}{(}\PYG{n}{file}\PYG{p}{)}\PYG{p}{)}
\PYG{n}{shuffle}\PYG{p}{(}\PYG{n}{d0}\PYG{o}{.}\PYG{n}{rows}\PYG{p}{)}
\PYG{n}{known}  \PYG{o}{=} \PYG{n}{Data}\PYG{p}{(}\PYG{p}{[}\PYG{n}{d0}\PYG{o}{.}\PYG{n}{cols}\PYG{o}{.}\PYG{n}{names}\PYG{p}{]} \PYG{o}{+} \PYG{n}{d0}\PYG{o}{.}\PYG{n}{rows}\PYG{p}{[}\PYG{p}{:}\PYG{l+m+mi}{50}\PYG{p}{]}\PYG{p}{)}
\PYG{n}{search} \PYG{o}{=} \PYG{n}{Data}\PYG{p}{(}\PYG{p}{[}\PYG{n}{d0}\PYG{o}{.}\PYG{n}{cols}\PYG{o}{.}\PYG{n}{names}\PYG{p}{]} \PYG{o}{+} \PYG{n}{d0}\PYG{o}{.}\PYG{n}{rows}\PYG{p}{[}\PYG{l+m+mi}{50}\PYG{p}{:}\PYG{p}{]}\PYG{p}{)}
\PYG{n}{oracle} \PYG{o}{=} \PYG{k}{lambda} \PYG{n}{r}\PYG{p}{:} \PYG{n}{oracleNearest}\PYG{p}{(}\PYG{n}{known}\PYG{p}{,} \PYG{n}{r}\PYG{p}{)}

\PYG{k}{def}\PYG{+w}{ }\PYG{n+nf}{oracleNearest}\PYG{p}{(}\PYG{n}{data}\PYG{p}{,} \PYG{n}{row}\PYG{p}{)}\PYG{p}{:} \PYG{esc}{\label{oracleNearest}}
  \PYG{n}{near} \PYG{o}{=} \PYG{n}{nearest}\PYG{p}{(}\PYG{n}{data}\PYG{p}{,} \PYG{n}{row}\PYG{p}{)}
  \PYG{k}{for} \PYG{n}{col} \PYG{o+ow}{in} \PYG{n}{data}\PYG{o}{.}\PYG{n}{cols}\PYG{o}{.}\PYG{n}{ys}\PYG{p}{:} \PYG{n}{row}\PYG{p}{[}\PYG{n}{col}\PYG{o}{.}\PYG{n}{at}\PYG{p}{]} \PYG{o}{=} \PYG{n}{near}\PYG{p}{[}\PYG{n}{col}\PYG{o}{.}\PYG{n}{at}\PYG{p}{]}
  \PYG{k}{return} \PYG{n}{disty}\PYG{p}{(}\PYG{n}{data}\PYG{p}{,} \PYG{n}{row}\PYG{p}{)}


\PYG{k}{def}\PYG{+w}{ }\PYG{n+nf}{nearest}\PYG{p}{(}\PYG{n}{data}\PYG{p}{,} \PYG{n}{row}\PYG{p}{)}\PYG{p}{:}
  \PYG{k}{return} \PYG{n+nb}{min}\PYG{p}{(}\PYG{n}{data}\PYG{o}{.}\PYG{n}{rows}\PYG{p}{,}
             \PYG{n}{key}\PYG{o}{=}\PYG{k}{lambda} \PYG{n}{r}\PYG{p}{:} \PYG{n}{distx}\PYG{p}{(}\PYG{n}{data}\PYG{p}{,} \PYG{n}{row}\PYG{p}{,} \PYG{n}{r}\PYG{p}{)}\PYG{p}{)}
\end{Verbatim}
\end{mintbg}
\endgroup
\endminipage\par\noindent
Finally, we can present \textsf{oneplus1}.   It runs the (1+1) loop
(mutate, score, accept, track the best) and nothing else. The
tactics used by \textsf{mutate} and \textsf{accept} are passed in
from \textsf{sa} and \textsf{ls}.
In the following code,
\textsf{restart=0} means ``never restart'' and \textsf{restart=100}
means ``restart if no improvements after 100 mutations'':
\par\nopagebreak\minipage{\linewidth}
\begingroup\mintfvset
\VerbatimPygments{\PYG}{\PYG}
\begin{mintbg}{codebg}
\begin{Verbatim}[commandchars=\\\{\},codes={\catcode`\$=3\catcode`\^=7\catcode`\_=8\relax}]
\PYG{k}{def}\PYG{+w}{ }\PYG{n+nf}{oneplus1}\PYG{p}{(}\PYG{n}{data}\PYG{p}{,} \PYG{n}{mutate}\PYG{p}{,} \PYG{n}{accept}\PYG{p}{,} \PYG{n}{oracle}\PYG{p}{,} \PYG{esc}{\label{oneplus1}}
             \PYG{n}{budget}\PYG{o}{=}\PYG{l+m+mi}{1000}\PYG{p}{,} \PYG{n}{restart}\PYG{o}{=}\PYG{l+m+mi}{0}\PYG{p}{)}\PYG{p}{:}
  \PYG{n}{h}\PYG{p}{,} \PYG{n}{best}\PYG{p}{,} \PYG{n}{best\PYGZus{}e} \PYG{o}{=} \PYG{l+m+mi}{0}\PYG{p}{,} \PYG{k+kc}{None}\PYG{p}{,} \PYG{l+m+mf}{1E32}
  \PYG{n}{s}\PYG{p}{,} \PYG{n}{e}\PYG{p}{,} \PYG{n}{imp} \PYG{o}{=} \PYG{n}{choice}\PYG{p}{(}\PYG{n}{data}\PYG{o}{.}\PYG{n}{rows}\PYG{p}{)}\PYG{p}{[}\PYG{p}{:}\PYG{p}{]}\PYG{p}{,} \PYG{l+m+mf}{1E32}\PYG{p}{,} \PYG{l+m+mi}{0}
  \PYG{k}{while} \PYG{n}{h} \PYG{o}{\PYGZlt{}} \PYG{n}{budget}\PYG{p}{:}
    \PYG{k}{for} \PYG{n}{sn} \PYG{o+ow}{in} \PYG{n}{mutate}\PYG{p}{(}\PYG{n}{s}\PYG{p}{)}\PYG{p}{:}
      \PYG{n}{h} \PYG{o}{+}\PYG{o}{=} \PYG{l+m+mi}{1}
      \PYG{n}{en} \PYG{o}{=} \PYG{n}{oracle}\PYG{p}{(}\PYG{n}{sn}\PYG{p}{)}
      \PYG{k}{if} \PYG{n}{accept}\PYG{p}{(}\PYG{n}{e}\PYG{p}{,} \PYG{n}{en}\PYG{p}{,} \PYG{n}{h}\PYG{p}{,} \PYG{n}{budget}\PYG{p}{)}\PYG{p}{:}
        \PYG{n}{s}\PYG{p}{,} \PYG{n}{e} \PYG{o}{=} \PYG{n}{sn}\PYG{p}{,} \PYG{n}{en}
      \PYG{k}{if} \PYG{n}{en} \PYG{o}{\PYGZlt{}} \PYG{n}{best\PYGZus{}e}\PYG{p}{:}
        \PYG{n}{best}\PYG{p}{,} \PYG{n}{best\PYGZus{}e}\PYG{p}{,} \PYG{n}{imp} \PYG{o}{=} \PYG{n}{sn}\PYG{p}{[}\PYG{p}{:}\PYG{p}{]}\PYG{p}{,} \PYG{n}{en}\PYG{p}{,} \PYG{n}{h}
        \PYG{k}{yield} \PYG{n}{h}\PYG{p}{,} \PYG{n}{best\PYGZus{}e}\PYG{p}{,} \PYG{n}{best}
      \PYG{k}{if} \PYG{n}{restart} \PYG{o+ow}{and} \PYG{n}{h} \PYG{o}{\PYGZhy{}} \PYG{n}{imp} \PYG{o}{\PYGZgt{}} \PYG{n}{restart}\PYG{p}{:} \PYG{esc}{\label{restart}}
        \PYG{n}{s} \PYG{o}{=} \PYG{n}{choice}\PYG{p}{(}\PYG{n}{data}\PYG{o}{.}\PYG{n}{rows}\PYG{p}{)}\PYG{p}{[}\PYG{p}{:}\PYG{p}{]}
        \PYG{n}{e}\PYG{p}{,} \PYG{n}{imp} \PYG{o}{=} \PYG{l+m+mf}{1E32}\PYG{p}{,} \PYG{n}{h}
        \PYG{k}{break}
\end{Verbatim}
\end{mintbg}
\endgroup
\endminipage\par\noindent

\begin{table*}[t]
\centering\tiny
\caption{Four treatments across 20 randomly selected MOOT benchmarks. Cells show
mean (sd) scores (from Equation~\ref{eq:wins}) over \NEW{100 repeats}; budget = 1000 oracle calls. +r/-r denotes restart/retry enabled/disabled (respectively).
Panel~(a) applies the pre-registered $0.35\sigma$ clamp of
\S\ref{sec:opt-results}; panel~(b) scores the same runs with the
clamp removed. Both panels were regenerated from one set of runs
so that they are directly comparable. \NEW{These runs supersede the corresponding table of the
original submission; following the reviewer's request, the
repeat count was raised from 20 to 100, which shrinks the
standard error of every cell mean to sd/10.} Rows are sorted by the
\textsf{ls-r} mean of panel~(a).}
\label{tab:opt}
(a) clamped (pre-registered scoring):\\[2pt]
\begin{tabular}{rrrrrrrl}
ls-r & ls+r & sa-r & sa+r & \#rows & \#x & \#y & file \\
\midrule
 42 (45) & 91 (18) & 99 (6) & 99 (6) & 4608 & 21 & 2 & SS-U \\
 56 (50) & 91 (16) & 96 (11) & 96 (11) & 86058 & 11 & 2 & SS-X \\
 62 (50) & 97 (9) & 100 (0) & 100 (0) & 4653 & 38 & 1 & SQL\_AllMeasurements \\
 63 (38) & 93 (16) & 99 (6) & 99 (6) & 1023 & 11 & 2 & SS-P \\
 65 (21) & 79 (9) & 89 (11) & 89 (11) & 53662 & 17 & 2 & SS-N \\
 74 (21) & 96 (10) & 94 (11) & 94 (11) & 196 & 3 & 2 & SS-G \\
 77 (22) & 96 (10) & 96 (10) & 97 (9) & 196 & 3 & 2 & SS-F \\
 77 (46) & 100 (0) & 100 (0) & 100 (0) & 1152 & 16 & 1 & X264\_AllMeasurements \\
 81 (21) & 97 (9) & 96 (10) & 96 (10) & 196 & 3 & 2 & SS-D \\
 84 (52) & 100 (0) & 100 (0) & 100 (0) & 864 & 17 & 3 & SS-M \\
 89 (20) & 100 (2) & 100 (2) & 100 (2) & 259 & 4 & 2 & SS-H \\
 92 (29) & 100 (0) & 100 (0) & 100 (0) & 5184 & 12 & 2 & SS-T \\
 95 (21) & 100 (0) & 100 (0) & 100 (0) & 3840 & 6 & 1 & rs-6d-c3\_obj1 \\
 95 (20) & 100 (0) & 100 (0) & 100 (0) & 196 & 3 & 1 & wc+wc-3d-c4-obj1 \\
 95 (14) & 100 (0) & 100 (0) & 100 (0) & 3840 & 6 & 2 & SS-S \\
 95 (17) & 100 (0) & 100 (0) & 100 (0) & 3840 & 6 & 2 & SS-J \\
 95 (20) & 100 (0) & 100 (0) & 100 (0) & 196 & 3 & 1 & wc+rs-3d-c4-obj1 \\
 96 (18) & 100 (0) & 100 (0) & 100 (0) & 192 & 9 & 1 & Apache\_AllMeasurements \\
 96 (19) & 100 (0) & 100 (0) & 100 (0) & 196 & 3 & 1 & wc+sol-3d-c4-obj1 \\
 97 (16) & 100 (0) & 100 (0) & 100 (0) & 2866 & 6 & 1 & sol-6d-c2-obj1 \\
\midrule
 81 & 97 & 98 & 98 & & & & mean \\
\end{tabular}\\[8pt]
(b) same runs, clamp removed:\\[2pt]
\begin{tabular}{rrrrrrrl}
ls-r & ls+r & sa-r & sa+r & \#rows & \#x & \#y & file \\
\midrule
 40 (42) & 85 (16) & 93 (6) & 93 (6) & 4608 & 21 & 2 & SS-U \\
 53 (48) & 86 (13) & 90 (9) & 90 (9) & 86058 & 11 & 2 & SS-X \\
 54 (44) & 85 (8) & 91 (3) & 91 (3) & 4653 & 38 & 1 & SQL\_AllMeasurements \\
 62 (36) & 90 (15) & 96 (6) & 96 (6) & 1023 & 11 & 2 & SS-P \\
 65 (21) & 77 (6) & 84 (7) & 84 (7) & 53662 & 17 & 2 & SS-N \\
 72 (18) & 90 (10) & 88 (11) & 87 (11) & 196 & 3 & 2 & SS-G \\
 74 (20) & 91 (10) & 90 (10) & 91 (9) & 196 & 3 & 2 & SS-F \\
 73 (44) & 97 (2) & 98 (1) & 98 (1) & 1152 & 16 & 1 & X264\_AllMeasurements \\
 76 (18) & 90 (10) & 89 (10) & 90 (10) & 196 & 3 & 2 & SS-D \\
 33 (52) & 93 (14) & 98 (1) & 98 (1) & 864 & 17 & 3 & SS-M \\
 82 (17) & 95 (6) & 96 (5) & 96 (5) & 259 & 4 & 2 & SS-H \\
 84 (27) & 95 (3) & 97 (2) & 97 (2) & 5184 & 12 & 2 & SS-T \\
 88 (21) & 98 (2) & 99 (1) & 99 (1) & 3840 & 6 & 1 & rs-6d-c3\_obj1 \\
 93 (20) & 99 (1) & 99 (1) & 99 (1) & 196 & 3 & 1 & wc+wc-3d-c4-obj1 \\
 86 (13) & 94 (4) & 96 (3) & 96 (3) & 3840 & 6 & 2 & SS-S \\
 91 (16) & 97 (2) & 97 (2) & 97 (2) & 3840 & 6 & 2 & SS-J \\
 92 (20) & 99 (2) & 99 (2) & 99 (2) & 196 & 3 & 1 & wc+rs-3d-c4-obj1 \\
 80 (19) & 92 (7) & 96 (4) & 96 (4) & 192 & 9 & 1 & Apache\_AllMeasurements \\
 94 (19) & 99 (1) & 99 (1) & 99 (1) & 196 & 3 & 1 & wc+sol-3d-c4-obj1 \\
 91 (15) & 95 (1) & 95 (1) & 95 (1) & 2866 & 6 & 1 & sol-6d-c2-obj1 \\
\midrule
 74 & 92 & 94 & 95 & & & & mean \\
\end{tabular}
\end{table*} 
Note that:
\begin{itemize}
\item The \textsf{restart} parameter jumps the search back to a
fresh random starting row if no improvement has appeared in the
last \textsf{restart} evaluations.  
\item \textsf{sa} calls \textsf{oneplus1} with \textsf{restart=0}
which, as can be seen on line~\ref{restart}, disables restarts for
that algorithm.
\end{itemize}
Note also that, as promised at the start of this section, not only
do these two optimizers share much of the same codebase, but this
code also reuses large parts of the clustering code:
\begin{itemize}
\item Our optimizers needed a surrogate oracle to assess new
mutants. We supplied this with a nearest neighbor algorithm based
on the clustering code  described in \S\ref{kluster}.
\item When mutating numerics, our optimizer uses distribution
knowledge of the standard deviation to decide what counts as a
reasonable mutation size. This information was already collected
by the \cls{Num} class, which we were using for the surrogates.
\end{itemize}


\subsection{Experiments with (1+1)}\label{sec:opt-results}

Code similarities aside, another reason to use EZR is that it is
very easy to reorganize such a small code base to conduct insightful
experiments. For example, in this section we comment
on the relative merits of restart/retry mechanisms.

Why is it interesting to explore this issue?
In the 1990s, LS with all its variants (GSAT~\cite{gsat},
WalkSAT~\cite{walksat}, GRASP~\cite{grasp}, Iterated LS~\cite{ils})
represented a dominant paradigm in metaheuristics research, with
a steady drumbeat of papers claiming clear advantage over older
methods like SA. A reader who
read that literature would expect LS to dominate here.

However, our results with EZR on the MOOT data
(Table~\ref{mootdata}) tell a different story. This section
details a recurring observation: many long-standing truisms
in optimization dissolve when confronted with modern
empirical benchmarks.

To compare LS and SA, we need data and some goal
within that data.
For data, we selected 20 data sets at random from the MOOT data
 (recall
that examples of those tables were displayed in
Table~\ref{moot}).

Turning now to what goal to explore, within any MOOT table
of data there is one row with
the smallest \textsf{disty}. We call this the {\em reference
optimum} $d^*$.\footnote{According to the
empirical algorithms literature~\cite{McGeoch2012,cohen1995empirical},
the
{\em reference optimum} is the best solution observed in the dataset. This is not necessarily the true global optimum
since in most real engineering problems,
the true optimum is unknown, so the reference optimum is the
strongest available standard.} A baseline $d_0$ is set to
the median \textsf{disty} of the rows in that table.
In order to compare optimization results across multiple
MOOT data sets, we can report the best row found as some
ratio of $d_0 - d^*$:
\begin{equation}\label{eq:wins}
  \text{score} = 100 \cdot \left( 1 - \frac{\text{disty(best)} - d^*}{d_0 - d^*} \right)
\end{equation}
Here, {\em score}$=0$ means the optimizer did no better than the
median row, {\em score}$=50$ means it closed half the gap, and
{\em score}$=100$ means it found the reference
optimum.

One consequence of this design should be stated up front.
Since our surrogate oracle copies $y$-values from the nearest
labelled row (line~\ref{oracleNearest}), any candidate's score
snaps to the goals of some existing row; hence the achievable
optimum is exactly $d^*$, the best {\em existing} row. This
favors methods that (in effect) sample among rows over methods
that construct genuinely novel points. Accordingly, in this
paper Equation~\ref{eq:wins} is read as a {\em relative}
comparison between treatments that all share the same oracle,
not as a claim of absolute attainment.

(Aside: for readers familiar with reinforcement learning,
we note that Equation~\ref{eq:wins} is the
complement of {\em regret}, normalized by the achievable
improvement range.)

\subsubsection{Statistical decisions}
Before running the experiments reported in this paper, we
committed to the following analysis rules and applied them
uniformly to every comparison in Sections~\ref{sec:opt},
\ref{sec:active}, and \ref{sec:textmine}:
\begin{itemize}
\item All reported \textsf{wins} scores use the formula in
Equation~\ref{eq:wins} above. Lower-is-better goals
(suffix ``$-$'') and higher-is-better goals (``$+$'') are
already normalized by \textsf{disty}, which folds them into
a single 0--100 scale.
\item Differences smaller than Sawilowsky's threshold of
$0.35\sigma$~\cite{sawilowsky2009new} are clamped to zero
before scoring. Here $\sigma$ is estimated robustly from the
(90th $-$ 10th)/2.56 percentile range of \textsf{disty}.
This clamp is implemented once in the \textsf{wins}
function (see line~\ref{wins}) and reused across every
experiment in this paper. Since this clamp folds
near-optimal results into a score of 100, it can mask
differences near the top of the scale. Therefore
Table~\ref{tab:opt} reports each result twice: once with the
pre-registered clamp, and once with the clamp removed.
\item Every randomized comparison is reported as the median
over 20 or more independent repeats with shuffled row
orders. Tables additionally report standard deviation.
\item We intentionally do not perform per-cell significance
tests (Wilcoxon, Cliff's delta). Pre-registering the
$0.35\sigma$ threshold and the 20-repeat aggregation rule
serves the same function (rejecting trivially small
differences) without requiring multiple-comparison
corrections across the dozens of cells in
Table~\ref{tab:opt} and Figure~\ref{check}. Readers who
prefer significance tests can recompute them from our
public logs at
\url{https://github.com/timm/bitstex/blob/main/runs/runs.txt}.
\end{itemize}

For each of the 20 MOOT benchmarks used in this study, we shuffle the
rows, set aside 50 as the surrogate's known set, and run
LS and SA with restart/retry enabled/disabled
for a budget of 1000 oracle calls. We repeated
this 20 times per (file, treatment) pair, recording the score of
the best mutant found under both the clamped and
unclamped scoring.

 Table~\ref{tab:opt} reports mean and standard
deviation across the 20 repeats.
In those results, three patterns are noteworthy:
\begin{itemize}
\item
\textit{Restart is a patch for LS, not an upgrade over SA.}
Removing restart from LS is catastrophic: the \textsf{ls-r} score mean
drops to \NEW{81, and on the worst rows it lands at 42, 56, 62, with
standard deviations of 45, 50, 50.}   Adding restart back fixes
LS (\textsf{ls+r} mean of 97), but does not push it past SA. SA
without restart already averages \NEW{98}, and adding restart to SA
does not move the mean.
To say that another way,
SA's stochastic acceptance is doing
the same job as restart.
\item
\textit{The variance is the story, not the mean.} The \textsf{ls-r}
column has standard deviations of \NEW{38--52} on its hardest
rows. The other three columns sit at \NEW{0--18} sd on those same rows.
For a practitioner, this means \textsf{ls-r} is not just worse on
average, it is unpredictable in a way the others are not.
\item
\textit{The ceiling is a scoring artifact; the conclusions
are not.} In the clamped panel of Table~\ref{tab:opt}, on \NEW{11} of
the 20 datasets every treatment except \textsf{ls-r} scores
100 (sd~0); i.e., that panel has limited power to discriminate
among the three good treatments. Two mechanisms produce this
ceiling: the $0.35\sigma$ clamp folds near-optimal results into
100, and (as noted above) the surrogate makes $d^*$ exactly
attainable. The unclamped panel removes the first mechanism, and
the ceiling with it: no cell remains at 100 (sd~0), and the means
shift from
\NEW{\[81\;,\;97\;,\;98\;,\;98\]}
to 
\NEW{\[74\;,\;92\;,\;94\;,\;95\]}
(for \textsf{ls-r}, \textsf{ls+r}, \textsf{sa-r}, \textsf{sa+r}).
Every conclusion of this section survives, and one grows
stronger: unclamped, \textsf{ls-r} on SS-M falls to \NEW{33 (sd~52)};
i.e., the clamp was masking the severity of raw local search,
not manufacturing the gap. Meanwhile \textsf{ls+r} still does
not pass SA, and restart on SA remains a rounding error (per-file
columns agree within one point).
\end{itemize}
The headline for a practitioner is as follows:
 reset/retry should not be used uncritically.
Rather, in new domains, it would be useful to comparatively
assess the merits of reset/retry against other approaches.
If you are running local search,
you need restart, and even with it you do not pull ahead of SA.
If you are running SA, restart is a rounding error. The simplest
working choice is \textsf{sa-r}: simulated annealing with the
default \textsf{restart=0}, the same configuration the algorithm
has had since 1983.

These results come with the obvious caveat: it only holds
for the data studied here. Further experimentation, on more data,
is required.  Tools like EZR encourage such experimentation
due to their simplicity.

In summary: (a)~optimization is not its own
separate algorithmic kingdom (since it shares so much with regression and classification),
and (b)~1983 SA still holds its ground against
the 1990s LS variants.

\section{Active Learning in 80 Lines}\label{sec:active}

This section tests  truisms about learning
from labeled data. 
When learning some model $y=f(x)$ from examples $(x,y)_1, (x,y)_2,$ etc.,
it is often assumed that:
\begin{itemize}
\item
The  more labels for the $y$ attributes, the better the models;
\item 
Using 
more  $x$ features yields better models;
\item
Retraining models is always  a slow process.
\end{itemize}
This section offers counterexamples to all these points.

\subsection{Why Active Learning?}
The prior results with (1+1) algorithms
assumed  that  we could  obtain   $y$ labels of up to 1000 rows.  
In SE, that assumption is unwise:
\begin{itemize}
\item
{\bf Labeling by human experts}  is possible but slow and error-prone when rushed~\cite{easterby1980design}, often taking hours for just a few cases~\cite{KingtonAlison2009,lustosa2024learning,valerdi2010heuristics}.
\item
{\bf Historical logs} provide large label sets but can be unreliable; e.g., 90\% of technical debt ``false positives'' were incorrect~\cite{yu2020identifying}; similar issues occur in security~\cite{wu2021data}, static analysis~\cite{kang2022detecting}, and defect data~\cite{shepperd2013data}.
\item
{\bf Automated labeling} also faces limits: regex-based heuristics are crude~\cite{kamei2012large}; LLMs are only assistive (but not authoritative).
Even in domains with naturally occurring oracles (e.g. compile with certain Makefile settings; run the full test suite), these can be extremely slow. For example, an exhaustive
exploration of the 11 parameters of the \textsc{x264} video encoder took 1000+  hours to terminate~\cite{DBLP:conf/wosp/ValovPGFC17}. Such costs mean that, in practice, the number of evaluations is often limited to just a few dozen~\cite{DBLP:journals/tse/Nair0MSA20,DBLP:conf/icse/0003XC021}.
\end{itemize}
 When we cannot trust many labels, we must do what we can with very few labels. Enter {\em active learning}~\cite{settles2008analysis}. 
 
 A {\em passive learner} trains on whatever data it is given (the Naive Bayes and
 $k$-means algorithms described above are passive learners). An {\em active
learner}, on the other hand,  picks its own training set. Specifically, an active learner reflects on the
model built so far to find the  most informative    example to label next. 
It then updates its model and the process repeats. In practice,
this means the active learner labels very few rows since it is 
skipping over all the  examples that are noisy, redundant, or
uninformative.  

\subsection{Taming the Cost of Model Update}

The cost of active learning is in the model updates. New labels arrive
one at a time. After each arrival, the model has to be updated. If
the model is an ensemble of regression trees (as in
SMAC3~\cite{hutter2011sequential}), this update means rebuilding all the trees.

EZR's active learner 
was inspired by SMAC3. For every part of SMAC3, we asked if   analogous
functionality could be achieved in a simpler way.
For example,   SMAC maintains many models in its ensemble.
Why not try just two models?  We tried:
\begin{itemize}
\item A \cls{Data} called \textsf{best} holding the top $\sqrt{N}$
(where ``best'' is defined by \textsf{disty} from line~\ref{disty}), where $N$ is the total number of
labeled rows so far.
\item A second \cls{Data} called \textsf{rest} holds the
remaining $N - \sqrt{N}$.
\item When a new label arrives, it enters \textsf{best}. If
  \textsf{best} grows larger than  $\sqrt{N+1}$, the worst row in
\textsf{best} is demoted to \textsf{rest}.
\end{itemize}
EZR implements this demotion using the same \textsf{add} function used
above for $k$-means and our Naive Bayes classifier:
\par\nopagebreak\minipage{\linewidth}
\begingroup\mintfvset
\VerbatimPygments{\PYG}{\PYG}
\begin{mintbg}{codebg}
\begin{Verbatim}[commandchars=\\\{\},codes={\catcode`\$=3\catcode`\^=7\catcode`\_=8\relax}]
\PYG{k}{def}\PYG{+w}{ }\PYG{n+nf}{rebalance}\PYG{p}{(}\PYG{n}{best}\PYG{p}{,} \PYG{n}{rest}\PYG{p}{,} \PYG{n}{lab}\PYG{p}{)}\PYG{p}{:}
  \PYG{k}{if} \PYG{n+nb}{len}\PYG{p}{(}\PYG{n}{best}\PYG{o}{.}\PYG{n}{rows}\PYG{p}{)} \PYG{o}{\PYGZgt{}} \PYG{n}{sqrt}\PYG{p}{(}\PYG{n+nb}{len}\PYG{p}{(}\PYG{n}{lab}\PYG{o}{.}\PYG{n}{rows}\PYG{p}{)}\PYG{p}{)}\PYG{p}{:}
    \PYG{n}{best}\PYG{o}{.}\PYG{n}{rows}\PYG{o}{.}\PYG{n}{sort}\PYG{p}{(} \PYG{esc}{\label{sorting}}
     \PYG{n}{key}\PYG{o}{=}\PYG{k}{lambda} \PYG{n}{row}\PYG{p}{:}\PYG{n}{disty}\PYG{p}{(}\PYG{n}{lab}\PYG{p}{,}\PYG{n}{row}\PYG{p}{)}\PYG{p}{)} \PYG{c+c1}{\PYGZsh{} for \PYGZdq{}disty\PYGZdq{} see line \ref{disty}}
    \PYG{n}{add}\PYG{p}{(}\PYG{n}{rest}\PYG{o}{.}\PYG{n}{cols}\PYG{p}{,}
       \PYG{n}{sub}\PYG{p}{(}\PYG{n}{best}\PYG{o}{.}\PYG{n}{cols}\PYG{p}{,}
         \PYG{n}{best}\PYG{o}{.}\PYG{n}{rows}\PYG{p}{[}\PYG{o}{\PYGZhy{}}\PYG{l+m+mi}{1}\PYG{p}{]}\PYG{p}{)}\PYG{p}{)}
    \PYG{n}{rest}\PYG{o}{.}\PYG{n}{rows}\PYG{o}{.}\PYG{n}{append}\PYG{p}{(}\PYG{n}{best}\PYG{o}{.}\PYG{n}{rows}\PYG{o}{.}\PYG{n}{pop}\PYG{p}{(}\PYG{p}{)}\PYG{p}{)}
\end{Verbatim}
\end{mintbg}
\endgroup
\endminipage\par\noindent
The only computationally expensive part of this loop
is the call to \textsf{best.rows.sort} on line \ref{sorting}. But even this
is not particularly expensive since \textsf{best} is bounded to $\sqrt{N}$;
i.e.\ 
this sort only has to rearrange a handful of items.  As to the larger \textsf{rest} sets,  EZR never needs to resort those rows.

\begin{figure}
\begin{center}
\includegraphics[width=3in]{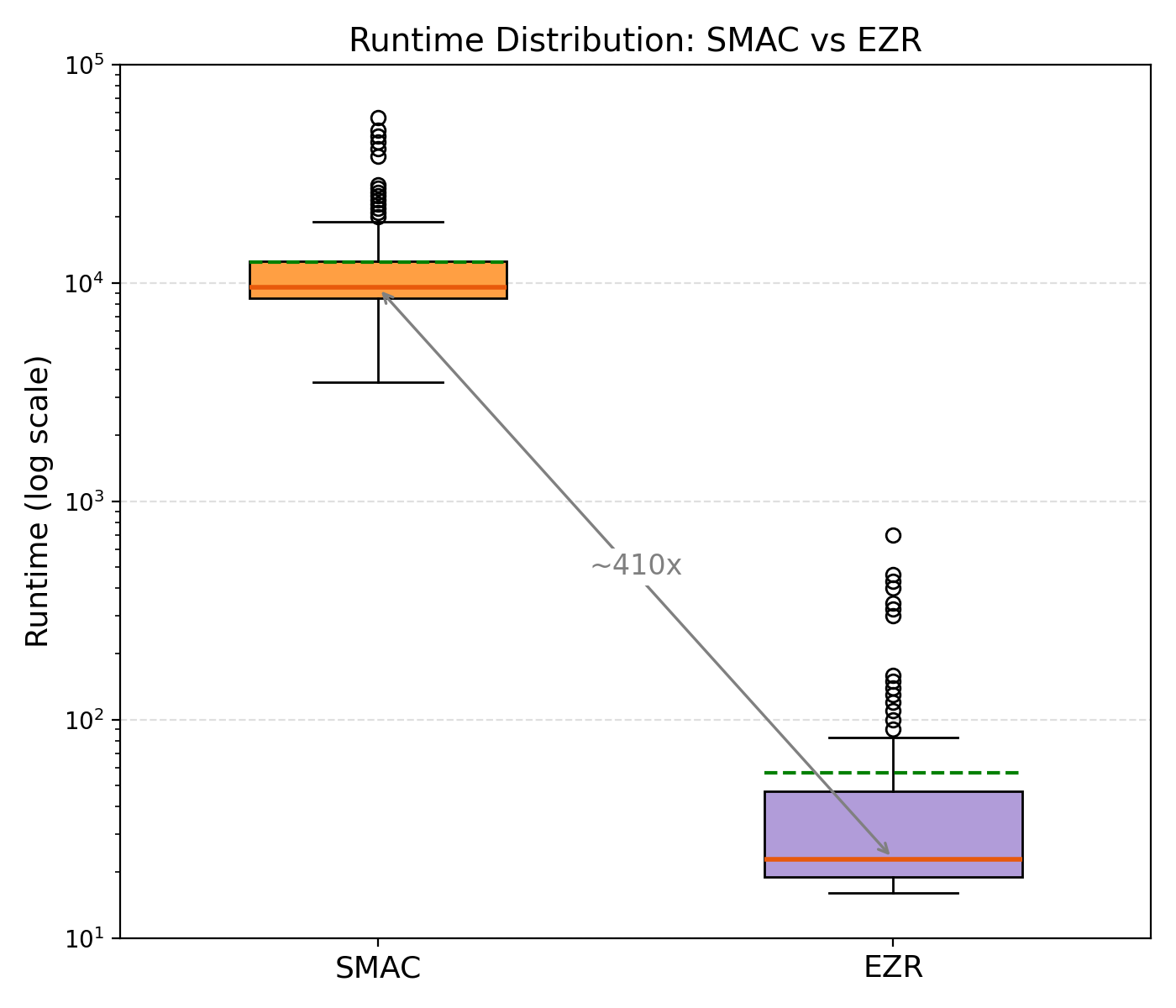}
\end{center}
\caption{SMAC3 runs 410 times slower than  EZR (on the  data of Table~\ref{tab:opt}). 
From~\cite{ganguly2026lowgodatalightse}.}\label{runtimes}
\end{figure}

Ganguly et al.~\cite{ganguly2026lowgodatalightse} report that this use of \textsf{rebalance}
makes EZR run
 over 400 times faster than SMAC3 \NEW{(see Figure~\ref{runtimes})}.  
The reason for this is simple: 
 SMAC3 has to rebuild
all its trees each time a new row is labeled.
EZR, on the other hand, just jumps a row from
{\em best} to {\em rest}.
As seen above (at line \ref{add}), adding or subtracting a row for a \cls{Data} requires very little computation.

Having tamed the cost of model update, we can now define the active learning loop that calls
\textsf{rebalance}.

 \subsection{From Active Learning to Models}
In order to test the generality
 of a learning process,
 it is best to learn a model
 from some training data, then apply it to
 test data.
 Hence, in order to test the code presented
 in the last section,
 we have to commit
 to a specific  model format (in order to transfer knowledge from   training to testing). But in that process, what model format should be used?



 One model format supported by EZR is the {\em tiny regression tree}
 such as shown in 
Figure~\ref{htree}.
That tree was grown using the code from \S\ref{trees}.
These trees are tiny since they are learned   from just the few dozen most
informative examples selected by EZR's active learner.
We use this regression tree format since a recent study by
 Rayegan et al.~\cite{Amiraliminimaldata} showed that these trees can rank features   (via mean decrease in impurity) just as well as other explanation
 algorithms such as  SHAP and ReliefF. Further, when it ranks features,
 EZR needs only  150 labels while SHAP and ReliefF need
 hundreds to thousands of rows.
 
Figure~\ref{htree} shows a software process control task where some manager must 
build a team with (e.g.):
\begin{itemize}
\item
The appropriate programmer capability ({\em pcap});
\item
In order to deliver a system where upper management
is enforcing some   {\em time} crunch on  completion;
\item
While working
in an environment with some level of process maturity ({\em  pmat})
\end{itemize}
Figure~\ref{htree} was built by EZR studying  a data set with 10,000 rows
generated by running what-if queries
on Boehm's COCOMO software models. These models have been extensively
researched and applied, particularly in the United States by the
Department of Defense~\cite{Boehm,boehm2000cost,Boehm:2006,cocomo}.

The left-hand-side of that figure
reports the mean \textsf{disty} of
the rows that fall into different  parts of the tree:
\begin{itemize}
\item
The mean \textsf{disty} of the
untreated rows is 0.52 (see \rc{1}). 
\item The most useful branch of the tree
collects rows whose \textsf{disty} is 0.45
(see \rc{2}).
\end{itemize}
This gap of $0.52-0.45 = 0.07$ looks too
small to be useful. However:
\begin{itemize}
\item
If we look closer at the raw $y$ values,
we see that   even small
changes in \textsf{disty}
can have great business significance.
\item
The
0.52 rows have median DEFECTS=4698, EFFORT=771, KLOC=113.
\item The
0.45 rows have median DEFECTS=4381, EFFORT=379, KLOC=162.
\item That is to say,
the constraints  in the branch down to 
\rc{2}  delivers $(162-113)/113 = 43\%$ more code with
$379/771 = 49\%$ of the effort, with defects roughly unchanged.
 \end{itemize}
\begin{figure}[!t]
{\fontsize{6pt}{7.2pt}\selectfont
\begin{alltt}
disty                                  DEFECTS-   EFFORT-   KLOC- 
-------------------------------------------------------------------
.52                                |    4698  |    771  |  113 \rc{1}
.50    time != xh                  |          |         |
.49    |  pmat != l                |          |         |
.48    |  |  ltex != n             |          |         |
.45    |  |  |  pcap == vh         |    4381  |    379  |  162 \rc{2}
.48    |  |  |  pcap != vh         |          |         |
.48    |  |  |  |  pvol != h       |          |         |
.48    |  |  |  |  |  stor == n    |          |         |
.47    |  |  |  |  |  |  pcap == h |          |         |
.48    |  |  |  |  |  |  pcap != h |          |         |
.48    |  |  |  |  |  stor != n    |          |         |
.50    |  |  |  |  pvol == h       |          |         |
.52    |  |  ltex == n             |          |         |
.51    |  |  |  rely != h          |          |         |
.54    |  |  |  rely == h          |          |         |
.55    |  pmat == l                |          |         |
.48    |  |  rely == n             |          |         |
.60    |  |  rely != n             |          |         |
.54    |  |  |  plex == h          |          |         |
.66    |  |  |  plex != h          |          |         |
\end{alltt}
}
\caption{EZR's regression trees (built using
code from \S\ref{trees}) predict \textsf{disty}.
As a method of focusing on essential details, the  $y$ values
are hidden except on two rows.}\label{htree}
\end{figure}
Our tiny regression trees are built from rows selected by an active 
learner. That learner, \textsf{acquire}, relies on two helpers: 
\textsf{warm\_start} (to seed the pool) and a \textsf{score} function 
(to pick what to label next). We describe each, then \textsf{acquire} 
itself.

\textsf{warm\_start} labels \textsf{the.learn.start} random rows
(default 4), sorts them by \textsf{disty}, and splits them into
\textit{best} and \textit{rest}. The leftover pool is returned for
later scoring.
\par\nopagebreak\minipage{\linewidth}
\begingroup\mintfvset
\VerbatimPygments{\PYG}{\PYG}
\begin{mintbg}{codebg}
\begin{Verbatim}[commandchars=\\\{\},codes={\catcode`\$=3\catcode`\^=7\catcode`\_=8\relax}]
\PYG{n}{the}\PYG{o}{.}\PYG{n}{learn}\PYG{o}{.}\PYG{n}{start} \PYG{o}{=} \PYG{l+m+mi}{4} \PYG{c+c1}{\PYGZsh{} config}

\PYG{k}{def}\PYG{+w}{ }\PYG{n+nf}{warm\PYGZus{}start}\PYG{p}{(}\PYG{n}{data}\PYG{p}{,} \PYG{n}{rows}\PYG{p}{,} \PYG{n}{label}\PYG{p}{)}\PYG{p}{:} \PYG{esc}{\label{warmstart}}
  \PYG{n}{lab} \PYG{o}{=} \PYG{n}{clone}\PYG{p}{(}\PYG{n}{data}\PYG{p}{,} \PYG{n}{rows}\PYG{p}{[}\PYG{p}{:}\PYG{n}{the}\PYG{o}{.}\PYG{n}{learn}\PYG{o}{.}\PYG{n}{start}\PYG{p}{]}\PYG{p}{)}
  \PYG{n}{lab}\PYG{o}{.}\PYG{n}{rows}\PYG{o}{.}\PYG{n}{sort}\PYG{p}{(}\PYG{n}{key}\PYG{o}{=}\PYG{k}{lambda} \PYG{n}{row}\PYG{p}{:} \PYG{n}{disty}\PYG{p}{(}\PYG{n}{lab}\PYG{p}{,} \PYG{n}{label}\PYG{p}{(}\PYG{n}{data}\PYG{p}{,} \PYG{n}{row}\PYG{p}{)}\PYG{p}{)}\PYG{p}{)}
  \PYG{n}{n} \PYG{o}{=} \PYG{n+nb}{int}\PYG{p}{(}\PYG{n}{sqrt}\PYG{p}{(}\PYG{n+nb}{len}\PYG{p}{(}\PYG{n}{lab}\PYG{o}{.}\PYG{n}{rows}\PYG{p}{)}\PYG{p}{)}\PYG{p}{)}
  \PYG{k}{return} \PYG{p}{(}\PYG{n}{lab}\PYG{p}{,}
          \PYG{n}{clone}\PYG{p}{(}\PYG{n}{data}\PYG{p}{,} \PYG{n}{lab}\PYG{o}{.}\PYG{n}{rows}\PYG{p}{[}\PYG{p}{:}\PYG{n}{n}\PYG{p}{]}\PYG{p}{)}\PYG{p}{,}
          \PYG{n}{clone}\PYG{p}{(}\PYG{n}{data}\PYG{p}{,} \PYG{n}{lab}\PYG{o}{.}\PYG{n}{rows}\PYG{p}{[}\PYG{n}{n}\PYG{p}{:}\PYG{p}{]}\PYG{p}{)}\PYG{p}{,}
          \PYG{n}{rows}\PYG{p}{[}\PYG{n}{the}\PYG{o}{.}\PYG{n}{learn}\PYG{o}{.}\PYG{n}{start}\PYG{p}{:}\PYG{p}{]}\PYG{p}{)}
\end{Verbatim}
\end{mintbg}
\endgroup
\endminipage\par\noindent
EZR allows  customization of the 
\textsf{score} function. For example,
here is a function that ranks unlabeled rows by the difference between their 
distance to the \textit{best} and \textit{rest} centroids. 
\par\nopagebreak\minipage{\linewidth}
\begingroup\mintfvset
\VerbatimPygments{\PYG}{\PYG}
\begin{mintbg}{codebg}
\begin{Verbatim}[commandchars=\\\{\},codes={\catcode`\$=3\catcode`\^=7\catcode`\_=8\relax}]
\PYG{k}{def}\PYG{+w}{ }\PYG{n+nf}{acquireWithCentroid}\PYG{p}{(}\PYG{n}{data}\PYG{p}{,} \PYG{n}{best}\PYG{p}{,} \PYG{n}{rest}\PYG{p}{,} \PYG{n}{row}\PYG{p}{)}\PYG{p}{:} \PYG{esc}{\label{acqqqq}}
  \PYG{k}{return} \PYG{p}{(}\PYG{n}{distx}\PYG{p}{(}\PYG{n}{data}\PYG{p}{,} \PYG{n}{row}\PYG{p}{,} \PYG{n}{mids}\PYG{p}{(}\PYG{n}{best}\PYG{p}{)}\PYG{p}{)} \PYG{o}{\PYGZhy{}}
          \PYG{n}{distx}\PYG{p}{(}\PYG{n}{data}\PYG{p}{,} \PYG{n}{row}\PYG{p}{,} \PYG{n}{mids}\PYG{p}{(}\PYG{n}{rest}\PYG{p}{)}\PYG{p}{)}\PYG{p}{)}
\end{Verbatim}
\end{mintbg}
\endgroup
\endminipage\par\noindent
In practice, this code is not slow. Recall that
\textsf{mids} caches centroids until new data invalidates them 
(line~\ref{zap}); hence the centroid
can be  built once  then reused across all unlabeled rows.

(Aside: to stress the main 
message of this paper, observe how the active learning optimizer reuses the 
clustering architecture described above).

\textsf{acquire} stitches everything together. After \textsf{warm\_start}, 
it loops up to \textsf{the.learn.budget} times.
In each loop it uses \textsf{score} to guess how much unlabeled rows
belong to {\em best} or {\em rest}. These guesses
are used to (a)~sort the unlabeled rows and (b)~select the most
promising unlabeled example, (c)~which is then labeled
before (d)~being passed to \textsf{rebalance} in order to
update \textit{best} and
\textit{rest}.
\par\nopagebreak\minipage{\linewidth}
\begingroup\mintfvset
\VerbatimPygments{\PYG}{\PYG}
\begin{mintbg}{codebg}
\begin{Verbatim}[commandchars=\\\{\},codes={\catcode`\$=3\catcode`\^=7\catcode`\_=8\relax}]
\PYG{n}{the}\PYG{o}{.}\PYG{n}{few} \PYG{o}{=} \PYG{l+m+mi}{128}   \PYG{c+c1}{\PYGZsh{} config}
\PYG{n}{the}\PYG{o}{.}\PYG{n}{learn}\PYG{o}{.}\PYG{n}{budget} \PYG{o}{=} \PYG{l+m+mi}{50} \PYG{c+c1}{\PYGZsh{} config}

\PYG{k}{def}\PYG{+w}{ }\PYG{n+nf}{acquire}\PYG{p}{(}\PYG{n}{data}\PYG{p}{,} \PYG{n}{score}\PYG{o}{=}\PYG{n}{acquireWithCentroid}\PYG{p}{,} \PYG{esc}{\label{acquiring}}
                  \PYG{n}{label}\PYG{o}{=}\PYG{k}{lambda} \PYG{n}{\PYGZus{}}\PYG{p}{,}\PYG{n}{row}\PYG{p}{:}\PYG{n}{row}\PYG{p}{)}\PYG{p}{:}
  \PYG{n}{rows} \PYG{o}{=} \PYG{n}{data}\PYG{o}{.}\PYG{n}{rows}\PYG{p}{[}\PYG{p}{:}\PYG{p}{]}
  \PYG{n}{shuffle}\PYG{p}{(}\PYG{n}{rows}\PYG{p}{)}
  \PYG{n}{lab}\PYG{p}{,} \PYG{n}{best}\PYG{p}{,} \PYG{n}{rest}\PYG{p}{,} \PYG{n}{unlab} \PYG{o}{=} \PYG{n}{warm\PYGZus{}start}\PYG{p}{(}
    \PYG{n}{data}\PYG{p}{,} \PYG{n}{rows}\PYG{p}{[}\PYG{p}{:}\PYG{n}{the}\PYG{o}{.}\PYG{n}{few}\PYG{p}{]}\PYG{p}{,} \PYG{n}{label}\PYG{p}{)} \PYG{esc}{\label{fewing}}
  \PYG{k}{for} \PYG{n}{\PYGZus{}} \PYG{o+ow}{in} \PYG{n+nb}{range}\PYG{p}{(}\PYG{n}{the}\PYG{o}{.}\PYG{n}{learn}\PYG{o}{.}\PYG{n}{budget}\PYG{p}{)}\PYG{p}{:}
    \PYG{k}{if} \PYG{o+ow}{not} \PYG{n}{unlab}\PYG{p}{:} \PYG{k}{break}
    \PYG{n}{pick}\PYG{p}{,} \PYG{o}{*}\PYG{n}{unlab} \PYG{o}{=} \PYG{n+nb}{sorted}\PYG{p}{(}\PYG{n}{unlab}\PYG{p}{,}
                     \PYG{n}{key}\PYG{o}{=}\PYG{k}{lambda} \PYG{n}{row}\PYG{p}{:}\PYG{n}{score}\PYG{p}{(}\PYG{n}{lab}\PYG{p}{,}\PYG{n}{best}\PYG{p}{,}\PYG{n}{rest}\PYG{p}{,}\PYG{n}{row}\PYG{p}{)}\PYG{p}{)}
    \PYG{n}{add}\PYG{p}{(}\PYG{n}{lab}\PYG{p}{,} \PYG{n}{add}\PYG{p}{(}\PYG{n}{best}\PYG{p}{,} \PYG{n}{label}\PYG{p}{(}\PYG{n}{data}\PYG{p}{,} \PYG{n}{pick}\PYG{p}{)}\PYG{p}{)}\PYG{p}{)} \PYG{esc}{\label{labelling}}
    \PYG{n}{rebalance}\PYG{p}{(}\PYG{n}{best}\PYG{p}{,} \PYG{n}{rest}\PYG{p}{,} \PYG{n}{lab}\PYG{p}{)}
  \PYG{n}{lab}\PYG{o}{.}\PYG{n}{rows}\PYG{o}{.}\PYG{n}{sort}\PYG{p}{(}
    \PYG{n}{key}\PYG{o}{=}\PYG{k}{lambda} \PYG{n}{row}\PYG{p}{:} \PYG{n}{disty}\PYG{p}{(}\PYG{n}{lab}\PYG{p}{,} \PYG{n}{row}\PYG{p}{)}\PYG{p}{)}
  \PYG{k}{return} \PYG{n}{lab} \PYG{esc}{\label{acqured}}
\end{Verbatim}
\end{mintbg}
\endgroup
\endminipage\par\noindent
Three details:
\begin{itemize}
\item
On  line~\ref{fewing}, 
the \textsf{few} caps the number of unlabeled rows searched
 per iteration; larger values did not improve performance.
\item
On line~\ref{labelling},
the 
\textsf{label} function is 
a hook for an oracle that assigns $y$ values.
For MOOT data, $y$ labels are already present on the row data, so it just returns
the   row. But for other domains, \textsf{label} could (e.g.) call some
simulator to collect new data.
\item
Other active learners have adaptive sampling policies where, early in the run,
they focus on   most uncertain
examples (e.g.\ where the likelihoods of {\em best} and {\em rest} are nearly equal). Later in the run, that same learner  might favor  certainty sampling (i.e. find unlabeled examples that are most
likely to be {\em best}). After two years of exploring that kind of adaptation, and after extensive
experimentation with the MOOT data, we found   nothing   better than
a
greedy search guided by line~\ref{acqqqq}.
\end{itemize}

 \subsection{Testing Active Learning}
The \textsf{acquire} function described above returns a set of labeled
rows. How to test if those conclusions are useful? 

In any realistic 
deployment, a human inspects a short list of recommendations before 
committing to one.
We can evaluate our active learners the same way. This is the standard 
``at-$k$'' practice from information retrieval (precision@$k$, 
hit@$k$, top-$k$ accuracy);
i.e., rather than asking ``did the model rank 
the single best item first?'', we ask ``is a good item somewhere in 
the model's top $k$ picks?''.

In EZR,  $k=$\textsf{the.learn.check} is set to, say, 5.
This \textsf{check} parameter  encodes how much we trust the model: 
\textsf{check}=1 means we trust it enough to just apply its
top recommendation, without checking anything else.
Larger 
values of \textsf{check} are appropriate if we are not 100\% certain about a model.

This  \textsf{check}  is applied in two scenarios:
\begin{itemize}
\item
The \textsf{w1} scenario runs and checks the {\em training data}.
 Given that  we are only allowed to label $M\ll N$ of the rows,
 \textsf{w1} asks,   how good are those $M$ labels at finding the best of the $N$ training
examples? 
\item
\textsf{w2} runs its checks on the {\em testing data}.
Here, we import a model from some past training session and use it
to check some test data. 
\end{itemize}
There is an important difference between these two scenarios.
\begin{itemize}
\item
 Anyone applying the
\textsf{w1} check must do all the active learning, and then all the checking. 
Hence   the \textsf{w1} check requires labeling \textsf{budget}  +  \textsf{check} 
\item
But the \textsf{w2} is far cheaper. Here, some   team has built a model
using some \textsf{budget} which a second team
is  now reusing in a novel situation.
Hence     \textsf{w2} check   just   requires \textsf{check}s.
\end{itemize}
 \begin{figure*}[!b]
\noindent
\begin{tabular}{@{}>{\raggedleft\arraybackslash}m{0.35\textwidth}@{\hspace{1em}}m{4in}@{}}
\textsf{w1} results (test on a 50\% training set):\newline
(all values computed using line~\ref{wins})\newline
  & \includegraphics[width=4in]{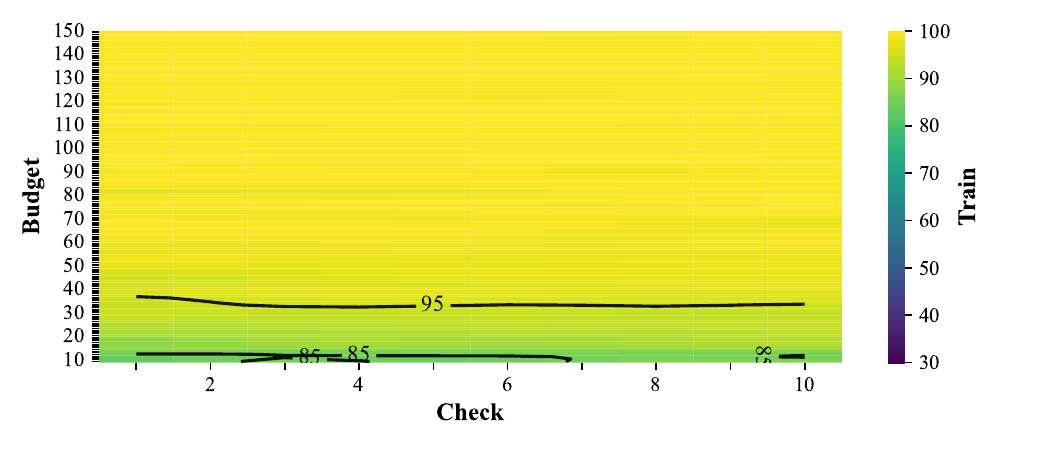} \\[1ex]
\textsf{w2} results (test on a separate 50\% hold-out set):\newline
(all values computed using line~\ref{wins})\newline
  & \includegraphics[width=4in]{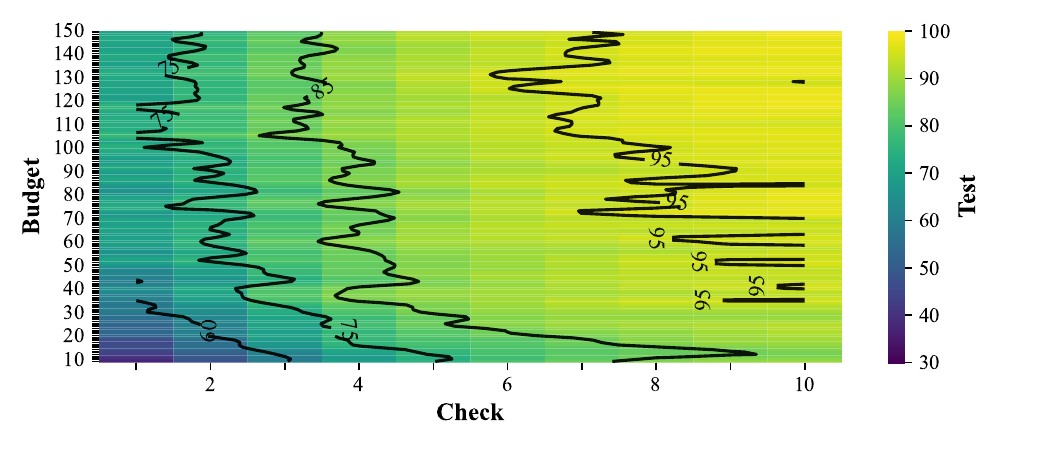}
\end{tabular}
\caption{\textsf{w1} and \textsf{w2} results in 100,000 trials.
For each trial:
(a)~pick a MOOT data set at random;
(b)~pick \textsf{budget} and \textsf{check} at random from
$1 \le B \le 150$ and $1\le C \le 10$;
(c)~apply active learning and report the wins
calculated from line~\ref{wins}.
 Contour lines mark \textsf{wins} of 60, 75,
85, 95. Each cell aggregates the median over many trials
at that (\textsf{budget}, \textsf{check}) pair, smoothed by a
small Gaussian blur from neighboring cells
(where cells $c_1,c_2,c_3$ at distance
1, 2, 3 contribute a weight $(71c_1 + 25c_2 + 4c_3)/100$).
 Contours flatten
toward the upper right, indicating diminishing returns from
larger budgets. Raw data and scripts:
\url{https://github.com/timm/bitstex/blob/main/data/runs.log}
(generated by \texttt{etc/runs.sh} and plotted by
\texttt{etc/plot2.py} in
\url{https://github.com/timm/ezr}).}
\label{check}
\end{figure*}

\subsubsection{Calculating \textsf{w1} and \textsf{w2}}
To implement those checks, EZR splits each data set in half.
The first half (capped at \textsf{the.few} = 128 rows) goes to
the active learner as the training pool. The second half is
kept separate, and never touched during acquisition.
\par\nopagebreak\minipage{\linewidth}
\begingroup\mintfvset
\VerbatimPygments{\PYG}{\PYG}
\begin{mintbg}{codebg}
\begin{Verbatim}[commandchars=\\\{\},codes={\catcode`\$=3\catcode`\^=7\catcode`\_=8\relax}]
\PYG{k}{def}\PYG{+w}{ }\PYG{n+nf}{ready}\PYG{p}{(}\PYG{n}{file}\PYG{p}{)}\PYG{p}{:}
  \PYG{n}{d} \PYG{o}{=} \PYG{p}{(}\PYG{n}{file} \PYG{k}{if} \PYG{n}{Data} \PYG{o}{==} \PYG{n+nb}{type}\PYG{p}{(}\PYG{n}{file}\PYG{p}{)}
       \PYG{k}{else} \PYG{n}{Data}\PYG{p}{(}\PYG{n}{csv}\PYG{p}{(}\PYG{n}{file}\PYG{p}{)}\PYG{p}{)}\PYG{p}{)}
  \PYG{n}{random}\PYG{o}{.}\PYG{n}{shuffle}\PYG{p}{(}\PYG{n}{d}\PYG{o}{.}\PYG{n}{rows}\PYG{p}{)}
  \PYG{n}{half} \PYG{o}{=} \PYG{n+nb}{len}\PYG{p}{(}\PYG{n}{d}\PYG{o}{.}\PYG{n}{rows}\PYG{p}{)} \PYG{o}{/}\PYG{o}{/} \PYG{l+m+mi}{2}
  \PYG{k}{return} \PYG{p}{(}\PYG{n}{d}\PYG{p}{,}
          \PYG{n}{clone}\PYG{p}{(}\PYG{n}{d}\PYG{p}{,} \PYG{n}{d}\PYG{o}{.}\PYG{n}{rows}\PYG{p}{[}\PYG{p}{:}\PYG{n}{half}\PYG{p}{]}\PYG{p}{[}\PYG{p}{:}\PYG{n}{the}\PYG{o}{.}\PYG{n}{few}\PYG{p}{]}\PYG{p}{)}\PYG{p}{,}
          \PYG{n}{d}\PYG{o}{.}\PYG{n}{rows}\PYG{p}{[}\PYG{n}{half}\PYG{p}{:}\PYG{p}{]}\PYG{p}{)}
\end{Verbatim}
\end{mintbg}
\endgroup
\endminipage\par\noindent
The controller for this process is \textsf{test\_acquire}. This runs 
20 times, collecting the best row found by the active learner. Each 
repeat (a)~shuffles the data; (b)~gives the first half to the active 
learner; (c)~lets it spend its budget; (d)~grows a tree from the 
labeled rows; and (e)~scores the result using \textsf{w1,w2}  on lines~\ref{w1} 
and~\ref{w2}:
\par\nopagebreak\minipage{\linewidth}
\begingroup\mintfvset
\VerbatimPygments{\PYG}{\PYG}
\begin{mintbg}{codebg}
\begin{Verbatim}[commandchars=\\\{\},codes={\catcode`\$=3\catcode`\^=7\catcode`\_=8\relax}]
\PYG{n}{the}\PYG{o}{.}\PYG{n}{learn}\PYG{o}{.}\PYG{n}{check}  \PYG{o}{=} \PYG{l+m+mi}{5}    \PYG{c+c1}{\PYGZsh{} config}
\PYG{n}{the}\PYG{o}{.}\PYG{n}{learn}\PYG{o}{.}\PYG{n}{budget} \PYG{o}{=} \PYG{l+m+mi}{50}   \PYG{c+c1}{\PYGZsh{} config}

\PYG{k}{def}\PYG{+w}{ }\PYG{n+nf}{test\PYGZus{}acquire}\PYG{p}{(}\PYG{n}{file}\PYG{o}{=}\PYG{n}{egopt1}\PYG{p}{)}\PYG{p}{:}
  \PYG{n}{d0} \PYG{o}{=} \PYG{n}{Data}\PYG{p}{(}\PYG{n}{csv}\PYG{p}{(}\PYG{n}{file}\PYG{p}{)}\PYG{p}{)}
  \PYG{n}{w1}\PYG{p}{,} \PYG{n}{w2} \PYG{o}{=} \PYG{n}{Num}\PYG{p}{(}\PYG{p}{)}\PYG{p}{,} \PYG{n}{Num}\PYG{p}{(}\PYG{p}{)}
  \PYG{n}{win} \PYG{o}{=} \PYG{n}{wins}\PYG{p}{(}\PYG{n}{d0}\PYG{p}{)}  \PYG{c+c1}{\PYGZsh{} for \PYGZbs{}\PYGZdq{}wins\PYGZbs{}\PYGZdq{} see line \ref{wins}}
  \PYG{k}{for} \PYG{n}{\PYGZus{}} \PYG{o+ow}{in} \PYG{n+nb}{range}\PYG{p}{(}\PYG{l+m+mi}{20}\PYG{p}{)}\PYG{p}{:}
    \PYG{n}{d}\PYG{p}{,} \PYG{n}{d\PYGZus{}train}\PYG{p}{,} \PYG{n}{test\PYGZus{}rows} \PYG{o}{=} \PYG{n}{ready}\PYG{p}{(}\PYG{n}{d0}\PYG{p}{)}
    \PYG{n}{lab} \PYG{o}{=} \PYG{n}{acquire}\PYG{p}{(}\PYG{n}{d\PYGZus{}train}\PYG{p}{)}
    \PYG{n}{t}   \PYG{o}{=} \PYG{n}{treeGrow}\PYG{p}{(}\PYG{n}{d\PYGZus{}train}\PYG{p}{,} \PYG{n}{lab}\PYG{o}{.}\PYG{n}{rows}\PYG{p}{)}
    \PYG{n}{guess} \PYG{o}{=} \PYG{n+nb}{sorted}\PYG{p}{(}\PYG{n}{test\PYGZus{}rows}\PYG{p}{,}
              \PYG{n}{key}\PYG{o}{=}\PYG{k}{lambda} \PYG{n}{r}\PYG{p}{:} \PYG{n}{mid}\PYG{p}{(}\PYG{n}{treeLeaf}\PYG{p}{(}\PYG{n}{t}\PYG{p}{,} \PYG{n}{r}\PYG{p}{)}\PYG{o}{.}\PYG{n}{ynum}\PYG{p}{)}\PYG{p}{)}
    \PYG{n}{add}\PYG{p}{(}\PYG{n}{w1}\PYG{p}{,} \PYG{n}{win}\PYG{p}{(}\PYG{n+nb}{min}\PYG{p}{(} \PYG{esc}{\label{w1}}
                  \PYG{n}{lab}\PYG{o}{.}\PYG{n}{rows}\PYG{p}{[}\PYG{p}{:}\PYG{n}{the}\PYG{o}{.}\PYG{n}{learn}\PYG{o}{.}\PYG{n}{check}\PYG{p}{]}\PYG{p}{,} \PYG{esc}{\label{check1}}
                 \PYG{n}{key}\PYG{o}{=}\PYG{k}{lambda} \PYG{n}{r}\PYG{p}{:} \PYG{n}{disty}\PYG{p}{(}\PYG{n}{d\PYGZus{}train}\PYG{p}{,} \PYG{n}{r}\PYG{p}{)}\PYG{p}{)}\PYG{p}{)}\PYG{p}{)}
    \PYG{n}{add}\PYG{p}{(}\PYG{n}{w2}\PYG{p}{,} \PYG{n}{win}\PYG{p}{(}\PYG{n+nb}{min}\PYG{p}{(} \PYG{esc}{\label{w2}}
                 \PYG{n}{guess}\PYG{p}{[}\PYG{p}{:}\PYG{n}{the}\PYG{o}{.}\PYG{n}{learn}\PYG{o}{.}\PYG{n}{check}\PYG{p}{]}\PYG{p}{,} \PYG{esc}{\label{check2}}
                 \PYG{n}{key}\PYG{o}{=}\PYG{k}{lambda} \PYG{n}{r}\PYG{p}{:} \PYG{n}{disty}\PYG{p}{(}\PYG{n}{d\PYGZus{}train}\PYG{p}{,} \PYG{n}{r}\PYG{p}{)}\PYG{p}{)}\PYG{p}{)}\PYG{p}{)}
  \PYG{k}{return} \PYG{n}{w1}\PYG{p}{,}\PYG{n}{w2}
\end{Verbatim}
\end{mintbg}
\endgroup
\endminipage\par\noindent

Since we want to compare results across
many data sets,
our \textsf{win} function used
Equation~\ref{eq:wins}. Further,
to avoid spurious conclusions based on trivially
small distances, we
clamp small improvements to 0. ``Small'' here is Sawilowsky's
threshold of $0.35\sigma$~\cite{sawilowsky2009new}, with $\sigma$
estimated robustly via  the standard method of   (90th $-$ 10th)/2.56 percentile range
of \textsf{disty}.
\par\nopagebreak\minipage{\linewidth}
\begingroup\mintfvset
\VerbatimPygments{\PYG}{\PYG}
\begin{mintbg}{codebg}
\begin{Verbatim}[commandchars=\\\{\},codes={\catcode`\$=3\catcode`\^=7\catcode`\_=8\relax}]
\PYG{k}{def}\PYG{+w}{ }\PYG{n+nf}{wins}\PYG{p}{(}\PYG{n}{data}\PYG{p}{)}\PYG{p}{:} \PYG{esc}{\label{wins}}
  \PYG{n}{ys} \PYG{o}{=} \PYG{n+nb}{sorted}\PYG{p}{(}\PYG{n}{disty}\PYG{p}{(}\PYG{n}{data}\PYG{p}{,} \PYG{n}{row1}\PYG{p}{)} \PYG{k}{for} \PYG{n}{row1} \PYG{o+ow}{in} \PYG{n}{data}\PYG{o}{.}\PYG{n}{rows}\PYG{p}{)}
  \PYG{n}{ten} \PYG{o}{=} \PYG{n+nb}{len}\PYG{p}{(}\PYG{n}{ys}\PYG{p}{)} \PYG{o}{/}\PYG{o}{/} \PYG{l+m+mi}{10}
  \PYG{n}{lo}  \PYG{o}{=} \PYG{n}{ys}\PYG{p}{[}\PYG{l+m+mi}{0}\PYG{p}{]}
  \PYG{n}{med} \PYG{o}{=} \PYG{n}{ys}\PYG{p}{[}\PYG{l+m+mi}{5}\PYG{o}{*}\PYG{n}{ten}\PYG{p}{]}
  \PYG{n}{sd}  \PYG{o}{=} \PYG{p}{(}\PYG{n}{ys}\PYG{p}{[}\PYG{l+m+mi}{9}\PYG{o}{*}\PYG{n}{ten}\PYG{p}{]} \PYG{o}{\PYGZhy{}} \PYG{n}{ys}\PYG{p}{[}\PYG{n}{ten}\PYG{p}{]}\PYG{p}{)} \PYG{o}{/} \PYG{l+m+mf}{2.56}
  \PYG{k}{def}\PYG{+w}{ }\PYG{n+nf}{f}\PYG{p}{(}\PYG{n}{row2}\PYG{p}{)}\PYG{p}{:}
    \PYG{n}{x} \PYG{o}{=} \PYG{n}{disty}\PYG{p}{(}\PYG{n}{data}\PYG{p}{,} \PYG{n}{row2}\PYG{p}{)}
    \PYG{k}{if} \PYG{n}{x} \PYG{o}{\PYGZlt{}} \PYG{n}{lo} \PYG{o}{+} \PYG{l+m+mf}{0.35}\PYG{o}{*}\PYG{n}{sd}\PYG{p}{:} \PYG{n}{x} \PYG{o}{=} \PYG{n}{lo}
    \PYG{k}{return} \PYG{n+nb}{max}\PYG{p}{(}\PYG{o}{\PYGZhy{}}\PYG{l+m+mi}{100}\PYG{p}{,}
          \PYG{n+nb}{int}\PYG{p}{(}\PYG{l+m+mi}{100} \PYG{o}{*} \PYG{p}{(}\PYG{l+m+mi}{1} \PYG{o}{\PYGZhy{}} \PYG{p}{(}\PYG{n}{x} \PYG{o}{\PYGZhy{}} \PYG{n}{lo}\PYG{p}{)}\PYG{o}{/}\PYG{p}{(}\PYG{n}{med} \PYG{o}{\PYGZhy{}} \PYG{n}{lo} \PYG{o}{+} \PYG{l+m+mf}{1e\PYGZhy{}32}\PYG{p}{)}\PYG{p}{)}\PYG{p}{)}\PYG{p}{)}
  \PYG{k}{return} \PYG{n}{f}
\end{Verbatim}
\end{mintbg}
\endgroup
\endminipage\par\noindent

\subsubsection{Parameter Effects}
Using the \textsf{w1} and \textsf{w2} scores,
we can comment on the validity of   EZR's default active
learning parameters of  \[(\textsf{budget},\textsf{check}) = (50,5).\]
In this section, we test if better settings (e.g. a larger training budget)
would lead to fundamentally better results.

 Figure~\ref{check} presents the results of a study where   MOOT data sets were selected
at random, then explored with \textsf{budget},\textsf{check} between 
  $[1,150]$ and  $[1,10]$. This process was repeated    100{,}000 times.

\begin{figure*}[!b]
\begin{center}
\includegraphics[width=\linewidth]{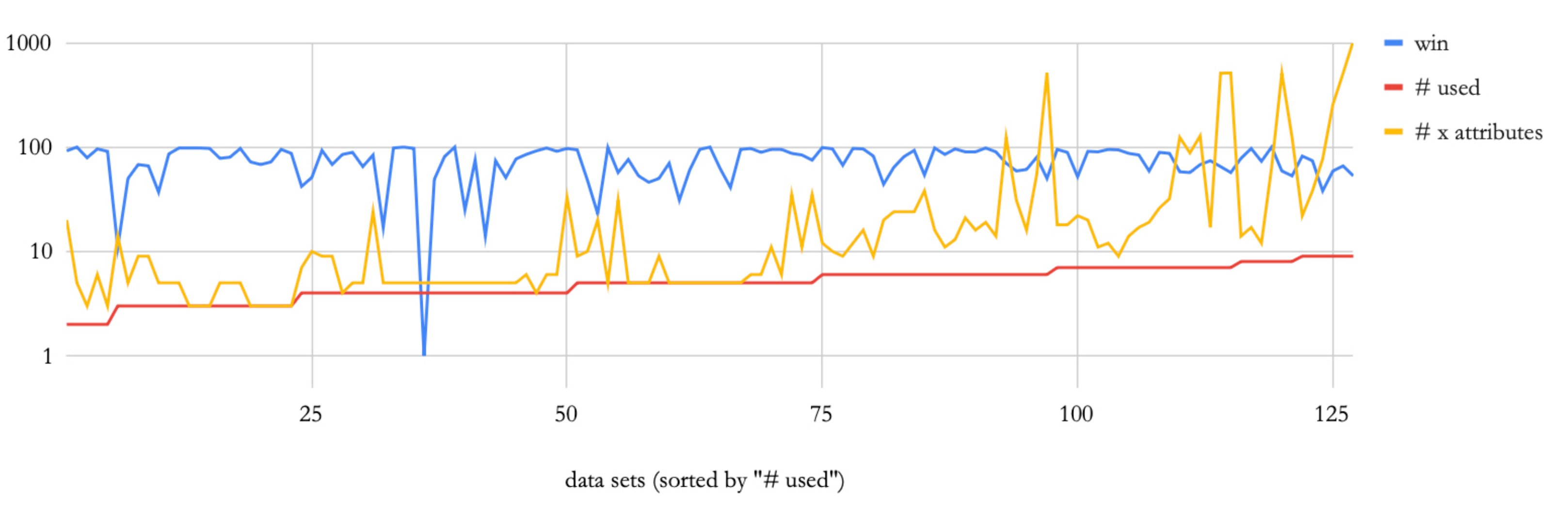}
\end{center}
\caption{Across 120+  MOOT tasks, the
\color{darkyellow}{\bf yellow line}\color{black}~shows the
number of $x$ attributes available (rising to over a thousand).
The \color{red}{\bf red line}\color{black}~shows the number of
unique attributes \IT{}'s trees actually used (under ten,
throughout). The \color{blue}{\bf blue line}\color{black}~shows
the \textsf{w2} (calculated using \textsf{wins}   from line~\ref{wins}) of those trees,
hovering near 100 across the entire range. Note that the  blue line does
{\bf not} slope downward from left to right.
That is, it is not true that performance drops if  EZR ignores more and more of the attributes.}
\label{used}
\end{figure*}

The top panel of Figure~\ref{check} reports \textsf{w1}: i.e. 
it is  scored on rows drawn
from its own training pool. 
The bottom panel of Figure~\ref{check} reports the  \textsf{w2}
results where we score on
  held-out rows     not seen during acquisition.
Recall from the above discussion that \textsf{w2} occurs when
a second team is reusing a model developed by another team
(and in  this \textsf{w2} scenario, this second team
only has to label \textsf{check} number of rows).

There are several noteworthy effects in     Figure~\ref{check}:
\begin{itemize}
\item The \textsf{w1} scores
are less than the \textsf{w2} scores. This is to be expected:
testing on training data tends to overestimate the performance on as-yet-unseen future testing data. 
\item At EZR's default values of (\textsf{budget},\textsf{check}) = (50,5),
we see our active learners usually achieve above an  85\% \textsf{win}.
That is to say, reusing someone else's model and only labeling five new examples
is enough to get most of the way to the reference optimum. 
\item
If we increase either of  (\textsf{budget},\textsf{check}) = (50,5), then we move to the top-right
of the bottom panel. In that region, we can see improvement of    the performance score on hold-out data.

\item
That said, the improvements above (50,5) are modest.   As we move towards the top right of the bottom panel, the performance contours start flattening
out as  performance grows to 100 (which is the maximum score possible on these data sets).
\end{itemize}
From this we conclude that for safety-critical and mission-critical applications,
it can be useful to move beyond (\textsf{budget},\textsf{check}) = (50,5): specifically,
{\em more checks} seem more useful than {\em more training budget}. And checking beyond (say) 7 items
seems not particularly useful. 
But for many engineering applications, EZR's current default settings would  suffice.

\subsubsection{Model Size}

Returning to Figure~\ref{htree}, that tree 
had an interesting feature that was not previously discussed. 
That tree splits its data on 8 different  attributes.
This is noteworthy since that data  set has 24 unique attributes;
i.e. EZR  built that model by ignoring most of the attributes.

This is not a one-off result: in the usual case,
EZR ignores most of the attributes.
Figure~\ref{used} shows how many attributes were used
in the tiny regression trees built after active
learning on   120+ MOOT tasks. 
 For example, in the {\em auto93} data set,
the goal is to maximize miles-per-gallon and acceleration without
increasing the weight of the car.

For these optimization tasks, Figure~\ref{used} shows three lines:

\begin{itemize}
\item The \color{darkyellow}{\bf yellow line}\color{black}~ is
the number of $x$ attributes available in each task. It rises
from a few dozen to over a thousand as we move right across
MOOT.
\item The \color{red}{\bf red line}\color{black}~ is the number
of unique attributes that actually appear in any of the trees
\IT{} grew on each task. It stays under ten throughout.
\item The \color{blue}{\bf blue line}\color{black}~ is the
\textsf{w2} 
\textsf{wins} score (from line~\ref{wins}) of those trees. 
This line  hovers near 100 across the entire range.
\item
This  \color{blue}{\bf blue line}\color{black}~line does not slope downward as we move from left to right. 
That is, 
for the  120+ MOOT tasks, fewer than ten variables suffice to
build effective predictive models, even when the data set
contains hundreds or thousands of attributes.
\end{itemize}
This {\em feature selection} result
has a long history in AI. Pearson described principal
components in 1902~\cite{pca}. Chang showed in 1974 that a
handful of exemplar rows can replace a full nearest-neighbor
training set without loss of classification
accuracy~\cite{chang74}. Johnson and Lindenstrauss proved in
1984 that high-dimensional data can be projected to far fewer
dimensions while preserving pairwise
distances~\cite{johnson1984extensions}. Kohavi and John reported
in 1997 that 80\% of features can usually be
discarded~\cite{kohavi97}. Zhu's 2005 semi-supervised learning
survey describes high-dimensional data as typically lying on a
much lower-dimensional manifold~\cite{zhu2005semi}. Settles
argued in 2009 that an active learner querying only the most
informative rows can match a learner trained on the full
set~\cite{settles2009active}. Our own work on SE ``keys'' across
2003--2021 found that a handful of parameters governs the
behavior of many software
models~\cite{me03a,me07a,me21a,menzies2008implications}.

What is new in \IT{} is that we extend this feature-selection effect
to
multi-objective optimization. Most of the prior work cited above
was about classification or regression. \IT{}, on the other hand, grows trees that
predict \textsf{disty}, which means the trees are doing
multi-objective ranking. The same effect (under ten variables
suffice) holds in that setting too.

\subsubsection{Comparison with State-of-the-Art}\label{sota}

In other publications, we have compared the \textsf{w2} scores generated
with state-of-the-art explanation and optimization algorithms.
For full disclosure: both of the studies summarized below
(Rayegan et al.~\cite{Amiraliminimaldata} and Ganguly et
al.~\cite{ganguly2026lowgodatalightse}) are \NEW{co-authored by authors of this paper}, as is the MOOT benchmark itself.
They corroborate, but do not independently confirm, the claims
of this section; see \S\ref{sec:ttv}.

Rayegan et al.~\cite{Amiraliminimaldata} compared \IT{}'s trees against
standard AI explanation algorithms
(SHAP, LIME, and Anchors) on standard XAI benchmarks. The metrics collected
were {\em actionability} (does the explanation tell you what to
change?) and {\em faithfulness} (does following the explanation
actually move the prediction?). \IT{}'s trees matched or
exceeded the named baselines on both metrics.
This is a useful property in safety-relevant settings. A
practitioner reading a SHAP report cannot easily tell whether
the surrogate has departed from the underlying model. Reading a
\IT{} tree, the practitioner is reading the actual decision
boundary. If the tree says \texttt{pcap == vh} buys 49\% effort
reduction, then \texttt{pcap == vh} buys 49\% effort reduction.

Ganguly et al.~\cite{ganguly2026lowgodatalightse}    compare \IT{}'s
optimizations against state-of-the-art optimizers including SMAC3~\cite{hutter2011sequential}
and others.
As mentioned above, they found that  EZR ran over 400 times faster than SMAC3. 
They also found that EZR delivered competitive results using a fraction
of the data needed by other methods.

So, in summary,
on the MOOT data, fewer labels,
fewer features, and lighter model updates each match or beat
more complete alternatives. And the software that achieves all
this is a very small extension to the EZR substrate used
for classification, regression and optimization.

\section{Text Mining}\label{sec:textmine}

Some of the above results have been published before (e.g. all the results
in \S\ref{sota}). 
Hence, to cap off this paper,
 we offer a new text mining result. Specifically, we successfully
 apply all the above machinery
 to text mining and relevancy filtering of documents.
 
 In the current literature,  such text mining is often
 explored using a RAG-like approach (retrieval-augmented
 generation, described in the next section). 
This requires LLM-grade embedding spaces, billion-parameter encoders,
and similarity queries against a vector database. 

Older
approaches explored text mining and relevancy filtering using
support vector machines~\cite{yu2018total,yu2018finding}. This section shows
that 30 lines of complementary Bayes~\cite{rennie2003tackling},
on the substrate from \S\ref{sec:peek}, suffice for this task.
These results are strong enough to raise a question: just
how simple is text mining? Perhaps it is time to conduct more
experiments comparing LLM-based pipelines to much simpler
alternatives.

 \subsection{Retrieval-Augmented Generation} 
 RAG~\cite{lewis2020rag} was a large language model
 technique used extensively when LLMs
 had smaller context
 windows. To avoid exhausting the space
 inside a context window,
   queries to an LLM were 
augmented (on a query-by-query basis)
with a small number of relevant chunks
from (e.g.)  documents stored on disk. 

A RAG system has two halves: a post-query dialog generator and a
 search method for relevant documents. While EZR (as yet)
has no dialog generation method, we show below
that finding relevant documents is surprisingly easy using
a Naive Bayes trick built
around EZR.

\subsection{Yu et al. and FASTREAD}

In this section, we compare our results to prior work 
by
Yu et al.~\cite{yu2018total}.
That work
explored methods for 
finding relevant text documents.
They explored
``primary study selection'' for
systematic literature reviews (SLR)
in software engineering. 
That is, the goal here is 
to read
the research literature
to find the $M\ll N$
papers that are most relevant
to some research query.
 Reading
every candidate takes weeks or months~\cite{yu2017feature,yu2018finding}. 
To reduce the number of papers that must be read, Yu's tool 
(called FASTREAD) used a linear SVM with
some sampling tricks:
\begin{itemize}
\item 
When initializing his SVM classifiers for {\em positive examples},
they began with a small number of ``warm start'' papers offered by domain experts. 

\item To
find a set of initial {\em negative examples}, Yu used {\em presumptive sampling}:
  an equal number of randomly-selected papers, declared negative on
  the assumption that, when the target class is rare, a random draw
  is overwhelmingly unlikely to belong to it.
  \item After initialization, Yu et al.\ used their own active learning
  methods to incrementally decide what to sample next.
\end{itemize}
Their approach yielded a 95\% recall on Yu's four corpora after labeling
roughly \NEW{300--630} papers out of 1,700--8,900. 
Yu et al.\ report that that was a
20--50\% improvement over the prior state of the
art.

For the rest of this paper,  we  ask if EZR can perform the same task using, say,
Naive Bayes and the {\em best} and {\em rest} models
used in our active learning.  
In this study ``best'' means most relevant to the human conducting an SLR.

\subsection{Rennie et al. and Complementary Bayes}
Standard classifiers like Naive Bayes have problems with text mining. 
Rennie et al.~\cite{rennie2003tackling} diagnose the problem as one
of ``bursty'' low frequency counts inherent in the sparse data used to train text miners.
Relevancy filtering in text mining is a particularly challenging low frequency problem since, within 10,000s of documents, there may only be (say) 50 relevant documents;
i.e., we are asking a classifier to predict for a class with a one-in-200 frequency (or rarer). 
To solve this,
Rennie et al. 
inverted the normal classification algorithm:
\begin{itemize}
\item
When training their {\em complementary Bayes classifier} (CNB), a new row updates the \cls{Data} distributions
for all the classes to which it does {\em not} belong. 
\item
Then, when classifying, 
instead of
reporting what class is most likely,    {\em complementary Bayes}
reports the class that a row is least likely to {\em not} belong.
\end{itemize}
We inserted this complementary Bayes classifier into our active learner as follows.
Recall that 
the \textsf{acquire} function of line~\ref{acquiring} accepts a \textsf{score} function
that guesses which unlabeled row is more likely to be  {\em best}. The following relevancy filter  hooks into the likelihood calculation of complementary Bayes.
\par\nopagebreak\minipage{\linewidth}
\begingroup\mintfvset
\VerbatimPygments{\PYG}{\PYG}
\begin{mintbg}{codebg}
\begin{Verbatim}[commandchars=\\\{\},codes={\catcode`\$=3\catcode`\^=7\catcode`\_=8\relax}]
\PYG{k}{def}\PYG{+w}{ }\PYG{n+nf}{acquireWithBayes}\PYG{p}{(}\PYG{n}{data}\PYG{p}{,} \PYG{n}{best}\PYG{p}{,} \PYG{n}{rest}\PYG{p}{,} \PYG{n}{row}\PYG{p}{)}\PYG{p}{:}
  \PYG{n}{n} \PYG{o}{=} \PYG{n+nb}{len}\PYG{p}{(}\PYG{n}{best}\PYG{o}{.}\PYG{n}{rows}\PYG{p}{)} \PYG{o}{+} \PYG{n+nb}{len}\PYG{p}{(}\PYG{n}{rest}\PYG{o}{.}\PYG{n}{rows}\PYG{p}{)}
  \PYG{k}{return} \PYG{n}{likes}\PYG{p}{(}\PYG{n}{rest}\PYG{p}{,} \PYG{n}{row}\PYG{p}{,} \PYG{n}{n}\PYG{p}{,} \PYG{l+m+mi}{2}\PYG{p}{)} \PYG{o}{\PYGZhy{}} \PYG{n}{likes}\PYG{p}{(}\PYG{n}{best}\PYG{p}{,} \PYG{n}{row}\PYG{p}{,} \PYG{n}{n}\PYG{p}{,} \PYG{l+m+mi}{2}\PYG{p}{)}
\end{Verbatim}
\end{mintbg}
\endgroup
\endminipage\par\noindent
We tested if this active learner approach can solve
the same problems   explored by Yu et al.\ (see Table~\ref{slr}).
Yu's test cases  come from 
four real-world published SE SLRs:
\begin{itemize}
\item
Hall et al.'s work~\cite{hall2012} on fault
prediction;
\item
Wahono~\cite{wahono2015systematic}'s study  on defect prediction;
\item
Radjenovi\'c et al.'s study~\cite{radjenovic2013software} on software fault
metrics;
\item
Kitchenham et al.'s study~\cite{kitchenham2013systematic} 
tertiary SLR study.
\end{itemize}
In each case, these humans (all expert
in the field of SE) read thousands of papers to select a few dozen
relevant to their research. 
The data sets of Table~\ref{slr} were extracted from the careful 
logs
they kept of their reading process.

\begin{table}[!t]
\caption{Yu's four SLR corpora~\cite{yu2018total}. In each,
only a small minority of candidates
are relevant. The Kitchenham corpus has two relevance levels:
132 papers passed title-and-abstract review, of which 45
survived content review.}
\label{slr}
\begin{center}
\footnotesize
\begin{tabular}{lrr}
\hline
Dataset & \#Candidate & \#Relevant \\
\hline
Wahono       & 7002 & 62 \\
Hall         & 8911 & 104 \\
Radjenovi\'c & 6000 & 48 \\
Kitchenham   & 1704 & 132 \\
\hline
\end{tabular}
\end{center}
\end{table}

\begin{figure*}[!t]
\centering
\includegraphics[width=.9\linewidth]{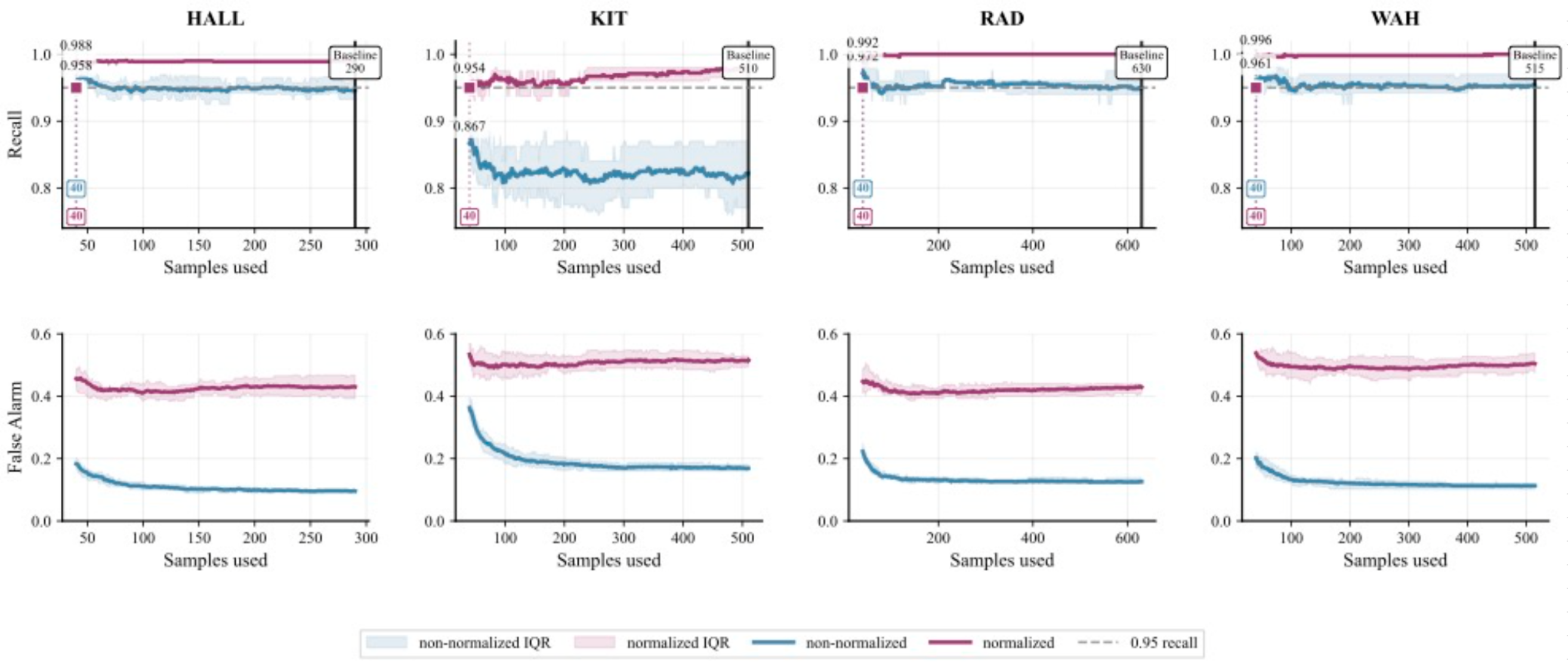}
\caption{Recall (top row) and false alarm (bottom row) as a
function of labels acquired, for non-normalized CNB (blue)
and normalized CNB (red) on Yu's four SLR corpora. \NEW{Warm-start
= 40 labeled samples for all runs (20 known-relevant + 20
presumptive negatives; the ``40'' marks).} Each curve is the median over 20 repeats,
with shaded IQR. Vertical bars on the recall panels mark the
labeling budget at which Yu et al.~\cite{yu2018finding} reached
the same recall with FASTREAD. Numerical labels at the left
edge of each recall panel mark the recall achieved
immediately after the warm start.}
\label{slrfig}
\end{figure*}
\subsection{Results}
To assess these results, we tracked
 two metrics:
 \begin{itemize}
 \item

\textit{Recall} is $\mathit{tp}/(\mathit{tp} + \mathit{fn})$:
the fraction of relevant papers found. 
{\em Higher} recalls are {\em better} since it means we are finding
more relevant papers.
  Yu's study achieved a recall of  95\%.
\item
\textit{False alarm} (sometimes called pf, the false positive
rate) is $\mathit{fp}/(\mathit{fp} + \mathit{tn})$: the fraction
of irrelevant papers incorrectly returned as relevant. 
{\em Lower} false alarms are {\em better} since, otherwise,
our human experts are wasting time reading irrelevancies.
Yu's study did not report false alarm.
\end{itemize}

\definecolor{slateblue}{RGB}{106,142,162}

\definecolor{ruby}{RGB}{177,98,143}

 Figure~\ref{slrfig} shows results of 20 trials where,
each time, the order of the training rows is shuffled
(in this case, each ``row'' was a paper's title and abstract).
Each run was initialized with a warm start of \NEW{40 labeled
papers: 20 known-relevant papers plus 20 presumptive negatives
(randomly drawn and declared negative without reading, per the
presumptive sampling described above). The choice of 20
positives} was informed by asking academics in our field ``for any particular area, how many dozen papers do you usually
know?''

After that,
complementary Bayes selected one paper at a time (the unlabeled paper
it ranked most strongly as ``yes''), and recall and false alarm were
updated after each new label. \NEW{The x-axis of
Figure~\ref{slrfig} counts total labeled samples (the 40
warm-start labels plus one per acquisition), and each run
continues to the number of labels Yu et al.\ spent on that
corpus (the ``Baseline'' mark at the right of each panel:
290--630).}

In  Figure~\ref{slrfig}, 
solid lines show the 50\% percentile, and the shaded
regions show the 25th to 75th range:
\begin{itemize}
\item
The four columns show results from our four 
corpora.
\item
The top row
of plots shows recall.
\item
The second row shows  false alarm rates
\item The 
 \color{ruby}{\bf red lines}\color{black}~ shows results using
 Rennie et al.'s recommendation of normalizing a likelihood
 by dividing it by the sum of all likelihoods.
 \item
 The  \color{slateblue}{\bf blue lines}\color{black}~
 shows results with disabled normalization.
 \end{itemize}
Note that the false alarm rate is far lower
when normalization is disabled; i.e., that part of
the Rennie et al.\ result should not be applied
uncritically to new data.
Clearly,  measuring {\em both} recall and false
alarms is informative for relevancy filtering.
 Yu et al.\ did not measure false alarms, and we would now
 call that a methodological error.
  
The most striking result is the early plateau. High recall
and low false alarm are achievable in all four datasets
after 100 labels or fewer, where Yu et al.\ required
\NEW{300--630} labels to reach the same recall.

The Kitchenham result deserves an explicit caveat. On that
corpus our recall plateaus near 82\%, below Yu's 95\%
target. This is a genuine tradeoff, not a win:
complementary Bayes converges faster \NEW{(after just the
40-label warm start, recall is already 87\%, settling to 82\%;
i.e., 108 of the 132 relevant papers)} but caps at a lower
ceiling than FASTREAD eventually reaches. Which side of
this tradeoff matters depends on the use case. {\em
Faster convergence at lower ceiling} is the right
preference when (a)~the labeling effort is the binding
constraint (e.g.\ a domain expert can read only a few dozen
papers before deadline), (b)~the SLR conductor wants a
quick first cut to refine the query, or (c)~partial recall
in days is more valuable than complete recall in weeks.
{\em Higher ceiling at higher cost} is the right preference
when the SLR is part of a regulatory submission, a
systematic medical review, or any setting where missing a
relevant paper is more expensive than reading an irrelevant
one. For those cases, FASTREAD remains the better choice.

On the other three corpora (Hall, Wahono, Radjenovi\'c) no
such ceiling effect appears: complementary Bayes converges
faster {\em and} matches or exceeds Yu's recall. The
Kitchenham corpus differs structurally: its 132 relevant
papers are split into two relevance levels (132 passed
title-and-abstract review; 45 survived content review),
which thins the positive signal and may be what bounds the
Naive Bayes ceiling. A reviewer who needs near-100\%
recall on a Kitchenham-like corpus should expect to pay
more labels than complementary Bayes alone requires.

In summary, compared to Yu's methods, EZR+complementary Bayes:
\begin{itemize}
\item Needs fewer labels (100 vs.\ \NEW{300--630});
\item Uses a simpler classifier (Naive Bayes rather than a
  linear SVM);
\item Uses a smaller codebase (based on EZR, complementary
Bayes is just three dozen lines of code);
\item Reports both recall and false alarm, which exposed the
  finding that one of Rennie's recommended preprocessing
  steps is counterproductive in this domain.
\end{itemize}

That last item carries the methodological lesson of this
section. Yu's data was fine, but the choice of metrics was
incomplete, as revealed by an investigation with simpler tools
(here, \IT{}).
We suspect that
similar gaps exist elsewhere in the empirical SE literature,
masked by tools too slow to encourage re-checks. Compact
toolkits like EZR make those re-checks cheap, and we suggest
the community should run them.

A broader point bears on the main argument of this paper.
Relevancy filtering in RAG is increasingly framed as a task for large
language models, using embedding spaces of billions of
parameters and similarity queries against a vector database.
Our experiment achieved comparable or better results with 30
lines of complementary Bayes riding on the same
\cls{Num}/\cls{Sym}/\cls{Data} substrate used throughout
EZR. We wonder how many other methods from the LLM community could be greatly simplified by simpler methods.



\section{Threats to Validity}\label{sec:ttv}
Feldt et al.~\cite{feldt2010validity} note that a critical
element of any empirical study is to analyze and mitigate
threats to the validity of the results. We therefore consider
the threats that we have alleviated, followed by the external,
construct, internal, and conclusion validity of this work.

\subsection{Alleviated Threats}
Two threats from our prior work have been mitigated here.
First, our earlier \IT{} experiments drew on a narrow MOOT
subset; in this paper, \NEW{the active-learning results of
\S\ref{sec:active} are} reported across all 124
tasks of Table~\ref{mootdata}, spanning configuration tuning,
performance prediction, defect prediction, test selection, and
cost estimation\NEW{; the optimizer study of \S\ref{sec:opt}
uses 20 randomly drawn MOOT tasks, and the text-mining study of
\S\ref{sec:textmine} uses Yu et al.'s four SLR corpora, which
are not MOOT tasks}. Second, prior text-mining
studies~\cite{yu2018total} reported only recall. Following
that gap, we report both recall {\em and} false alarm in
\S\ref{sec:textmine}, which surfaced the finding that
Rennie's recommended normalization step
\cite{rennie2003tackling} inflates false alarm on SLR data
and should not be applied uncritically.

\subsection{External Validity}
External validity concerns whether our results generalize
beyond the scope of this study.

\textbf{Sampling bias.} The six counter-findings of
Table~\ref{truisms} were derived from MOOT, which holds 124
tabular SE tasks with explicit $x$ and $y$ columns where $y$
is expensive to obtain. Drawn from ICSE, FSE, TSE, IST, EMSE,
TOSEM, and ASE, MOOT is broad but not universal;
i.e. the 
conclusions reached here may not transfer to:
\begin{itemize}
\item generation tasks (translation, summarization, code
synthesis), where LLM-scale data is genuinely required;
\item perception tasks (vision, speech), where representation
learning dominates;
\item safety-critical certification, where formal methods
rather than empirical optimization are the relevant tool.
\end{itemize}
To mitigate this threat to validity,
all our methods, code, and raw experimental logs
are public:
\IT{} source at
\url{https://github.com/timm/ezr};
this paper's reproduction artifacts at
\url{https://github.com/timm/bitstex/blob/main/data/runs.log}.
\NEW{The version of \texttt{ezr.py} studied in this paper is
release \textsf{v0.9.4}, archived on Zenodo under the
version-specific DOI \textsf{10.5281/zenodo.22554751}; it holds
less than 1000 lines across two files (a core \texttt{ezr.py}
plus \texttt{cli.py} demos and self-tests; see
Appendix~\ref{sec:repro}). The line numbers in our listings
trace to the earlier single-file release \textsf{v0.5.0} (412
lines, 44 functions; DOI \textsf{10.5281/zenodo.17118009}),
whose functions carry forward into \textsf{v0.9.4}. The concept
DOI \textsf{10.5281/zenodo.11183058} resolves to all archived
releases.} \IT{} requires Python
$\geq$~3.12 and has no dependencies beyond the standard library.
Other researchers can repeat, refute, or extend our results.

\textbf{Researcher bias.} The two replication studies cited
throughout this paper (Rayegan et al.~\cite{Amiraliminimaldata};
Ganguly et al.~\cite{ganguly2026lowgodatalightse}) are both
co-authored by \NEW{authors of this paper}, as is the MOOT
benchmark~\cite{menzies2025moot} on which all experiments run.
These works therefore corroborate our results but are not
independent evidence; the abstract and \S\ref{sota} attribute
them accordingly. Genuinely independent replication remains
open, and all artifacts needed for it are public (see below).

\textbf{Learner bias.} Any single study can examine only a subset
of available algorithms. \IT{} implements Naive Bayes,
$k$-means, $k$-means++, classification/regression trees,
simulated annealing, local search, and complementary Bayes.
Other learners (random forests, gradient-boosted trees,
neural networks, transformer-based predictors) may shift the
comparisons.
To mitigate this threat to validity,
as per the No-Free-Lunch
theorem~\cite{wolpert1997nfl} (no learner dominates
universally) we make no such claim for \IT{}. Rather, we frame
\IT{} as a strong, simple baseline against which
heavyweight alternatives can be assessed.

\textbf{Comparison bias.} It is infeasible to compare against every
heavyweight alternative. Our choice of comparators may have
favored \IT{} inadvertently.
To mitigate for this threat to validity,
our comparators were selected from those
that prior studies in the same domain named as
representative: SMAC3 and TPE for
optimization~\cite{hutter2011sequential,bergstra2011tpe};
SHAP, LIME, Anchors, and ReliefF for
explanation~\cite{lundberg2017unified,ribeiro2016should,
ribeiro2018anchors,robnik1997adaptation}; FASTREAD for SLR
relevance filtering~\cite{yu2018total}. Other comparators
(HyperOpt, Optuna, BoTorch) remain untested.

\subsection{Construct Validity}
Construct validity concerns whether our chosen metrics
measure what we claim they measure.

\textbf{Evaluation bias.} Our scoring rule (Equation~\ref{eq:wins})
is a normalized regret measure that clamps trivially small
differences \mbox{($<0.35\sigma$)} to zero, following
Sawilowsky~\cite{sawilowsky2009new}. Alternative metrics
(area under the regret curve, Pareto hypervolume,
$\epsilon$-indicator, raw $y$-distance to heaven) could
reorder some treatments.
To mitigate this threat, for text mining, we report both recall
{\em and} false alarm. Prior
work~\cite{yu2018total} reported only recall. Replications
by Rayegan et al.~\cite{Amiraliminimaldata}
and Ganguly et al.~\cite{ganguly2026lowgodatalightse}
(both co-authored by \NEW{authors of this paper}; see ``Researcher
bias'' above) use different methodologies and reach compatible
conclusions, which mitigates (but does not eliminate) this
threat.

\subsection{Internal Validity}
Internal validity asks whether alternative causes might
explain our observations.

\textbf{Encoding bias.} \IT{} was implemented by the authors, and
any implementation may contain bugs.
To mitigate this threat,
our code base is small \NEW{(under 1000 lines)} and
publicly available
(\url{https://github.com/timm/ezr}). Two replication
studies~\cite{Amiraliminimaldata,ganguly2026lowgodatalightse}
(co-authored by \NEW{authors of this paper}) have re-run \IT{}'s
algorithms against their own pipelines and reached compatible
conclusions, which bounds (but, given the shared authorship,
does not eliminate) the probability that our results are an
artifact of encoding error.

\textbf{Surrogate bias.} The (1+1) optimizer experiments of
\S\ref{sec:opt} use a 50-row nearest-neighbor surrogate as
oracle, following Pfisterer et
al.~\cite{pfisterer2022yahpo} and Zela et
al.~\cite{zela2020surrogate}. The surrogate is itself an
estimate; its accuracy bounds the accuracy of every
treatment ranking we report. Two consequences:
\begin{itemize}
\item Treatments whose true performance differs by less
than the surrogate's noise floor cannot be reliably
distinguished by our experiment.
\item A more accurate oracle (full ground truth, larger
hold-out set, a model trained per-task) could move
individual cells in Table~\ref{tab:opt}.
\end{itemize}
To mitigate this threat, we use a nearest-neighbor surrogate
that is the de facto standard for such
studies~\cite{pfisterer2022yahpo, zela2020surrogate} (our
setup matches theirs). The 50-row known set is held constant
across all treatments in each run, so any surrogate bias
affects all treatments equally within a comparison.
\NEW{Evidence that the surrogate is adequate is in
Table~\ref{tab:opt} itself: every \textsf{win} score is
computed on the {\em true} \textsf{disty} of the row the
search finally selected, not on the surrogate. In the clamped
panel~(a), \textsf{sa-r} reaches a mean win of 98 out of 100
across the 20 randomly selected MOOT tasks and lands on 100 for
13 of them. Under the unclamped scoring of panel~(b) no cell
sits at 100 and the \textsf{sa-r} mean is 94, so this evidence
is suggestive rather than decisive: a surrogate that drives the
search to within a few points of the reference optimum on most
tasks is doing its job, but panel~(b) shows that optimum is
approached, not exactly attained.} 
\subsection{Conclusion Validity}
Conclusion validity asks whether our statistical treatment
supports the conclusions drawn. We report medians and
standard deviations across repeats, with Sawilowsky's
$0.35\sigma$ clamp suppressing trivial effects. We did not
apply per-cell significance tests (Wilcoxon, Cliff's delta).
A future extension would tighten this with effect-size
reporting per Arcuri and
Briand~\cite{arcuri2011practical}.

\section{Closing remarks}\label{sec:disc}

This paper set out to defend the claims
that (a)~reading code can find simplifications
to that code and (b)~new problems need not always be
tackled with more complex algorithms and more data. We offered
\NEW{less than 1000 lines} of Python that:
\begin{itemize}
\item Built a simple substrate (\cls{Data}, \cls{Num},
\cls{Sym}, and \textsf{add}) in \S\ref{sec:peek};
\item Used that substrate to implement, in just a few lines,
Naïve Bayes, $k$-means, and $k$-means++ (\S\ref{sec:famous});
\item Then classification and regression trees
(\S\ref{trees});
\item Then two optimizers, simulated annealing and local
search (\S\ref{sec:opt});
\item Then an active learner (\S\ref{sec:active});
\item And a text-mining relevance filter
(\S\ref{sec:textmine}).
\end{itemize}

Along the way we found that:
\begin{itemize}
\item In its 1983 default configuration, simulated annealing
can beat 1990s local-search variants, despite a decade of
papers claiming the newer variants had surpassed it
(\S\ref{sec:opt});
\item Our active learner reached the reference optimum with
fewer than 100 labels and fewer than 10 features (even when
thousands of features were available); a companion study
co-authored by \NEW{authors of this paper} reports its model update running
\NEW{over 400 times} faster than SMAC3~\cite{ganguly2026lowgodatalightse}
(\S\ref{sec:active});
\item Text-mining tasks that are now routinely handed to
massive language models can be implemented by 30 lines of
complementary Bayes on the substrate above, and the same
experiment exposed a measurement gap that the prior
heavyweight evaluation had missed (\S\ref{sec:textmine}).
\end{itemize}
These are not improvements at the margin; they are reductions
of one to three orders of magnitude in code, dependencies,
labels, and compute, along each of the ten axes of
Table~\ref{tab:light}.

We obtained these results by reading code. \IT{} grew across
several years of looking at AI toolkits, noticing similarities,
and deleting what those similarities made redundant. This
methodological position matters because it is now under attack.
An LLM asked separately for Naive Bayes, $k$-means, and
$k$-means++ produces three independent implementations with
three separate data structures and three separate maintenance
burdens. The substrate-sharing pattern that collapses them into
30 lines only becomes visible when a human reads the three
implementations side by side. The opportunity cost of skipping
that reading is invisible at the time of skipping and pays out
later as needless complexity.

What does this imply beyond tabular SE optimization? The
abstract of this paper deliberately scoped its claims to
that domain, because that is where we ran experiments and
where the comparators (SMAC, SHAP, FASTREAD) live. The
broader speculation is that AI's next step might not be
larger systems but smaller, unified ones, compact enough to
be configured, critiqued, taught, and maintained by their
users. We do not prove this here; we only note that, in the
one corner of AI we examined closely, the smaller-and-unified
approach matched or beat the larger-and-specialized one
along all ten axes of Table~\ref{tab:light}. Whether the
same pattern holds for generation, perception, or
safety-critical certification is an open question; the
methods of \IT{} (read, refactor, measure against simple
baselines) are at least worth trying.

The reverse statement to our results is also worth making.
The SE empirical literature contains many examples where the
published advantage of a heavyweight method shrank or
disappeared once re-tested against a carefully built simple
baseline~\cite{Fu17,fu2016tuning,agrawal2019dodge,Nair17a,
tantithamthavorn2016automated,grinsztajn2022why,Hou24}. \IT{}
is one more entry in that pattern. We invite the same scrutiny
applied to \IT{} itself. All our artifacts are public at
\url{https://github.com/timm/ezr} and
\url{https://github.com/timm/bitstex}; we would readily support any and all
refutation attempts.

For practitioners, the immediate implication is procedural:
before deploying a heavyweight pipeline, run a simple baseline
first. If the baseline matches the heavyweight on your data,
the heavyweight is technical debt. If it does not, you have
learned something specific about what your problem requires
that no general benchmark could have told you. Either outcome
is a win that costs hours, not weeks.

In conclusion, just because some tasks are hard does not mean
all tasks are hard. The challenge we put to the community is
this: have we really checked what is genuinely complex, and
what is, on careful reading, actually very simple?

\appendix

\section{Reproduction notes}\label{sec:repro}

This appendix offers the fastest route to checking that this
paper's code runs. The commands below were tested September
2026 on macOS (Apple Silicon); total time, including
installation, is under two minutes.

Install the latest \IT{} from PyPI (Python $\geq$~3.12), then
fetch the MOOT data:
\par\nopagebreak\minipage{\linewidth}
\begingroup\mintfvset
\VerbatimPygments{\PYGD}{\PYGD}
\begin{mintbg}{codebg}
\begin{Verbatim}[commandchars=\\\{\},codes={\catcode`\$=3\catcode`\^=7\catcode`\_=8\relax}]
pip\PYGD{+w}{ }install\PYGD{+w}{ }ezr
git\PYGD{+w}{ }clone\PYGD{+w}{ }http://github.com/timm/moot
\end{Verbatim}
\end{mintbg}
\endgroup
\endminipage\par\noindent
(This fetches \textsf{v0.9.4}, the version studied here,
archived under DOI \textsf{10.5281/zenodo.22554751}; the
line numbers in our listings trace to the earlier single-file
release \textsf{v0.5.0}, DOI
\textsf{10.5281/zenodo.17118009}, per \S\ref{sec:ttv}.)

For a fast smoke test, skip the slower experiments (the
active-learning sweep over all 124 tasks, and the text mining of
\S\ref{sec:textmine}) and run the per-task demos, each of which
finishes in well under a second. In the following, the shell
variable \textsf{D} can point to any csv file in the
\textsf{moot/optimize} subtree; here we
set it to the Apache task of Table~\ref{mootdata}:
\par\nopagebreak\minipage{\linewidth}
\begingroup\mintfvset
\VerbatimPygments{\PYGD}{\PYGD}
\begin{mintbg}{codebg}
\begin{Verbatim}[commandchars=\\\{\},codes={\catcode`\$=3\catcode`\^=7\catcode`\_=8\relax}]
\PYGD{n+nv}{D}\PYGD{o}{=}moot/optimize/config/Apache\PYGDZus{}AllMeasurements.csv
ezr\PYGD{+w}{ }tree\PYGD{+w}{ }\PYGD{n+nv}{\PYGDZdl{}D}\PYGD{+w}{      }\PYGD{c+c1}{\PYGDZsh{} regression tree (cf. Fig. 2)}
ezr\PYGD{+w}{ }search\PYGD{+w}{ }sa\PYGD{+w}{ }\PYGD{n+nv}{\PYGDZdl{}D}\PYGD{+w}{ }\PYGD{c+c1}{\PYGDZsh{} simulated annealing (S5)}
ezr\PYGD{+w}{ }acquire\PYGD{+w}{ }\PYGD{n+nv}{\PYGDZdl{}D}\PYGD{+w}{   }\PYGD{c+c1}{\PYGDZsh{} active learning (S6)}
\end{Verbatim}
\end{mintbg}
\endgroup
\endminipage\par\noindent
(From a repository checkout, without \textsf{pip}, the same
commands run as \mbox{\textsf{python3 -B cli.py tree \$D}},
etc.; \textsf{ezr.py} itself is a library with no command-line
interface.)
The tree command prints a tree such as:
\par\nopagebreak\minipage{\linewidth}
\begingroup\mintfvset
\VerbatimPygments{\PYGD}{\PYGD}
\begin{mintbg}{codebg}
\begin{Verbatim}[commandchars=\\\{\},codes={\catcode`\$=3\catcode`\^=7\catcode`\_=8\relax}]
                    ,0.47 ,(192), \PYGDZob{}Performance\PYGDZhy{}=1427.66\PYGDZcb{}
keepAlive == 0      ,0.24 ,( 96), \PYGDZob{}Performance\PYGDZhy{}=1087.19\PYGDZcb{}
\PYGD{esc}{  hostnameLookups != 0 ,0.17 ,( 48), {Performance-=973.75}}
  \PYGD{err}{|}  accessLog != 0    ,0.15 ,( 24), \PYGDZob{}Performance\PYGDZhy{}=928.75\PYGDZcb{}
\end{Verbatim}
\end{mintbg}
\endgroup
\endminipage\par\noindent
the search command prints an energy trace such as:
\par\nopagebreak\minipage{\linewidth}
\begingroup\mintfvset
\VerbatimPygments{\PYGD}{\PYGD}
\begin{mintbg}{codebg}
\begin{Verbatim}[commandchars=\\\{\},codes={\catcode`\$=3\catcode`\^=7\catcode`\_=8\relax}]
 evals  energy
     1    0.92
     2    0.91
     4    0.30
    16    0.11
\end{Verbatim}
\end{mintbg}
\endgroup
\endminipage\par\noindent
and the acquire command prints the best rows found under a
50-label budget:
\par\nopagebreak\minipage{\linewidth}
\begingroup\mintfvset
\VerbatimPygments{\PYGD}{\PYGD}
\begin{mintbg}{codebg}
\begin{Verbatim}[commandchars=\\\{\},codes={\catcode`\$=3\catcode`\^=7\catcode`\_=8\relax}]
:budget 50 :check 5
  :win 100 :d2h 0.162 [1, 1, 0, 0, 1, 1, 1, 0, 0, 840]
  :win 100 :d2h 0.179 [1, 1, 0, 1, 1, 1, 1, 0, 1, 870]
\end{Verbatim}
\end{mintbg}
\endgroup
\endminipage\par\noindent
Finally, \textsf{ezr test\_all} runs the toolkit's entire
self-test suite (core primitives, trees, clustering, search,
acquisition, classification, text mining, statistics); on our
machine this takes about eight seconds and ends with
\textsf{Done. fails=0}.

Due to platform differences (random number generators, floating
point, Python version), the numbers printed on other machines
may differ slightly from those shown above; what should be
stable is the shape of the output: trees of that form, energy
traces that decrease, and \textsf{win} scores near 100 on this
task.

\bibliography{bib3,main}

\end{document}